\documentclass[manuscript,screen,nonacm]{acmart}
\AtBeginDocument{%
  }

\setcopyright{acmlicensed}
\copyrightyear{2018}
\acmYear{2018}
\acmDOI{XXXXXXX.XXXXXXX}
\acmConference[Conference acronym 'XX]{Make sure to enter the correct
  conference title from your rights confirmation emai}{June 03--05,
  2018}{Woodstock, NY}
\acmISBN{978-1-4503-XXXX-X/18/06}

\newcommand{\repo}{\url{https://tinyurl.com/AttitudesAffirmativeAlgorithms}}

\newcommand{\ftest}[3]{$F$(#1,#2) = #3}
\newcommand{\pvalue}[1]{$p $ \textless #1}

\newcommand{\etasq}[1]{$\eta_{p}^{2}$ = #1}
\newcommand{\msd}[2]{$M$ = #1, $SD$ = #2}

\newcommand{\csq}[2]{$\chi^2$(#1) = #2}

\usepackage{enumitem}
\usepackage{titlesec}
\usepackage{csquotes}
\usepackage{multirow}

\setlist[itemize]{left=0.3cm, parsep=0pt, topsep=0.5pt}

\titlespacing*{\section}{0pt}{*0.6}{*0.6}
\titlespacing*{\subsection}{0pt}{*0.6}{*0.6}
\titlespacing*{\subsubsection}{0pt}{*0.6}{*0.6}

\SetBlockEnvironment{displayquote}
\renewenvironment{displayquote}
  {\begin{quote}\itshape}
  {\end{quote}}

\begin{document}

\title{Laypeople's Attitudes Towards Fair, Affirmative, and Discriminatory Decision-Making Algorithms}

\author{Gabriel Lima}
\affiliation{%
  \institution{Max Planck Institute for Security and Privacy}
  \country{Germany}}
\email{gabriel.lima@mpi-sp.org}

\author{Nina Grgi\'{c}-Hla\v{c}a}
\affiliation{%
  \institution{Max Planck Institute for Software Systems \& Max Planck Institute for Research on Collective Goods}
  \country{Germany}
}
\email{nghlaca@mpi-sws.org}

\author{Markus Langer}
\affiliation{%
  \institution{University of Freiburg}
  \country{Germany}}
\email{markus.langer@psychologie.uni-freiburg.de}

\author{Yixin Zou}
\affiliation{%
  \institution{Max Planck Institute for Security and Privacy}
  \country{Germany}}
\email{yixin.zou@mpi-sp.org}


\renewcommand{\shortauthors}{Lima et al.}

\begin{abstract}

\emph{Affirmative algorithms} have emerged as a potential answer to algorithmic discrimination, seeking to redress past harms and rectify the source of historical injustices. We present the results of two experiments ($N$$=$$1193$) capturing laypeople's perceptions of affirmative algorithms---those which explicitly prioritize the historically marginalized---in hiring and criminal justice. We contrast these opinions about affirmative algorithms with folk attitudes towards algorithms that prioritize the privileged (i.e., discriminatory) and systems that make decisions independently of demographic groups (i.e., fair). We find that people---regardless of their political leaning and identity---view fair algorithms favorably and denounce discriminatory systems. In contrast, we identify disagreements concerning affirmative algorithms: liberals and racial minorities rate affirmative systems as positively as their fair counterparts, whereas conservatives and those from the dominant racial group evaluate affirmative algorithms as negatively as discriminatory systems. We identify a source of these divisions: people have varying beliefs about who (if anyone) is marginalized, shaping their views of affirmative algorithms. We discuss the possibility of bridging these disagreements to bring people together towards affirmative algorithms.

\end{abstract}

\ccsdesc[500]{Applied computing~Psychology}
\ccsdesc[500]{Human-centered computing~Empirical studies in HCI}

\keywords{Affirmative Algorithms, Algorithmic Fairness, Systemic Injustice, Algorithmic Discrimination}

\maketitle

\section{Introduction}
Research has revealed numerous instances of algorithms deployed in high-risk domains discriminating against minority groups~\cite{reutersamazon,propublicastory,politicodutch,markup2021crime,obermeyer2019dissecting}. These findings have prompted research on \emph{algorithmic fairness}, which seeks to make algorithms more fair and thus mitigate these potential harms by training algorithms to satisfy fairness criteria. Yet, some scholars argue (and have shown~\cite{zhang2024structural,jorgensen2023not,lee2024algorithms}) that this prevailing approach to algorithmic fairness is incomplete because it fails to address the underlying source of the discrimination that algorithms replicate and perpetuate~\cite{hong2023prediction,zimmermann2022proceed,fazelpour2022algorithmic,lin2022artificial,green2022escaping,hanna2020towards,weinberg2022rethinking}.

Many have instead argued for \emph{affirmative algorithms}~\cite{minow2023equality,ho2020affirmative,theus2023striving, davis2021algorithmic}. Proponents posit that algorithms should acknowledge that they are embedded in contexts marked by systemic injustice and thus actively address and rectify the underlying source of these injustices. For instance, instead of attempting to equalize outcomes across demographic groups, affirmative algorithms could be optimized to redress historically discriminatory policies~\cite{so2022beyond,so2023race} and use different decision thresholds for marginalized groups~\cite{zhang2022affirmative}.

We present the results of two experiments ($N$$=$$1193$) capturing people's perceptions of affirmative algorithms (i.e., those which explicitly prioritize historically marginalized groups) deployed to assist humans making hiring and bail decisions. Furthermore, we contrast people's opinions about affirmative algorithms with lay attitudes towards decision-making systems that favor the historically privileged (i.e., discriminatory) and algorithms that make decisions independently of individuals' demographic groups (i.e., fair in the sense of not showing disparate treatment).

Considering that affirmative algorithms are only justifiable when society is systemically unjust, we also explore whether explicit telling participants that the domain is unjust impacts their opinions about affirmative algorithms. Finally, inspired by work showing that people disagree about affirmative action depending on their political beliefs~\cite{sidanius1996racism,clawson2003support} and identity~\cite{clawson2003support,klineberg2003ethnic,golden2001reactions,allen2003examining,golden2001reactions,kravitz1993attitudes,truxillo2000roles}, we investigate if similar disagreements emerge in the case of affirmative algorithms, focusing on potential divisions across political learning and participants' racial and gender identity.

We find that people---no matter their political leaning and racial and gender identity---rate fair algorithms positively and denounce discriminatory algorithms that prioritize historically privileged groups. In contrast, we observe diverging opinions concerning affirmative algorithms. Liberal participants preferred affirmative algorithms in the hiring domain, viewing them as positively as fair algorithms, whereas conservatives considered them as unfair as discriminatory systems. In the bail decision-making domain, we observe that those belonging to racial minority groups favor affirmative algorithms as much as their fair counterparts, while members of the dominant racial group evaluate affirmative systems as negatively as discriminatory algorithms.

We also identify a potential source of these disagreements: people have varying attitudes towards affirmative algorithms because they differ in who---if anyone---they consider historically marginalized. We also discuss how our contrasting results between decision-making domains may emerge due to the identity axes most salient in each context: debates concerning injustice in criminal justice often focus on race~\cite{hinton2018unjust,michelle2010new}, whereas hiring discrimination largely considers multiple identity axes~\cite{rosette2018intersectionality,crenshaw2013mapping,mishel2016discrimination,sears2024lgbtq}.

Our findings show that individuals care about algorithmic fairness, although they disagree on what fairness actually entails. While some acknowledge that certain groups have been historically marginalized and thus believe that algorithms could prioritize them in certain contexts, others denounce the idea as another form of discrimination. Alongside our result that explicitly telling people that the decision-making contexts are unfair have no effect on opinions concerning affirmative algorithms, we discuss whether it is possible to bridge the observed disagreements concerning affirmative algorithms if society is to strive for ``affirmative algorithmic futures''~\cite{theus2023striving} that remedy past harms and reform the unjust structures that algorithms surface and perpetuate.

\section{Background}
\subsection{Critiques of Algorithmic Fairness}

Algorithmic decision-making has become pervasive in many high-risk domains. Algorithms can, for instance, assist humans decide who should be hired~\cite{reutersamazon}, granted bail~\cite{propublicastory}, or investigated for fraud~\cite{politicodutch}. Although some argue algorithms can make decision-making more efficient and impartial~\cite{lepri2018fair}, there exist numerous instances in which algorithms were found to cause harm against minority groups. Reports indicate that hiring algorithms discriminate against women~\cite{reutersamazon}; disproportionately deny social benefits to ethnic minorities~\cite{politicodutch}; underdiagnose health risks to racial minorities~\cite{obermeyer2019dissecting}; and overestimate the likelihood that members of marginalized groups will commit crimes~\cite{propublicastory}.

These examples of how algorithms can be discriminatory have fueled research on \emph{algorithmic fairness}~\cite{hardt2016equality,grgic2018human,kusner2017counterfactual,zafar2017disp_mistreat,zafar2017disp_impact}. This line of work largely relies on defining fairness criteria and training algorithms to satisfy them~\cite{green2022escaping,weinberg2022rethinking}. There exist several fairness criteria~\cite{mitchell2021algorithmic,pessach2022review}, many of which are computationally incompatible with each other~\cite{chouldechova2018case,kleinberg2016inherent}.

Although computational algorithmic fairness has been the primary strategy to make algorithms more fair, scholars have argued---and shown~\cite{zhang2024structural,jorgensen2023not,lee2024algorithms}---that satisfying specific fairness criteria is not enough to prevent algorithmic discrimination. Critics of the approach often argue that algorithms should instead be ``affirmative''~\cite{zhang2022affirmative,minow2023equality,ho2020affirmative} or ``reparative''~\cite{davis2021algorithmic}. Proponents of affirmative algorithms posit that because society is systemically and structurally unjust against specific groups, fair algorithms can only obscure background injustice since they do not directly address the underlying issues that lead to the disproportionate outcomes algorithms surface~\cite{hong2023prediction,zimmermann2022proceed,zhang2022affirmative}. By acknowledging that algorithms are embedded within structural injustice~\cite{fazelpour2022algorithmic,lin2022artificial}, decision-making systems can be designed to account for the fact that the context in which they are deployed is unjust~\cite{green2022escaping,hanna2020towards,weinberg2022rethinking} and thus strive for ``affirmative futures''~\cite{theus2023striving} by correcting for these injustices.

Recent work demonstrates how affirmative algorithms could be implemented in the real world. Systems designed to assess risk in the criminal justice domain could lower risk scores for members of demographic groups that have been historically overpoliced~\cite{zhang2022affirmative}. In the context of housing, affirmative algorithms could be used to estimate the cost of reparations~\cite{so2022beyond} and optimize lending programs to redress historically discriminatory housing policies~\cite{so2023race}. Algorithms that predict cancer risk could also include demographic variables to adjust for historical underdiagnosis in minority groups, increasing accuracy and screening access~\cite{zink2024race}.

\subsection{People's Perceptions of Decision-Making Algorithms}

Another research agenda focuses on descriptive notions of algorithmic fairness, studying what people judge to be fair when algorithms are deployed~\cite{starke2022fairness}. By capturing people's perceptions of algorithms, one can ensure that their development and regulation are aligned with lay expectations~\cite{awad2020crowdsourcing,rahwan2018society}. This alignment is particularly relevant in the context of fairness, a domain that requires negotiation between several competing values and stakeholders~\cite{wong2020democratizing}.

Prior work has captured people's perceptions and expectations of algorithmic fairness in several high-risk domains~\cite{lee2018understanding, langer2019highly,grgic2018beyond, srivastava2019mathematical, harrison2020empirical,grgic2018human,plane2017exploring,kasinidou2021agree}. 
All in all, research suggests that perceptions of algorithmic fairness vary significantly across different domains~\cite{hannan2021gets,smith2020exploring} (see~\citet{starke2022fairness} for a review). Particularly relevant to our current study, prior work has explored opinions concerning \emph{fair} algorithms. In the context of loan decisions, people differentiate between different notions of fairness while showing a preference for decisions that account for each applicant's payback rate~\cite{saxena2019fairness,kasinidou2021agree}. \citet{cheng2021soliciting} found that laypeople prefer algorithms that equalize both true positive and false positive rates odds in the context of child maltreatment prediction, and~\citet{harrison2020empirical} identified a preference for bail decision-making algorithms that equalize false positive rates across different demographic groups.

The literature described above focuses on people's opinions concerning algorithmic decision-making and attempts to make it more equal across groups in several domains. In this paper, we explore people's perceptions of algorithms that do not try to equalize outcomes across groups but instead explicitly prioritize those who have been historically marginalized---what some would call affirmative algorithms~\cite{zhang2022affirmative,minow2023equality,ho2020affirmative}. 

We capture laypeople's perceptions of affirmative algorithms---a gap largely unexplored in prior research---and contrast them with lay opinions about systems that prioritize privileged groups and algorithms that equalize outcomes between different groups. More specifically, we explore how people judge an algorithm that prioritizes those who have been historically marginalized (hereafter, affirmative) and compare these views to lay opinions concerning systems that favor the historically privileged (discriminatory) or treat everyone equally irrespective of their demographics (i.e., fair with respect to disparate treatment). 

\subsection{Affirmative Algorithms as Attempts to Rectify Past Harms}

Affirmative algorithms aim to redress past and current harms towards individuals who have been and continue to be discriminated against. To justify their development and deployment, one must thus acknowledge that certain groups of people are systemically marginalized and more susceptible to harm, with affirmative algorithms aiming to compensate for these injustices~\cite{zhang2022affirmative,minow2023equality,ho2020affirmative}. Individuals must thus recognize that the domain in which the algorithm is deployed is unjust to support the deployment of affirmative algorithms. Otherwise, they might view an affirmative algorithm as another instance of a discriminatory system. Building upon calls for acknowledging that decision-making algorithms are deployed in domains marked by injustice~\cite{zhang2022affirmative,green2022escaping,fazelpour2020algorithmic,lin2022artificial}, we also explore how prompting individuals to think about the fairness of a particular domain---and telling them that the domain is systematically unfair---impacts their views of algorithms deployed in the same domain.

\subsection{Potential Disagreements in Perceptions of Fair and Affirmative Algorithms}

Affirmative algorithms can be seen as an algorithmic implementation of affirmative action, which refers to policies and practices that aim to increase participation of disadvantaged groups by, for instance, using group membership as selection criteria~\cite{anderson2010imperative}. Studies have explored people's attitudes towards affirmative action (see~\citet{crosby2006understanding} for a review), the results of which suggest that individuals may disagree on whether affirmative algorithms are justifiable.

Affirmative action is perceived as more acceptable for some groups (e.g., those with disabilities) than for others~\cite{kravitz1993attitudes,sniderman1993scar}. Support for affirmative action also varies based on the identity of the person making the judgment. Women~\cite{golden2001reactions,kravitz1993attitudes,truxillo2000roles} and people of color~\cite{clawson2003support,klineberg2003ethnic,golden2001reactions,allen2003examining} endorse affirmative action more than men and White individuals, respectively. There is also evidence that personal experiences with discrimination~\cite{slaughter2002black} and conservatism~\cite{sidanius1996racism,clawson2003support} shape people's attitudes.

Hence, people might disagree on affirmative action---and thus affirmative algorithms---depending on several factors. We explore whether similar disagreements emerge in the case of affirmative algorithm and examine one potential source of these disagreements. We hypothesize that individuals have varying attitudes towards affirmative algorithms because they disagree about who is historically marginalized, if anyone at all.

We leverage the moral judgment theory of~\citet{gray2025morality} as a lens to interpret our findings. They argue that moral judgments rely on comparisons with a template of an agent causing harm to a vulnerable victim, such that one potential source for varying moral judgments is disagreements about who is a vulnerable victim and who is not. Prior work demonstrates that conservatives and liberals disagree on who is vulnerable to harm~\cite{womick2024moral}. Liberals view minority groups as extremely vulnerable, endorsing practices that support minorities (e.g., affirmative action). In contrast, conservatives prioritize policies that treat everyone equally due to beliefs that everyone is equally vulnerable.

In our study, we also capture who participants consider marginalized and explore whether these opinions align with their attitudes towards algorithms. Inspired by prior work looking at differences across the political domain~\cite{womick2024moral} and evidence that political leaning is associated with attitudes towards affirmative action~\cite{clawson2003support,sidanius1996racism}, we examine whether liberals and conservatives disagree on how algorithmic outcomes should be distributed, investigating whether these disagreements can be explained by disagreements about which demographic groups they consider marginalized.

Prior work demonstrates that identity also shapes one's views on affirmative action~\cite{golden2001reactions,kravitz1993attitudes,truxillo2000roles,aberson2003beliefs,clawson2003support,klineberg2003ethnic,allen2003examining}. Hence, we also explore how participants' positionality in relation to gender and race influences their views of different algorithms. We focus on these two identity axes due to their centrality in algorithmic fairness~\cite{birhane2022forgotten}, as well as evidence that individuals from different racial~\cite{nelson2013marley,fang2022historical,kraus2017americans,kraus2022framing,callaghan2021testing} and gender groups~\cite{onyeador2023misperception} have varying perceptions of inequality.

\section{Methods}
\label{sec:methods}

We conducted two 4x3 between-participants vignette-based experiments. We manipulated whether participants were introduced to an algorithm that prioritized specific groups and, if so, which group (algorithm type: affirmative vs. discriminatory vs. fair vs. control). We also varied whether participants were prompted to evaluate the fairness of the decision-making domain and, if so, whether they were also explicitly told that the domain was systematically unjust (context: contextualized evaluation vs. decontextualized evaluation vs. no evaluation). Finally, we explored whether participants' political leaning, racial, and gender identities moderates their opinions about different types of algorithms.

In our experiments, we considered two high-stakes decision-making domains commonly studied in the algorithmic fairness literature: hiring~\cite{lee2018understanding, langer2019highly,nytresume,reutersamazon,ifow,harvardhiringalgorithms} and bail decisions~\cite{grgic2018beyond, srivastava2019mathematical, harrison2020empirical,grgic2018human,propublicastory,markup2021crime}. Our first experiment focused on the hiring domain, depicting an algorithm designed to assist companies decide who to interview. We chose the hiring domain because of its competitive nature: job candidates compete for one position, such that individuals can be clearly prioritized in relation to others. In our second experiment, participants were shown a vignette in which the algorithm assisted courts in making bail decisions. Bail decision-making is not a competitive setting, since decisions concerning a defendant do not directly impact other defendants. We chose this domain to replicate our experiment in a context that would not be typically considered for affirmative action due its non-competitive nature---though some have argued for affirmative algorithms in criminal justice~\cite{zhang2022affirmative}. The studies were approved by the first author's Ethical Review Board (ERB), and all of the data and scripts used for analysis are available online: \repo{}.

\begin{figure}[t!]
    \centering
    \includegraphics[width=\textwidth]{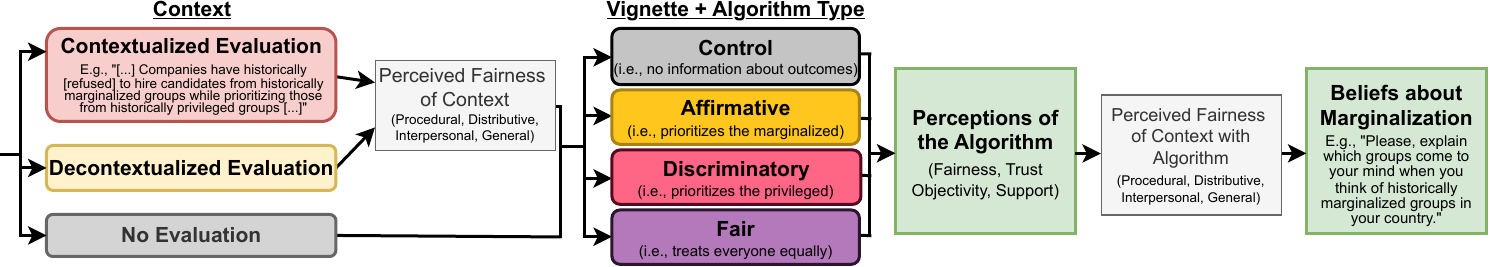}
    \vspace{-0.5cm}
    \caption{High-level overview of our methodology. Our vignette and manipulations are presented in Appendix~\ref{supp:methods}. We focus on participants' perceptions of the algorithm and beliefs about marginalization due to space constraints (see Appendix~\ref{supp:analysis} for supplementary analysis).}
    \Description{Flowchart depicting the experimental design of our study. Please refer to the Appendix for the exact phrasing of the vignette and manipulations.}
    \label{fig:methods}
\end{figure}

\subsection{Study Design}

Figure~\ref{fig:methods} presents an overview of our methodology; all study materials and questions are presented in Appendix~\ref{supp:methods}. After giving consent, participants were told that we were interested in their opinions concerning hiring (or bail) decisions in their country. Because bail decisions could have been less familiar to participants, those who took part in the second experiment also read an explanation of what bail decisions entail and what their potential outcomes are.

Participants were first randomly assigned to one of three \emph{context} treatment conditions: 

\begin{itemize}
    \item Participants assigned to the \emph{contextualized evaluation} manipulation read two paragraphs explaining that members of some demographic groups have been (and continue to be) historically marginalized and discriminated against, while others group have been historically privileged. For instance, in the hiring scenario, participants were told that companies have refused ``to hire candidates from historically marginalized groups while prioritizing those from historically privileged groups.'' Upon reading the two paragraphs, participants were asked their perceived fairness of the decision-making domain.
    \item Those assigned to the \emph{decontextualized evaluation} manipulation were \emph{not} shown any information concerning historical injustice. Instead, they were only asked to evaluate the fairness of the decision-making domain.
    \item Participants in the \emph{no evaluation} condition were not shown any information about historical injustice nor were they prompted to evaluate the fairness of the domain.
\end{itemize}

We expected participants in the \emph{contextualized evaluation} and \emph{decontextualized evaluation} conditions to rate affirmative algorithms more positively since they were prompted to consider whether the decision-making domain was fair---in line with calls for acknowledging that algorithms are embedded in systemic injustice~\cite{zhang2022affirmative,green2022escaping,fazelpour2020algorithmic,lin2022artificial}. When comparing these two conditions, we hypothesized that participants in the \emph{contextualized} condition would show even greater support given that the manipulation could also inform participants who believe the decision-making domain is fair.

All participants were then shown a vignette in which an algorithm is assisting companies (or courts) in making hiring (or bail) decisions. Everyone read the same baseline vignette, describing that the algorithm was trained on data from past human decisions. In the case of hiring, the algorithm scored job applicants ``based on their resume and cover letter, with higher scores indicating that the applicant is a better candidate'' for a position. Our vignette on bail decision-making stated that the algorithm scored defendants based on their current charges and past criminal records, such that scores refer ``to the risk of the defendant reoffending within two years, with higher scores indicating that the applicant is more likely to reoffend within two years.''\footnote{Our bail decision-making vignette was directly inspired by reports that risk algorithms are more likely to flag members of minority groups as potential re-offenders~\cite{propublicastory} and proposals to use different risk thresholds to assess defendants from distinct racial groups~\cite{zhang2022affirmative}.} The baseline vignette concluded by stating that the algorithm uses the scores to recommend who should be offered interviews (or released on bail).

Participants were then shown additional information depending on the \emph{algorithm type} condition to which they were randomly assigned:

\begin{itemize}
    \item Participants assigned to the \emph{affirmative} condition were told that the algorithm prioritized ``members of demographic groups that have been historically discriminated against.'' The manipulation also stated that, given individuals with similar qualifications (or charges and criminal records), the algorithm was more likely to recommend the historically marginalized for interviews (or release on bail).
    \item In contrast, those in the \emph{discriminatory} treatment read a vignette describing an algorithm that prioritized those who ``have been historically favored in [hiring/bail] decisions.'' The vignette stated that the system was more likely to recommend those from privileged groups for interviews (or release on bail) given similar individuals.
    \item Participants assigned to the \emph{fair} condition were instead told that the system treated everyone equally, such that it was ``equally likely to recommend candidates [...] regardless of their demographic group.''
    \item Finally, those in the \emph{control} condition were not shown any information about the algorithm's decisions.
\end{itemize}

Both vignettes used the same term to refer to marginalized and privileged groups and did not use the terms ``discriminatory'' and ``affirmative'' in the vignettes not to bias participants' responses given the weight that these specific terms may carry. The text was modified as little as possible between domains to mitigate noise. Upon reading the vignette, participants answered the questions concerning the algorithm and their beliefs concerning who is marginalized.

A few design choices are worth clarifying. Both vignettes mentioned ``marginalized'' and ``privileged'' groups without explicitly identifying any identity axes. We did so to explore which groups come to participants' mind when completing the study without restricting their judgments to a specific group. We also aimed to mitigate some biases that may emerge from focusing on specific groups. Prior work suggests that people disregard information that contradicts their prior beliefs~\cite{lord1979biased,edwards1996disconfirmation,harmon2019introduction, festinger1962theory,london2003job,ilgen1979consequences} or portrays them as being privileged due to their identity~\cite{knowles2014deny,branscombe2007racial,jost2005antecedents}. The \emph{decontextualized evaluation} context manipulation also aimed to mitigate these potential biases.

The discrimination portrayed in our vignettes is akin to the disparate treatment doctrine in United States (US) law~\cite{barocas2016big} and the direct discrimination principle in non-discrimination law in the European Union (EU)~\cite{binns2023legal}. Although most real-life instances of algorithmic discrimination would not be classified as disparate treatment violations in the US due to its requirement of intention to discriminate, many cases could be analyzed through the EU lens of direct discrimination since it examines the ``reasons or grounds for a decision''~\cite{binns2023legal}. We chose to frame our vignettes in this way to make the prioritization of a particular group abundantly clear rather than relying on participants' ability to estimate whether an algorithm imposes disparate impacts on different groups, mitigating potential confounding effects.\footnote{We also note that these two non-discrimination doctrines rely on \emph{protected groups}, which are legally protected from discrimination---in contrast, our studies did not focus on particular groups, as explained above. Finally, we acknowledge that while the affirmative algorithm in the vignette would likely be illegal in the US given recent decisions that banned affirmative action in university admissions~\cite{guardianaffirmatice}, similar systems could be more justifiable under EU law. Because non-discrimination EU law is more contextual and flexible by considering existing and historical disadvantages and injustices~\cite{wachter2021fairness,weerts2024neutrality}, affirmative action (or positive action in EU terms) through algorithmic means could potentially be allowed. Affirmative action and its legality vary widely around the world~\cite{gisselquist2023affirmative}; in this paper, we focus primarily on EU and US law.}

\subsection{Measures}

\subsubsection{Perceived Fairness of the Decision-Making Domain:}

Participants assigned to the \emph{contextualized evaluation} and \emph{decontextualized evaluation} context manipulations first evaluated the fairness of the decision-making domain. We captured fairness judgments across three dimensions: procedural~\cite{lee2019procedural,wang2020factors,colquitt2001dimensionality}, distributive~\cite{saxena2019fairness,srivastava2019mathematical}, and interpersonal fairness~\cite{rj1986interactional,schlicker2021expect}. We also asked participants to evaluate the fairness in a more general sense. 
Due to space constraints, we report the analysis of how perceived fairness varies between decision-making domains and depending on the \emph{context} manipulations in Appendix~\ref{supp:analysis}.

\subsubsection{Perceptions of the Algorithm:}

All participants evaluated the algorithm with respect to four aspects: fairness (0 = Not fair at all, 6 = Very fair), objectivity (-3 = Strongly disagree, 3 = Strongly agree), trust (0 = No trust at all, 6 = Extreme trust), and support (-3 = Strongly oppose, 3 = Strongly support). We measured perceived fairness~\cite{lee2018understanding} and objectivity~\cite{pethig2023biased} due to their relevance to algorithmic fairness. We also included a measure on trust~\cite{lee2018understanding} inspired by prior work viewing reported trust as one's willingness to be subjected to a decision-maker~\cite{knowles2023trustworthy}. Finally, we included a face-valid question on participants' support (or opposition) to each type of algorithm (see Appendix~\ref{supp:methods} for phrasing).

Participants were then asked to re-evaluate the perceived fairness of the decision-making domain imagining that the algorithm was deployed in their own country. We do not present an analysis of these measures due to space limitations.

\subsubsection{Beliefs About Marginalization:}

We then asked participants to ``explain which groups come to [their] mind when [they] think of historically marginalized groups in [their] country'' in an open-ended manner, which we operationalized as people's opinions concerning which groups are marginalized. In the same page, we also asked people to identify groups that they consider ``historically privileged'' and to explain affirmative, discriminatory, and fair algorithms in their own words. We disabled copy-pasting in these questions to mitigate large language model use~\cite{veselovsky2023prevalence}. 

\subsubsection{Demographics:}
Participants then answered several personal and demographic questions. Most relevant to our study, we captured people's political leaning using a 5-point scale (-2 = Conservative, 0 = Moderate, 2 = Liberal). Participants also voluntarily indicated their racial and gender identity (see Appendix~\ref{supp:methods} for the exact questions).

\subsection{Analysis Plan}

We employed ANOVA tests, treating the dependent variables as continuous. We analyzed the two studies separately and only report factors that are significant at the $\alpha$=.05 level in the main text for conciseness. Upon identifying a significant effect, we tested for pairwise differences using contrasts and applied Bonferroni corrections to account for multiple comparisons. All ANOVA and pairwise comparison tables are presented in Appendix~\ref{supp:anova}.

First, we examined the effect of our experimental manipulations on participants' judgments of algorithms, including both their main effects and two-way interaction. A significant main effect of our \emph{algorithm type} manipulation would indicate that people differentiate between different types of algorithm. Moreover, a significant interaction between the two factors would suggest that prompting people to evaluate the fairness of the domain---or explicitly telling them that the domain in unjust---has heterogenous effects on people's perceptions of different types of algorithms.

Second, we explored whether people's perceptions of different types of algorithms are moderated by their political leaning, racial group, and gender identity.\footnote{In the case political leaning, we report the results of ANCOVA models and conduct pairwise comparisons between the slopes of each \emph{algorithm type} condition since we treat political leaning as a continuous independent variable. We consider that there is no association between political leaning and opinions about a particular algorithm if the slope's 95\% confidence interval (CI) includes zero.} Although we also control for the main effects of these factors, we only report the results of the interaction term between the \emph{algorithm type} manipulation and participants' identity or political leaning in the main text due to our focus on moderation effects. 

We mapped participants' self-reported racial and gender identity for simplicity and better statistical power. Participants who self-identified as White were categorized as ``White,'' whereas all other participants were grouped as ``Non-White.'' This categorization is aligned with prior work that has found that people of color are more supportive of affirmative action than White individuals~\cite{clawson2003support,klineberg2003ethnic,golden2001reactions,allen2003examining} and research examining differences in perceptions of racial injustice between White and other racial groups~\cite{kraus2017americans,kraus2022framing,kuo2020high,goldsmith2004schools}. As for gender, we grouped participants into ``Men'' and ``Non-Men'' groups, aligning with work showing that women are more likely to endorse affirmative action than men~\cite{golden2001reactions,kravitz1993attitudes,truxillo2000roles} and mitigating biases from our small sample size of non-binary and trans individuals.\footnote{We acknowledge that the ``Non-White'' and ``Non-Men'' groups include a diverse groups of participants with a wide range of identities and experiences and discuss potential within-group differences in Appendix~\ref{supp:analysis} due to space limitations. We also propose that future work exploring differences between different groups could strategically recruit participants from these minority groups in order to achieve a more balanced sample.}

Finally, the first author manually inspected participants' open-ended descriptions of which groups they deem marginalized. The author identified the groups in the hiring domain as they emerged during the coding process and then applied the same categories to responses in the bail domain. We only report the results of groups that were mentioned by at least 10\% of the respondents in both studies in the main text (see Appendix~\ref{supp:analysis} for the complete analysis). We first compared the frequency at which each group was mentioned between the two studies using $\chi^2$ tests. We then employed logistic regressions to model the likelihood that a specific group was mentioned, including political leaning, racial group, and gender identity as independent variables.

\subsection{Participants}

A power analysis using G-Power~\cite{faul2009statistical} suggests that 4x3 ANOVA tests require 508 participants to detect a small-to-medium interaction effect (partial eta-squared \etasq{0.04}) at the significance level of 0.05 with 0.95 power. We thus recruited 600 participants per study to account for potential attention-check failures.

In total, we recruited 1200 participants through Prolific~\cite{palan2018prolific}. Participants were required to be US residents, fluent in English, and have completed at least 50 tasks on Prolific with at least 95\% approval rate. We sampled participants at different hours over several days to mitigate sampling biases that may occur due to time~\cite{casey2017intertemporal}.\footnote{For the hiring domain study, we initially recruited participants only for the \emph{no evaluation} and \emph{decontextualized evaluation} context treatments. Upon analysis, we hypothesized that the non-significant effect of our \emph{decontextualized evaluation} treatment---as reported below---could have emerged from its simplicity. A week later, we recruited additional respondents for the \emph{contextualized evaluation} condition to explore whether a more direct manipulation has a larger---and significant--effect. In the bail-decision-making study, all participants were recruited concurrently.} We discarded responses from those who failed an instructed-response question or failed to identity the decision-making domain in the end of the study. Our final sample comprised 1193 participants, out of which 598 read the hiring vignette and 595 took part in the bail decision study. All participants were paid 1.2 GBP (approximately 1.52 USD), with a median pay of 8.5 GBP per hour (approximately 10.76 USD per hour).

Our sample comprised 567 (47.53\%) women, 578 (48.45\%) men, and 48 (4.02\%) individuals who self-identified as non-binary, transgender, or preferred to not disclose. The average age of our participants was 39 years old ($SD = 12.6$). More than half of participants ($n = 799$, 66.97\%) self-identified as White, 172 (14.42\%) as Black/African American, 107 (8.97\%) as Asian, and 53 (4.44\%) as multi-racial. Our sample leaned slightly liberal (\msd{0.43}{1.40}). We did not identify significant differences in the gender and race distribution between different studies (i.e., p-values of $\chi^2$ tests are larger than 0.05). Participants who took part in the bail decisions study were slightly more conservative (diff = -0.211, $SE$ = 0.082, p < .05) and older (diff = 3.266, $SE$ = 0.729, p < .001).

\section{Results}
\label{sec:results}
\subsection{Perceptions of Affirmative, Discriminatory, and Fair Algorithms}
\label{sec:perceptions}

\begin{figure}[t!]
    \centering
    \includegraphics[width=\linewidth]{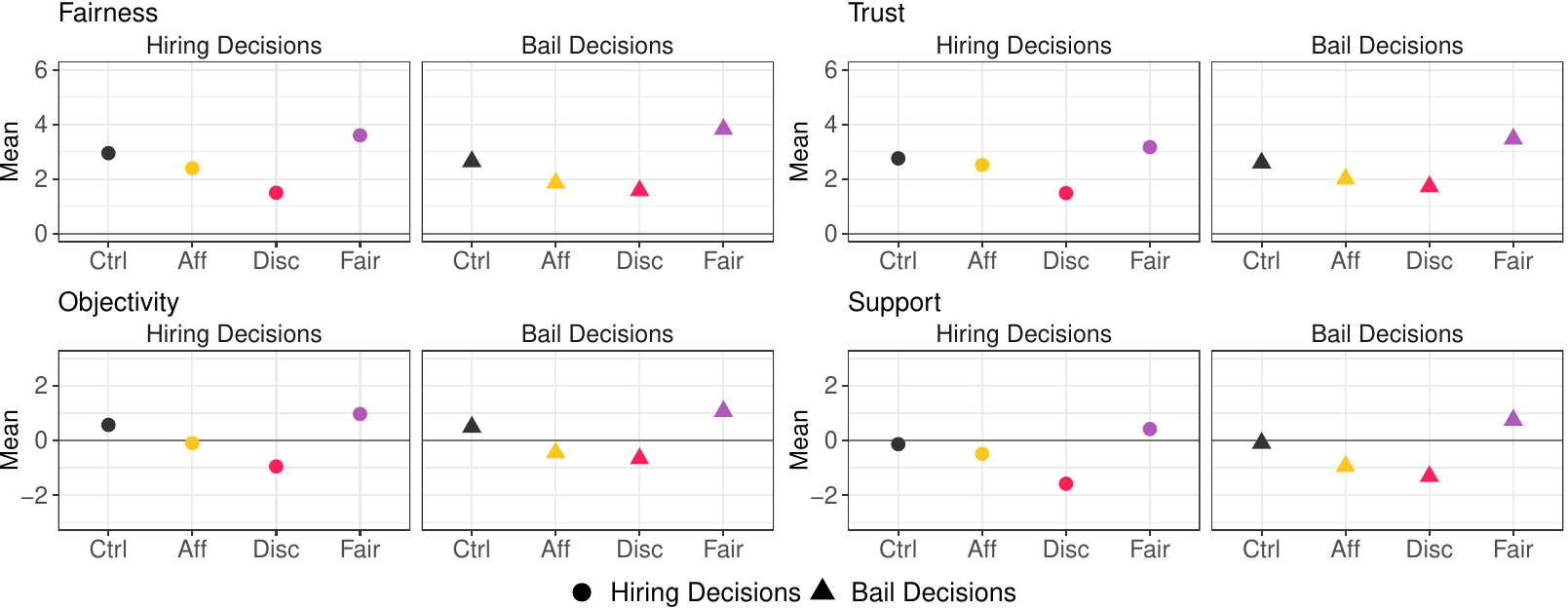}
    \vspace{-0.6cm}
    \caption{Participants' mean judgments of fairness, trust, objectivity, support concerning different types of algorithms deployed to assist hiring and bail decisions. The x-axis refers to the \emph{algorithm type} experimental manipulations: control (Ctrl), affirmative (Aff), discriminatory (Disc), and fair (Fair). Standard errors are included in the figure but are not visible due to their small values.} 
    \label{fig:algorithm_condition_study}
    \Description{
    This figure illustrates perceived fairness, trust, objectivity, and support across two domains: Hiring Decisions and Bail Decisions. The x-axis shows four algorithm conditions: Control, Affirmative, Discriminatory, and Fair. The y-axis represents the mean values for each metric.}
    
\end{figure}

Figure~\ref{fig:algorithm_condition_study} presents participants' mean perceptions of the algorithm depending on its type and the decision-making domain. In the hiring domain, we found a significant effect of the algorithm type in judgments of fairness (\ftest{3}{586}{45.63}, \pvalue{.001}, \etasq{0.19}), trust (\ftest{3}{586}{32.23}, \pvalue{.001}, \etasq{0.14}), objectivity (\ftest{3}{586}{45.80}, \pvalue{.001}, \etasq{0.19}) and support (\ftest{3}{586}{38.65}, \pvalue{.001}, \etasq{0.17}). In contrast, we observed no statistically significant effect of our context manipulations or their interactions with the algorithm type concerning any of the measures.

In the hiring domain, fair algorithms were judged more positively than their affirmative and discriminatory counterparts with respect to all measures. Moreover, the fair decision-making system was considered more fair and received more support than the algorithm in the control condition. The affirmative algorithm was considered more fair, trustworthy, objective, and received more support than its discriminatory counterpart, while being judged less fair and objective than the control algorithm. The discriminatory system was evaluated more negatively than all other decision-making systems concerning all factors.

In the bail domain, we found comparable effects of our algorithm type manipulation with respect to fairness (\ftest{3}{583}{61.41}, \pvalue{.001}, \etasq{0.24}), trust (\ftest{3}{583}{37.53}, \pvalue{.001}, \etasq{0.16}), objectivity (\ftest{3}{583}{42.87}, \pvalue{.001}, \etasq{0.18}), and support (\ftest{3}{583}{48.35}, \pvalue{.001}, \etasq{0.20}). Similarly to the case of hiring, our context manipulations and their interaction with the algorithm type was not significant across all measures.

As in the case of hiring, the fair algorithm was evaluated more positively than affirmative and discriminatory algorithms concerning all measures. Furthermore, the fair decision-making system was viewed more favorably than the algorithm in the control condition. However, in contrast to the case of hiring decisions, affirmative and discriminatory algorithms were rated similarly across all measures and more negatively than the system in the control condition.

\subsection{Varying Opinions About Algorithms Depending on Political Leaning and Identity}
\label{sec:leaning_identity}

\begin{figure}[t!]
    \centering
    \includegraphics[width=\linewidth]{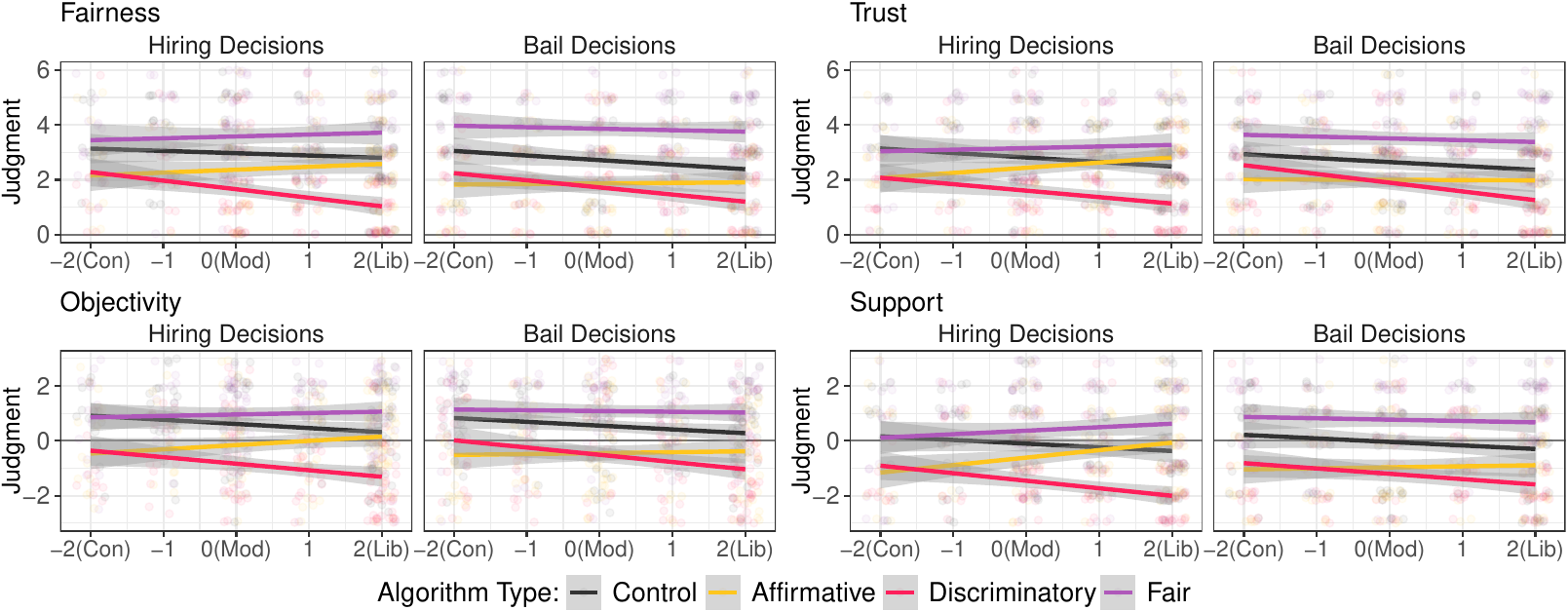}
    \vspace{-0.6cm}
    \caption{Participants' judgments of fairness, trust, objectivity, support concerning different types of algorithms depending on participant's political leaning. The x-axis represent political leaning on a 5-point scale, in which -2 refers to conservative (Con), 0 to moderate (Mod), and 2 to liberal (Lib).}
    \label{fig:algorithm_condition_politics}
    \Description{
    This figure explores perceived fairness, trust, objectivity, and support judgments in Hiring Decisions and Bail Decisions. The x-axis represents political ideology, ranging from -2 (Conservative) to +2 (Liberal). The y-axis shows judgment scores for each metric. The figure highlights how political ideology correlates with evaluations of four types of algorithm: Control, Affirmative, Discriminatory, and Fair.}
\end{figure}

\subsubsection{Political Leaning:}

Figure~\ref{fig:algorithm_condition_politics} shows the correlation between participants' political leaning and their views of each type of algorithm. In the hiring domain, we observed a significant moderation effect between political leaning and algorithm type in judgments of fairness (\ftest{3}{580}{3.67}, \pvalue{.05}, \etasq{0.02}), trust (\ftest{3}{580}{4.42}, \pvalue{.01}, \etasq{0.02}), objectivity (\ftest{3}{580}{3.86}, \pvalue{.01}, \etasq{0.02}), and support (\ftest{3}{580}{6.11}, \pvalue{.001}, \etasq{0.03}). We found no association between political leaning and opinions about algorithms described as fair or in the control condition. In contrast, we observed that the more liberal the participant was, the more negatively they judged discriminatory algorithms with respect to all measures. We identified the opposite trend in reported trust and support towards affirmative systems, such that the more liberal a participant was, the more they trusted and supported affirmative algorithms. In sum, liberals and conservatives agreed on their views of fair algorithms but disagreed about affirmative and discriminatory algorithms.

Focusing on judgments concerning bail decision-making algorithms, we did not identify any significant moderation effect. We note, however, that Figure~\ref{fig:algorithm_condition_politics} shows a trend similar to that found for hiring algorithms: liberals participants seem to report more positive views concerning affirmative algorithms than conservatives, while the two groups show similar opinions concerning fair systems. Looking into the association between political leaning for each type of algorithm separately, we observed that the more liberal a participant was, the more critical they were about discriminatory algorithms concerning all measures. We found no significant correlation between political leaning and judgments concerning affirmative systems.

\begin{figure}[t!]
    \centering
    \includegraphics[width=\linewidth]{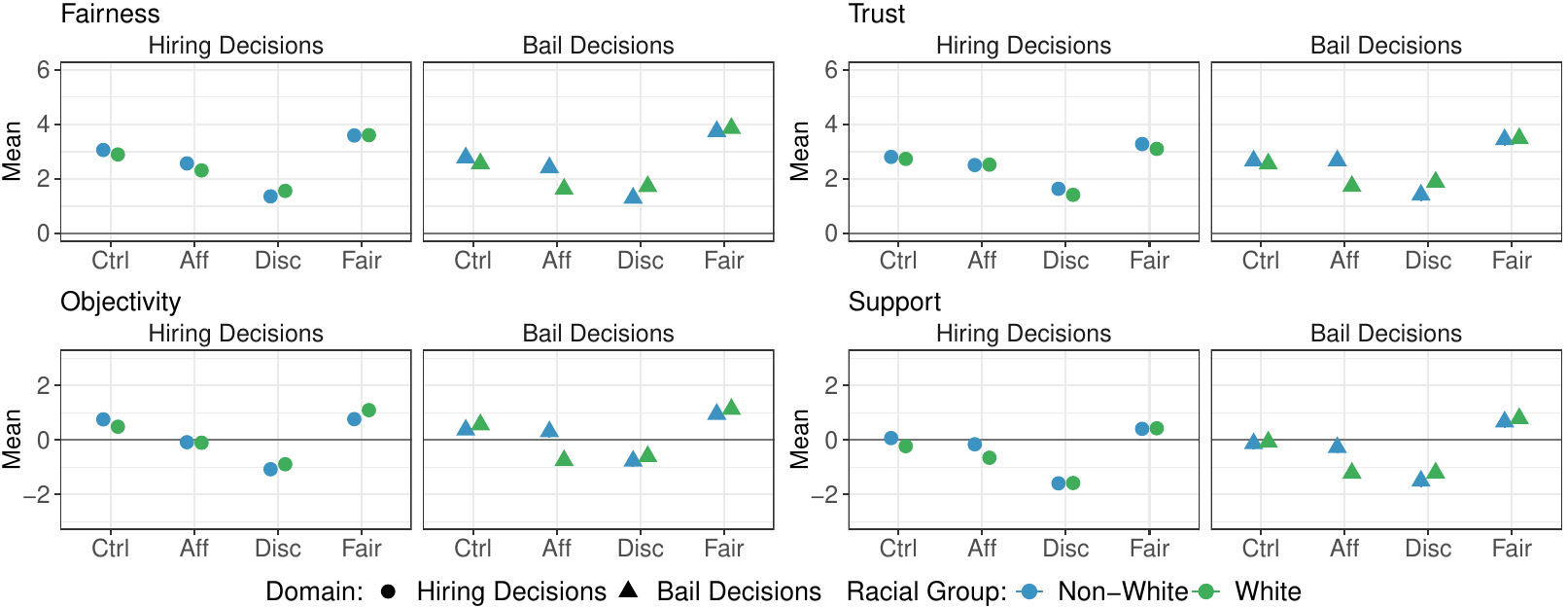}
    \vspace{-0.6cm}
    \caption{Participants' mean judgments of fairness, trust, objectivity, support concerning different types of algorithms depending on participant's racial group. The x-axis refers to the \emph{algorithm type} experimental manipulations: control (Ctrl), affirmative (Aff), discriminatory (Disc), and fair (Fair). Standard errors are included in the figure but are not visible due to their small values.}
    \label{fig:algorithm_condition_race}
    \Description{
    This figure shows the results of an analysis of perceived fairness, trust, objectivity, and support across two domains: Hiring Decisions and Bail Decisions. The x-axis represents four algorithm conditions: Control, Affirmative, Discriminatory, and Fair. The y-axis shows the mean values for each metric. The results are separated by racial groups, specifically Non-White and White, highlighting differences in perceptions based on racial identity across algorithmic conditions.}
\end{figure}

\subsubsection{Racial Group:}

Figure~\ref{fig:algorithm_condition_race} presents participants' mean judgments depending on the type of algorithm and participants' racial group. We did not observe any significant interaction between algorithm type and participants' racial group in judgments concerning hiring algorithms.

On the other hand, we found significant interactions between racial group and the type of algorithm in the bail decision-making context concerning fairness (\ftest{3}{587}{3.51}, \pvalue{.05}, \etasq{0.02}), trust (\ftest{3}{587}{4.76}, \pvalue{.01}, \etasq{0.02}), objectivity (\ftest{3}{587}{5.32}, \pvalue{.001}, \etasq{0.03}), and support (\ftest{3}{587}{3.75}, \pvalue{.05}, \etasq{0.02}). Pairwise comparisons between different racial groups only suggest differences in perceptions of affirmative algorithm. Participants categorized as Non-White considered affirmative algorithms more fair and objective and reported higher trust and support in them than White participants. We did not observe any differences between racial groups in perceptions of other algorithms.

\subsubsection{Gender Identity:}

We did not observe significant interactions between gender identity and algorithm type in any of the decision-making domains.

\subsection{Which Groups are Considered Marginalized}
\label{sec:perceptions_marg}

Finally, we explored whether disagreements concerning affirmative algorithms can be explained by people's varying conceptions of which groups (if any) are marginalized. We identified four groups that were mentioned as marginalized by more than 10\% of the participants: racial minorities (81.6\%), cisgender groups (23.1\%), low-income groups (15.6\%), and LGBTQ+ individuals (12.8\%). Racial groups include mentions of, for instance, African/Black Americans, Native Americans, and Asian Americans. Participants' responses were coded as mentioning cisgender groups when they referred to women (or men) as marginalized, i.e., when participants mentioned binary cisgender people. Low-income groups include responses that cited, for instance, ``poor people'' or ``people of lower class.'' Finally, LGBTQ+ groups included mentions to sexual orientation minorities, as well as trans and non-binary individuals.

We observed significant differences in who was identified as marginalized between the decision-making domains. Racial (\csq{1}{4.85}, \pvalue{.05}, Cramer's $V$ = 0.066) and low-income (\csq{1}{24.07}, \pvalue{.001}, $V$ = 0.144) groups were more likely to be mentioned in the bail decisions study. In contrast, cisgender groups (\csq{1}{37.69}, \pvalue{.001}, $V$ = 0.180) and LGBTQ+ individuals (\csq{1}{6.58}, \pvalue{.05}, $V$ = 0.077) were mentioned more frequently in hiring decision-making study.

Focusing on participants from the hiring study (see Appendix~\ref{supp:analysis} for statistical tests), liberal participants were more likely to identify racial, cisgender, low-income, and LGBTQ+ groups as marginalized. Men participants identified cisgender minorities and LGBTQ+ individuals as marginalized less frequently. There were no differences between racial groups.

In contrast, participants' racial identity had an effect on who they consider marginalized in the bail decision-making study (see Appendix~\ref{supp:analysis}). White participants were less likely to identify racial minorities as marginalized. As in the case of hiring decisions, the more liberal participants were, the more likely they were to identify racial, cisgender, and LGBTQ+ groups as marginalized. Men participants were also less likely to identify cisgender groups as marginalized.

\section{Discussion}
\label{sec:discussion}

Below, we first discuss people's preference for fair algorithms (\S\ref{sec:fair_good}). We then explore disagreements concerning affirmative and discriminatory systems (\S\ref{sec:aff_disc}) and explain how they may emerge from varying beliefs about who is marginalized (\S\ref{sec:marginalized}). Finally, we consider the effectiveness (or lack thereof) of manipulations that contextualize algorithms in injustice (\S\ref{sec:null_context}) before reflecting on limitations (\S\ref{sec:limitations}) and sharing some concluding remarks (\S\ref{sec:conclusion}).

\subsection{People's Support For Fair Algorithms}
\label{sec:fair_good}

Across both decision-making domains, people preferred fair algorithms over their affirmative and discriminatory counterparts, as well over algorithms whose outcomes were unknown. Our results are in line with prior work showing that people value algorithmic fairness highly~\cite{kieslich2022artificial} and research showing that fairness perceptions are correlated with satisfaction with algorithmic systems~\cite{shin2019role} and reported trust in them~\cite{shin2020user}.

Yet, we highlight that our studies focused on one particular notion of fairness: disparate treatment, meaning that similar individuals had the same chance to receive a positive decision (i.e., to be offered an interview or granted bail). It is possible that other notions of fairness would have received different degrees of support, as suggested by prior work~\cite{saxena2019fairness,kasinidou2021agree,cheng2021soliciting}. Future work could explore how different notions of fairness compare to the other types of algorithm.

\subsection{Varying Opinions About Affirmative and Discriminatory Algorithms}
\label{sec:aff_disc}

We identified disagreements concerning affirmative algorithms across political orientation and participants' racial identities. We also discuss differences in the decision-making domains that could lead to our observed disagreements.

\subsubsection{Disagreements Across the Political Spectrum:}

In the case of hiring decisions, liberal participants evaluated affirmative algorithms as positively as fair algorithms. In contrast, conservatives considered algorithms that prioritized the marginalized as unfair as systems that favored the historically privileged. These findings suggest that liberals considered discrimination that counters historical injustices justifiable, whereas conservative individuals viewed any kind of discrimination as unfair. Our findings align with prior work showing differences in perceived algorithmic fairness across political leaning~\cite{grgic2021dimensions} and suggesting a correlation between politics and support for affirmative action~\cite{sidanius1996racism,clawson2003support}.

We acknowledge that we did not find significant moderation effects of political leaning in the case of a bail decision-making algorithm. Nonetheless, we observed a similar trend such that the more liberal participants were, the more they denounced discrimination that favored privileged groups. Instead, as we discuss below, racial identity was a more determining factor for bail decisions.

We also observed that political leaning was not associated with people's views concerning the fair algorithm in either domains. In other words, people agreed on the importance of algorithmic fairness regardless of their political views. Similarly, we found that participants largely agreed that algorithms that prioritize the privileged are unfair. Disagreements between participants mainly emerged when the algorithm explicitly favored the historically marginalized. The takeaway is that our participants agreed about the importance of fairness, but disagreed on what fairness actually entails. Liberals seem to consider historical patterns of discrimination when evaluating affirmative systems, whereas conservatives prefer to evaluate algorithms independently of historical injustice. 

If affirmative algorithms are to be deployed, it is imperative to study whether these political disagreements can be bridged. Prior work suggests that sharing personal experiences has the potential to bridge political divides~\cite{kubin2021personal,kubin2023reducing}. It is possible that manipulations that describe personal experiences of (algorithmic) discrimination could make affirmative algorithms less controversial---in contrast to the null effect of our impersonal context manipulations. Future work could also explore potential mediators that lead to these disagreements (e.g., political dehumanization~\cite{kubin2023reducing}, beliefs in meritocracy~\cite{son2002meritocracy}) to design interventions that make affirmative algorithms more justifiable to all.

\subsubsection{Disagreements Across Different Racial Identities:}

Participants from racial minority groups were more supportive of affirmative algorithms in the bail decision-making domain. In contrast, we found no significant racial differences in perceptions of other bail algorithms or in opinions concerning any hiring algorithm. Our results are partially aligned with research showing that people of color (and those who have experienced discrimination~\cite{slaughter2002black}) are more likely to endorse affirmative action than White individuals~\cite{clawson2003support,klineberg2003ethnic,golden2001reactions,allen2003examining}. Our findings contrast those of~\citet{grgic2021dimensions}, who found that people from different racial groups agree on which features should be included in a bail algorithm. Yet, when evaluating affirmative algorithms, people from different racial groups disagree on whether they are justifiable.

We hypothesize that these racial differences emerge from the saliency of racial identity in the criminal justice domain. Although also relevant in hiring, racial identity is relatively more prominent in bail decision-making. This is particularly relevant in the United States (US)---where the study was conducted. Debates concerning injustice in the US criminal justice often relate to race~\cite{hinton2018unjust,michelle2010new}, whereas debates on hiring discrimination also focus on other identity axes, such as gender and sexual orientation~\cite{rosette2018intersectionality,crenshaw2013mapping,mishel2016discrimination,sears2024lgbtq}. As explored below, the centrality of race in criminal justice is also exemplified by the relative focus on racial groups as marginalized in the case of bail decision-making.

Finally, we also note that participants' gender identity did not influence their views of different types of algorithm. These findings are at odds with prior work finding that women are more likely to endorse affirmative action~\cite{golden2001reactions,kravitz1993attitudes,truxillo2000roles} and research suggesting varying opinions about algorithms deployed in the educational domain depending on gender~\cite{pierson2017gender}. Nonetheless, our results are aligned with work showing that gender does not impact perceptions of algorithmic fairness in bail decisions~\cite{grgic2021dimensions}. We call for future work to explore whether differences similar to what we observed with respect to race emerge in domains where gender discrimination is the most prominent axis of injustice.

\subsubsection{Differences Between Hiring and Bail Decisions:}

Another potential source of our contrasting results is the nature of the decision-making domain. Hiring decisions are a competitive scenario where multiple candidates compete for one position. In comparison, bail decisions are non-competitive in that they only directly impact the defendant. It is possible that because hiring is competitive, liberal participants recognized that some groups deserve some priority given historical patterns of discrimination. In the bail context, however, whether someone commits a crime relies more on individual decisions, making affirmative algorithms less justifiable. We also found some differences in perceived fairness of the domains: bail decisions were judged more unfair than hiring decisions (see Appendix~\ref{supp:analysis}). Future work could replicate our study using different contexts, as well as directly investigate whether people view affirmative and discriminatory algorithms differently in (non-)competitive decision-making domains.

\subsection{How Varying Opinions About Marginalization Explain Disagreements About Affirmative Algorithms}
\label{sec:marginalized}

Inspired by the theory that moral disagreements can emerge from varying beliefs about who is vulnerable to harm~\cite{gray2025morality}, we explored whether people's opinions about affirmative algorithms can be explained by varying conceptions about which groups are marginalized. We found that liberals are more likely to consider several groups marginalized when compared to conservatives---in line with liberals showing more positive attitudes towards affirmative systems. Opinions about who is marginalized are associated with what people expect to be done about marginalization: the more groups one acknowledges as vulnerable, the more one believes in and supports attempts to rectify these harms with algorithms.

We also observed differences between the decision-making domains. In line with our observed disagreements concerning affirmative algorithms across race, we also found that racial minorities were more likely to identify racial groups as marginalized in the bail decisions study. We suggest that racial minorities are more likely to identify race as a marginalizing factor due to widespread experiences of discrimination in the criminal justice domain~\cite{slaughter2002black}. Racial identity becomes more salient when evaluating bail decision-making and that also manifests when identifying who is marginalized, but mostly among those who are more likely to personally experience marginalization. Future work could explore this framework of perceptions of marginalization in other domains.

We note that bail decision-making can also cause harm to individuals who are unrelated to specific decisions. If an algorithm makes a wrong decision and grants bail to a defendant that later re-offends, it can impact others since they may become victims of a criminal offense. In contrast, a wrong hiring decision can only cause harm to a qualified candidate who was not called for an interview or the company who hired a ``bad'' applicant. It may be that in the context of bail decisions, other groups come to mind when identifying potential victims of wrong algorithmic decisions. Future work could also explore who people consider to be indirectly vulnerable to harm by algorithmic decisions.

Finally, the theory posing that moral disagreements emerge from varying conceptions of who is a vulnerable victim also acknowledge the role of the actor causing harm~\cite{gray2025morality}. Future work could explore whether manipulating the decision-maker has any impact on perceptions of affirmative action. For instance, studies could contrast opinions about algorithmic and human affirmative action or vary the term used to introduce the algorithmic system~\cite{langer2021look}.

\subsection{Contextualizing Algorithms in Systemic Injustice Has Null Effects on Perceptions of Algorithms}
\label{sec:null_context}

We did not observe significant effects of our context manipulations on people's perceptions of any type of algorithm. Explicitly telling people that algorithms are embedded in context marked by systems injustice or prompting individuals to evaluate the fairness of the domain did not impact their views on decision-making systems. Our findings are at odds with calls from scholars who argue that a more contextualized analysis of algorithmic decision-making should make individuals more critical of computational solutions to injustice~\cite{kasirzadeh2022algorithmic,lin2022artificial,green2020algorithmic,birhane2021algorithmic,fazelpour2022algorithmic}. 

The null effects of our context manipulations are particularly noteworthy in the case of affirmative algorithms. Prioritizing the historically marginalized is justifiable only when one acknowledges that they have been systematically disadvantaged. Instead, we found that one's political leaning and positionality with respect to the decision-making domain determine their views of affirmative systems. Potentially because one's identity and political beliefs are hard to change, manipulations like ours have little to no effect on laypeople's views of attempts to redress historical discrimination. If one believes in systemic injustice (which is associated with political leaning) or is part of a demographic group that is especially at risk, they are more likely to support the deployment of algorithms that counter historical harms.

Our findings also call for a reconsideration of how designers of algorithms may frame attempts to redress historical injustice. Because of the difficulty in changing people's opinions about algorithms that explicitly prioritize specific groups, designers could re-evaluate how they can develop and advertise systems that are aligned with people's desire for fairness while rectifying past harms. For instance, affirmative algorithms could be introduced in a way that highlights its justice rather than underscoring its preference for a particular group. Algorithms and the way that they are portrayed can be easily modified, whereas people's core beliefs and personal experiences cannot be easily changed.

\subsection{Limitations}
\label{sec:limitations}

It is possible that our results are the result of social desirability biases~\cite{grimm2010social}, a common limitation in survey studies that rely on self-report measures. In our case, social desirability could mean that participants reported more negative opinions, particularly when algorithms were portrayed as discriminatory or when participants were assigned to the context manipulations. Yet, we note that our context manipulations had null effects on people's judgments. Future studies could explore behavioral measures, e.g., people's interactions with algorithms, to mitigate these potential biases.

Our sample is also restricted to US residents. Hence, our results might not be generalizable to other communities, particularly considering that different countries have different laws and perspectives on affirmative action~\cite{gisselquist2023affirmative}. The racial and political distribution of our participants is also skewed towards White and liberal individuals. Future work could employ different recruitment strategies that oversample participants from other groups (see also Appendix~\ref{supp:analysis}).

\subsection{Concluding Remarks}
\label{sec:conclusion}

Our findings suggest that people care about algorithmic fairness; however, they disagree on what fairness actually entails. Liberal individuals and those belonging to racial minorities were more likely to acknowledge that some groups have been historically marginalized, thus supporting algorithms that explicitly prioritize these minority groups. In contrast, conservatives and those from the dominant racial group viewed affirmative algorithms as another instance of an unfair and unjustifiable discriminatory system, partially because of their weaker beliefs in historical injustice. Our work brings into question whether these disagreements can be bridged if society is to strive for ``affirmative futures'' that not only acknowledges that some have---and continue to be---excluded but also redresses these harms and rectifies the broken-down system that algorithms perpetuate~\cite{theus2023striving}.

\bibliographystyle{ACM-Reference-Format}
\bibliography{sample-base}


\begin{thebibliography}{121}


\ifx \showCODEN    \undefined \def \showCODEN     #1{\unskip}     \fi
\ifx \showDOI      \undefined \def \showDOI       #1{#1}\fi
\ifx \showISBNx    \undefined \def \showISBNx     #1{\unskip}     \fi
\ifx \showISBNxiii \undefined \def \showISBNxiii  #1{\unskip}     \fi
\ifx \showISSN     \undefined \def \showISSN      #1{\unskip}     \fi
\ifx \showLCCN     \undefined \def \showLCCN      #1{\unskip}     \fi
\ifx \shownote     \undefined \def \shownote      #1{#1}          \fi
\ifx \showarticletitle \undefined \def \showarticletitle #1{#1}   \fi
\ifx \showURL      \undefined \def \showURL       {\relax}        \fi
\providecommand\bibfield[2]{#2}
\providecommand\bibinfo[2]{#2}
\providecommand\natexlab[1]{#1}
\providecommand\showeprint[2][]{arXiv:#2}

\bibitem[Aberson and Haag(2003)]%
        {aberson2003beliefs}
\bibfield{author}{\bibinfo{person}{Christopher~L Aberson} {and} \bibinfo{person}{Sarah~C Haag}.} \bibinfo{year}{2003}\natexlab{}.
\newblock \showarticletitle{Beliefs about affirmative action and diversity and their relationship to support for hiring policies}.
\newblock \bibinfo{journal}{\emph{Analyses of Social Issues and Public Policy}} \bibinfo{volume}{3}, \bibinfo{number}{1} (\bibinfo{year}{2003}), \bibinfo{pages}{121--138}.
\newblock


\bibitem[Allen(2003)]%
        {allen2003examining}
\bibfield{author}{\bibinfo{person}{Rhonda~YW Allen}.} \bibinfo{year}{2003}\natexlab{}.
\newblock \showarticletitle{Examining the implementation of affirmative action in law enforcement}.
\newblock \bibinfo{journal}{\emph{Public Personnel Management}} \bibinfo{volume}{32}, \bibinfo{number}{3} (\bibinfo{year}{2003}), \bibinfo{pages}{411--418}.
\newblock


\bibitem[Anderson(2010)]%
        {anderson2010imperative}
\bibfield{author}{\bibinfo{person}{Elizabeth Anderson}.} \bibinfo{year}{2010}\natexlab{}.
\newblock \showarticletitle{The imperative of integration}.
\newblock In \bibinfo{booktitle}{\emph{The imperative of integration}}. \bibinfo{publisher}{Princeton University Press}.
\newblock


\bibitem[Angwin et~al\mbox{.}(2016)]%
        {propublicastory}
\bibfield{author}{\bibinfo{person}{Julia Angwin}, \bibinfo{person}{Madeleine Varner}, {and} \bibinfo{person}{Ariana Tobin}.} \bibinfo{year}{2016}\natexlab{}.
\newblock \bibinfo{title}{Machine Bias: There's Software Used Across the Country to Predict Future Criminals. And it's Biased Against Blacks}.
\newblock
\newblock
\newblock
\shownote{ProPublica. \url{https://tinyurl.com/5t3apr69}}.


\bibitem[Awad et~al\mbox{.}(2020)]%
        {awad2020crowdsourcing}
\bibfield{author}{\bibinfo{person}{Edmond Awad}, \bibinfo{person}{Sohan Dsouza}, \bibinfo{person}{Jean-Fran{\c{c}}ois Bonnefon}, \bibinfo{person}{Azim Shariff}, {and} \bibinfo{person}{Iyad Rahwan}.} \bibinfo{year}{2020}\natexlab{}.
\newblock \showarticletitle{Crowdsourcing moral machines}.
\newblock \bibinfo{journal}{\emph{Commun. ACM}} \bibinfo{volume}{63}, \bibinfo{number}{3} (\bibinfo{year}{2020}), \bibinfo{pages}{48--55}.
\newblock


\bibitem[Barocas and Selbst(2016)]%
        {barocas2016big}
\bibfield{author}{\bibinfo{person}{Solon Barocas} {and} \bibinfo{person}{Andrew~D Selbst}.} \bibinfo{year}{2016}\natexlab{}.
\newblock \showarticletitle{Big data's disparate impact}.
\newblock \bibinfo{journal}{\emph{Calif. L. Rev.}}  \bibinfo{volume}{104} (\bibinfo{year}{2016}), \bibinfo{pages}{671}.
\newblock


\bibitem[Binns et~al\mbox{.}(2023)]%
        {binns2023legal}
\bibfield{author}{\bibinfo{person}{Reuben Binns}, \bibinfo{person}{Jeremias Adams-Prassl}, {and} \bibinfo{person}{Aislinn Kelly-Lyth}.} \bibinfo{year}{2023}\natexlab{}.
\newblock \showarticletitle{Legal taxonomies of machine bias: Revisiting direct discrimination}. In \bibinfo{booktitle}{\emph{Proceedings of the 2023 ACM conference on fairness, accountability, and transparency}}. \bibinfo{pages}{1850--1858}.
\newblock


\bibitem[Birhane(2021)]%
        {birhane2021algorithmic}
\bibfield{author}{\bibinfo{person}{Abeba Birhane}.} \bibinfo{year}{2021}\natexlab{}.
\newblock \showarticletitle{Algorithmic injustice: a relational ethics approach}.
\newblock \bibinfo{journal}{\emph{Patterns}} \bibinfo{volume}{2}, \bibinfo{number}{2} (\bibinfo{year}{2021}).
\newblock


\bibitem[Birhane et~al\mbox{.}(2022)]%
        {birhane2022forgotten}
\bibfield{author}{\bibinfo{person}{Abeba Birhane}, \bibinfo{person}{Elayne Ruane}, \bibinfo{person}{Thomas Laurent}, \bibinfo{person}{Matthew S.~Brown}, \bibinfo{person}{Johnathan Flowers}, \bibinfo{person}{Anthony Ventresque}, {and} \bibinfo{person}{Christopher L.~Dancy}.} \bibinfo{year}{2022}\natexlab{}.
\newblock \showarticletitle{The forgotten margins of AI ethics}. In \bibinfo{booktitle}{\emph{Proceedings of the 2022 ACM Conference on Fairness, Accountability, and Transparency}}. \bibinfo{pages}{948--958}.
\newblock


\bibitem[Bogen(2019)]%
        {harvardhiringalgorithms}
\bibfield{author}{\bibinfo{person}{Miranda Bogen}.} \bibinfo{year}{2019}\natexlab{}.
\newblock \bibinfo{title}{All the Ways Hiring Algorithms Can Introduce Bias}.
\newblock
\newblock
\newblock
\shownote{Harvard Business Review. \url{https://tinyurl.com/4rew6v6d}}.


\bibitem[Branscombe et~al\mbox{.}(2007)]%
        {branscombe2007racial}
\bibfield{author}{\bibinfo{person}{Nyla~R Branscombe}, \bibinfo{person}{Michael~T Schmitt}, {and} \bibinfo{person}{Kristin Schiffhauer}.} \bibinfo{year}{2007}\natexlab{}.
\newblock \showarticletitle{Racial attitudes in response to thoughts of White privilege}.
\newblock \bibinfo{journal}{\emph{European Journal of Social Psychology}} \bibinfo{volume}{37}, \bibinfo{number}{2} (\bibinfo{year}{2007}), \bibinfo{pages}{203--215}.
\newblock


\bibitem[Callaghan et~al\mbox{.}(2021)]%
        {callaghan2021testing}
\bibfield{author}{\bibinfo{person}{Bennett Callaghan}, \bibinfo{person}{Leilah Harouni}, \bibinfo{person}{Cydney~H Dupree}, \bibinfo{person}{Michael~W Kraus}, {and} \bibinfo{person}{Jennifer~A Richeson}.} \bibinfo{year}{2021}\natexlab{}.
\newblock \showarticletitle{Testing the efficacy of three informational interventions for reducing misperceptions of the Black--White wealth gap}.
\newblock \bibinfo{journal}{\emph{Proceedings of the National Academy of Sciences}} \bibinfo{volume}{118}, \bibinfo{number}{38} (\bibinfo{year}{2021}), \bibinfo{pages}{e2108875118}.
\newblock


\bibitem[Casey et~al\mbox{.}(2017)]%
        {casey2017intertemporal}
\bibfield{author}{\bibinfo{person}{Logan~S Casey}, \bibinfo{person}{Jesse Chandler}, \bibinfo{person}{Adam~Seth Levine}, \bibinfo{person}{Andrew Proctor}, {and} \bibinfo{person}{Dara~Z Strolovitch}.} \bibinfo{year}{2017}\natexlab{}.
\newblock \showarticletitle{Intertemporal differences among MTurk workers: Time-based sample variations and implications for online data collection}.
\newblock \bibinfo{journal}{\emph{Sage Open}} \bibinfo{volume}{7}, \bibinfo{number}{2} (\bibinfo{year}{2017}), \bibinfo{pages}{2158244017712774}.
\newblock


\bibitem[Cheng et~al\mbox{.}(2021)]%
        {cheng2021soliciting}
\bibfield{author}{\bibinfo{person}{Hao-Fei Cheng}, \bibinfo{person}{Logan Stapleton}, \bibinfo{person}{Ruiqi Wang}, \bibinfo{person}{Paige Bullock}, \bibinfo{person}{Alexandra Chouldechova}, \bibinfo{person}{Zhiwei Steven~Steven Wu}, {and} \bibinfo{person}{Haiyi Zhu}.} \bibinfo{year}{2021}\natexlab{}.
\newblock \showarticletitle{Soliciting stakeholders’ fairness notions in child maltreatment predictive systems}. In \bibinfo{booktitle}{\emph{Proceedings of the 2021 CHI Conference on Human Factors in Computing Systems}}. \bibinfo{pages}{1--17}.
\newblock


\bibitem[Chouldechova et~al\mbox{.}(2018)]%
        {chouldechova2018case}
\bibfield{author}{\bibinfo{person}{Alexandra Chouldechova}, \bibinfo{person}{Diana Benavides-Prado}, \bibinfo{person}{Oleksandr Fialko}, {and} \bibinfo{person}{Rhema Vaithianathan}.} \bibinfo{year}{2018}\natexlab{}.
\newblock \showarticletitle{A case study of algorithm-assisted decision making in child maltreatment hotline screening decisions}. In \bibinfo{booktitle}{\emph{Conference on Fairness, Accountability and Transparency}}. PMLR, \bibinfo{pages}{134--148}.
\newblock


\bibitem[Clawson and Waltenburg(2003)]%
        {clawson2003support}
\bibfield{author}{\bibinfo{person}{Rosalee~A Clawson} {and} \bibinfo{person}{Eric~N Waltenburg}.} \bibinfo{year}{2003}\natexlab{}.
\newblock \showarticletitle{Support for a Supreme Court affirmative action decision: a story in black and white}.
\newblock \bibinfo{journal}{\emph{American Politics Research}} \bibinfo{volume}{31}, \bibinfo{number}{3} (\bibinfo{year}{2003}), \bibinfo{pages}{251--279}.
\newblock


\bibitem[Colquitt(2001)]%
        {colquitt2001dimensionality}
\bibfield{author}{\bibinfo{person}{Jason~A Colquitt}.} \bibinfo{year}{2001}\natexlab{}.
\newblock \showarticletitle{On the dimensionality of organizational justice: a construct validation of a measure.}
\newblock \bibinfo{journal}{\emph{Journal of applied psychology}} \bibinfo{volume}{86}, \bibinfo{number}{3} (\bibinfo{year}{2001}), \bibinfo{pages}{386}.
\newblock


\bibitem[Crenshaw(2013)]%
        {crenshaw2013mapping}
\bibfield{author}{\bibinfo{person}{Kimberl{\'e}~Williams Crenshaw}.} \bibinfo{year}{2013}\natexlab{}.
\newblock \showarticletitle{Mapping the margins: Intersectionality, identity politics, and violence against women of color}.
\newblock In \bibinfo{booktitle}{\emph{The public nature of private violence}}. \bibinfo{publisher}{Routledge}, \bibinfo{pages}{93--118}.
\newblock


\bibitem[Crosby et~al\mbox{.}(2006)]%
        {crosby2006understanding}
\bibfield{author}{\bibinfo{person}{Faye~J Crosby}, \bibinfo{person}{Aarti Iyer}, {and} \bibinfo{person}{Sirinda Sincharoen}.} \bibinfo{year}{2006}\natexlab{}.
\newblock \showarticletitle{Understanding affirmative action}.
\newblock \bibinfo{journal}{\emph{Annu. Rev. Psychol.}} \bibinfo{volume}{57}, \bibinfo{number}{1} (\bibinfo{year}{2006}), \bibinfo{pages}{585--611}.
\newblock


\bibitem[Dastin(2018)]%
        {reutersamazon}
\bibfield{author}{\bibinfo{person}{Jeffrey Dastin}.} \bibinfo{year}{2018}\natexlab{}.
\newblock \bibinfo{title}{Amazon scraps secret AI recruiting tool that showed bias against women}.
\newblock
\newblock
\newblock
\shownote{Reuters. \url{https://tinyurl.com/y64598bh}}.


\bibitem[Davis et~al\mbox{.}(2021)]%
        {davis2021algorithmic}
\bibfield{author}{\bibinfo{person}{Jenny~L Davis}, \bibinfo{person}{Apryl Williams}, {and} \bibinfo{person}{Michael~W Yang}.} \bibinfo{year}{2021}\natexlab{}.
\newblock \showarticletitle{Algorithmic reparation}.
\newblock \bibinfo{journal}{\emph{Big Data \& Society}} \bibinfo{volume}{8}, \bibinfo{number}{2} (\bibinfo{year}{2021}), \bibinfo{pages}{20539517211044808}.
\newblock


\bibitem[Edwards and Smith(1996)]%
        {edwards1996disconfirmation}
\bibfield{author}{\bibinfo{person}{Kari Edwards} {and} \bibinfo{person}{Edward~E Smith}.} \bibinfo{year}{1996}\natexlab{}.
\newblock \showarticletitle{A disconfirmation bias in the evaluation of arguments}.
\newblock \bibinfo{journal}{\emph{Journal of personality and social psychology}} \bibinfo{volume}{71}, \bibinfo{number}{1} (\bibinfo{year}{1996}), \bibinfo{pages}{5}.
\newblock


\bibitem[Fang and White(2022)]%
        {fang2022historical}
\bibfield{author}{\bibinfo{person}{Albert~H Fang} {and} \bibinfo{person}{Steven White}.} \bibinfo{year}{2022}\natexlab{}.
\newblock \showarticletitle{Historical information and beliefs about racial inequality}.
\newblock \bibinfo{journal}{\emph{Politics, Groups, and Identities}} (\bibinfo{year}{2022}), \bibinfo{pages}{1--22}.
\newblock


\bibitem[Faul et~al\mbox{.}(2009)]%
        {faul2009statistical}
\bibfield{author}{\bibinfo{person}{Franz Faul}, \bibinfo{person}{Edgar Erdfelder}, \bibinfo{person}{Axel Buchner}, {and} \bibinfo{person}{Albert-Georg Lang}.} \bibinfo{year}{2009}\natexlab{}.
\newblock \showarticletitle{Statistical power analyses using G* Power 3.1: Tests for correlation and regression analyses}.
\newblock \bibinfo{journal}{\emph{Behavior research methods}} \bibinfo{volume}{41}, \bibinfo{number}{4} (\bibinfo{year}{2009}), \bibinfo{pages}{1149--1160}.
\newblock


\bibitem[Fazelpour and Lipton(2020)]%
        {fazelpour2020algorithmic}
\bibfield{author}{\bibinfo{person}{Sina Fazelpour} {and} \bibinfo{person}{Zachary~C Lipton}.} \bibinfo{year}{2020}\natexlab{}.
\newblock \showarticletitle{Algorithmic fairness from a non-ideal perspective}. In \bibinfo{booktitle}{\emph{Proceedings of the AAAI/ACM Conference on AI, Ethics, and Society}}. \bibinfo{pages}{57--63}.
\newblock


\bibitem[Fazelpour et~al\mbox{.}(2022)]%
        {fazelpour2022algorithmic}
\bibfield{author}{\bibinfo{person}{Sina Fazelpour}, \bibinfo{person}{Zachary~C Lipton}, {and} \bibinfo{person}{David Danks}.} \bibinfo{year}{2022}\natexlab{}.
\newblock \showarticletitle{Algorithmic fairness and the situated dynamics of justice}.
\newblock \bibinfo{journal}{\emph{Canadian Journal of Philosophy}} \bibinfo{volume}{52}, \bibinfo{number}{1} (\bibinfo{year}{2022}), \bibinfo{pages}{44--60}.
\newblock


\bibitem[Festinger(1962)]%
        {festinger1962theory}
\bibfield{author}{\bibinfo{person}{Leon Festinger}.} \bibinfo{year}{1962}\natexlab{}.
\newblock \bibinfo{booktitle}{\emph{A theory of cognitive dissonance}}. Vol.~\bibinfo{volume}{2}.
\newblock \bibinfo{publisher}{Stanford university press}.
\newblock


\bibitem[Gisselquist et~al\mbox{.}(2023)]%
        {gisselquist2023affirmative}
\bibfield{author}{\bibinfo{person}{Rachel~M Gisselquist}, \bibinfo{person}{Simone Schotte}, {and} \bibinfo{person}{Min~J Kim}.} \bibinfo{year}{2023}\natexlab{}.
\newblock \bibinfo{booktitle}{\emph{Affirmative action around the world: insights from a new dataset}}.
\newblock Number 2023/59. \bibinfo{publisher}{WIDER Working Paper}.
\newblock


\bibitem[Golden et~al\mbox{.}(2001)]%
        {golden2001reactions}
\bibfield{author}{\bibinfo{person}{Heather Golden}, \bibinfo{person}{Steve Hinkle}, {and} \bibinfo{person}{Faye Crosby}.} \bibinfo{year}{2001}\natexlab{}.
\newblock \showarticletitle{Reactions to affirmative action: Substance and semantics}.
\newblock \bibinfo{journal}{\emph{Journal of Applied Social Psychology}} \bibinfo{volume}{31}, \bibinfo{number}{1} (\bibinfo{year}{2001}), \bibinfo{pages}{73--88}.
\newblock


\bibitem[Goldsmith(2004)]%
        {goldsmith2004schools}
\bibfield{author}{\bibinfo{person}{Pat~Antonio Goldsmith}.} \bibinfo{year}{2004}\natexlab{}.
\newblock \showarticletitle{Schools' racial mix, students' optimism, and the Black-White and Latino-White achievement gaps}.
\newblock \bibinfo{journal}{\emph{Sociology of education}} \bibinfo{volume}{77}, \bibinfo{number}{2} (\bibinfo{year}{2004}), \bibinfo{pages}{121--147}.
\newblock


\bibitem[Gray and Pratt(2025)]%
        {gray2025morality}
\bibfield{author}{\bibinfo{person}{Kurt Gray} {and} \bibinfo{person}{Samuel Pratt}.} \bibinfo{year}{2025}\natexlab{}.
\newblock \showarticletitle{Morality in Our Mind and Across Cultures and Politics}.
\newblock \bibinfo{journal}{\emph{Annual Review of Psychology}}  \bibinfo{volume}{76} (\bibinfo{year}{2025}).
\newblock


\bibitem[Green(2022)]%
        {green2022escaping}
\bibfield{author}{\bibinfo{person}{Ben Green}.} \bibinfo{year}{2022}\natexlab{}.
\newblock \showarticletitle{Escaping the impossibility of fairness: From formal to substantive algorithmic fairness}.
\newblock \bibinfo{journal}{\emph{Philosophy \& Technology}} \bibinfo{volume}{35}, \bibinfo{number}{4} (\bibinfo{year}{2022}), \bibinfo{pages}{90}.
\newblock


\bibitem[Green and Viljoen(2020)]%
        {green2020algorithmic}
\bibfield{author}{\bibinfo{person}{Ben Green} {and} \bibinfo{person}{Salom{\'e} Viljoen}.} \bibinfo{year}{2020}\natexlab{}.
\newblock \showarticletitle{Algorithmic realism: expanding the boundaries of algorithmic thought}. In \bibinfo{booktitle}{\emph{Proceedings of the 2020 conference on fairness, accountability, and transparency}}. \bibinfo{pages}{19--31}.
\newblock


\bibitem[Grgi{\'c}-Hla{\v{c}}a et~al\mbox{.}(2022)]%
        {grgic2021dimensions}
\bibfield{author}{\bibinfo{person}{Nina Grgi{\'c}-Hla{\v{c}}a}, \bibinfo{person}{Gabriel Lima}, \bibinfo{person}{Adrian Weller}, {and} \bibinfo{person}{Elissa~M Redmiles}.} \bibinfo{year}{2022}\natexlab{}.
\newblock \showarticletitle{Dimensions of diversity in human perceptions of algorithmic fairness}. In \bibinfo{booktitle}{\emph{Proceedings of the 2nd ACM Conference on Equity and Access in Algorithms, Mechanisms, and Optimization}}. \bibinfo{pages}{1--12}.
\newblock


\bibitem[Grgic-Hlaca et~al\mbox{.}(2018)]%
        {grgic2018human}
\bibfield{author}{\bibinfo{person}{Nina Grgic-Hlaca}, \bibinfo{person}{Elissa~M Redmiles}, \bibinfo{person}{Krishna~P Gummadi}, {and} \bibinfo{person}{Adrian Weller}.} \bibinfo{year}{2018}\natexlab{}.
\newblock \showarticletitle{Human perceptions of fairness in algorithmic decision making: A case study of criminal risk prediction}. In \bibinfo{booktitle}{\emph{proc. of the Web conference}}. \bibinfo{pages}{903--912}.
\newblock


\bibitem[Grgi{\'c}-Hla{\v c}a et~al\mbox{.}(2018)]%
        {grgic2018beyond}
\bibfield{author}{\bibinfo{person}{Nina Grgi{\'c}-Hla{\v c}a}, \bibinfo{person}{Muhammad~Bilal Zafar}, \bibinfo{person}{Krishna~P Gummadi}, {and} \bibinfo{person}{Adrian Weller}.} \bibinfo{year}{2018}\natexlab{}.
\newblock \showarticletitle{{Beyond Distributive Fairness in Algorithmic Decision Making: Feature Selection for Procedurally Fair Learning}}. In \bibinfo{booktitle}{\emph{{AAAI}}}.
\newblock


\bibitem[Grimm(2010)]%
        {grimm2010social}
\bibfield{author}{\bibinfo{person}{Pamela Grimm}.} \bibinfo{year}{2010}\natexlab{}.
\newblock \showarticletitle{Social desirability bias}.
\newblock \bibinfo{journal}{\emph{Wiley international encyclopedia of marketing}} (\bibinfo{year}{2010}).
\newblock


\bibitem[Hanna et~al\mbox{.}(2020)]%
        {hanna2020towards}
\bibfield{author}{\bibinfo{person}{Alex Hanna}, \bibinfo{person}{Emily Denton}, \bibinfo{person}{Andrew Smart}, {and} \bibinfo{person}{Jamila Smith-Loud}.} \bibinfo{year}{2020}\natexlab{}.
\newblock \showarticletitle{Towards a critical race methodology in algorithmic fairness}. In \bibinfo{booktitle}{\emph{Proceedings of the 2020 conference on fairness, accountability, and transparency}}. \bibinfo{pages}{501--512}.
\newblock


\bibitem[Hannan et~al\mbox{.}(2021)]%
        {hannan2021gets}
\bibfield{author}{\bibinfo{person}{Jacqueline Hannan}, \bibinfo{person}{Huei-Yen~Winnie Chen}, {and} \bibinfo{person}{Kenneth Joseph}.} \bibinfo{year}{2021}\natexlab{}.
\newblock \showarticletitle{Who gets what, according to whom? An analysis of fairness perceptions in service allocation}. In \bibinfo{booktitle}{\emph{Proceedings of the 2021 AAAI/ACM Conference on AI, Ethics, and Society}}. \bibinfo{pages}{555--565}.
\newblock


\bibitem[Hardt et~al\mbox{.}(2016)]%
        {hardt2016equality}
\bibfield{author}{\bibinfo{person}{Moritz Hardt}, \bibinfo{person}{Eric Price}, \bibinfo{person}{Nati Srebro}, {et~al\mbox{.}}} \bibinfo{year}{2016}\natexlab{}.
\newblock \showarticletitle{Equality of Opportunity in Supervised Learning}. In \bibinfo{booktitle}{\emph{NeurIPS}}.
\newblock


\bibitem[Harmon-Jones and Mills(2019)]%
        {harmon2019introduction}
\bibfield{author}{\bibinfo{person}{Eddie Harmon-Jones} {and} \bibinfo{person}{Judson Mills}.} \bibinfo{year}{2019}\natexlab{}.
\newblock \showarticletitle{An introduction to cognitive dissonance theory and an overview of current perspectives on the theory}.
\newblock In \bibinfo{booktitle}{\emph{Cognitive Dissonance, Second Edition: Reexamining a Pivotal Theory in Psychology}}. \bibinfo{publisher}{American Psychological Association}.
\newblock


\bibitem[Harrison et~al\mbox{.}(2020)]%
        {harrison2020empirical}
\bibfield{author}{\bibinfo{person}{Galen Harrison}, \bibinfo{person}{Julia Hanson}, \bibinfo{person}{Christine Jacinto}, \bibinfo{person}{Julio Ramirez}, {and} \bibinfo{person}{Blase Ur}.} \bibinfo{year}{2020}\natexlab{}.
\newblock \showarticletitle{An empirical study on the perceived fairness of realistic, imperfect machine learning models}. In \bibinfo{booktitle}{\emph{Proceedings of the 2020 conference on fairness, accountability, and transparency}}. \bibinfo{pages}{392--402}.
\newblock


\bibitem[Heikkilä(2022)]%
        {politicodutch}
\bibfield{author}{\bibinfo{person}{Melissa Heikkilä}.} \bibinfo{year}{2022}\natexlab{}.
\newblock \bibinfo{title}{Dutch scandal serves as a warning for Europe over risks of using algorithms}.
\newblock
\newblock
\newblock
\shownote{Politico. \url{https://tinyurl.com/yaye24zc}}.


\bibitem[Hinton et~al\mbox{.}(2018)]%
        {hinton2018unjust}
\bibfield{author}{\bibinfo{person}{Elizabeth Hinton}, \bibinfo{person}{LaShae Henderson}, {and} \bibinfo{person}{Cindy Reed}.} \bibinfo{year}{2018}\natexlab{}.
\newblock \showarticletitle{An unjust burden: The disparate treatment of Black Americans in the criminal justice system}.
\newblock \bibinfo{journal}{\emph{Vera Institute of Justice}} \bibinfo{volume}{1}, \bibinfo{number}{1} (\bibinfo{year}{2018}), \bibinfo{pages}{1--20}.
\newblock


\bibitem[Ho and Xiang(2020)]%
        {ho2020affirmative}
\bibfield{author}{\bibinfo{person}{Daniel~E Ho} {and} \bibinfo{person}{Alice Xiang}.} \bibinfo{year}{2020}\natexlab{}.
\newblock \showarticletitle{Affirmative algorithms: The legal grounds for fairness as awareness}.
\newblock \bibinfo{journal}{\emph{U. Chi. L. Rev. Online}} (\bibinfo{year}{2020}), \bibinfo{pages}{134}.
\newblock


\bibitem[Hong(2023)]%
        {hong2023prediction}
\bibfield{author}{\bibinfo{person}{Sun-ha Hong}.} \bibinfo{year}{2023}\natexlab{}.
\newblock \showarticletitle{Prediction as extraction of discretion}.
\newblock \bibinfo{journal}{\emph{Big Data \& Society}} \bibinfo{volume}{10}, \bibinfo{number}{1} (\bibinfo{year}{2023}), \bibinfo{pages}{20539517231171053}.
\newblock


\bibitem[IFOW(2022)]%
        {ifow}
\bibfield{author}{\bibinfo{person}{IFOW}.} \bibinfo{year}{2022}\natexlab{}.
\newblock \bibinfo{title}{All the Ways Hiring Algorithms Can Introduce Bias}.
\newblock
\newblock
\newblock
\shownote{Institute for the Future of Work. \url{https://tinyurl.com/4xwn6p79}}.


\bibitem[Ilgen et~al\mbox{.}(1979)]%
        {ilgen1979consequences}
\bibfield{author}{\bibinfo{person}{Daniel~R Ilgen}, \bibinfo{person}{Cynthia~D Fisher}, {and} \bibinfo{person}{M~Susan Taylor}.} \bibinfo{year}{1979}\natexlab{}.
\newblock \showarticletitle{Consequences of individual feedback on behavior in organizations}.
\newblock \bibinfo{journal}{\emph{Journal of applied psychology}} \bibinfo{volume}{64}, \bibinfo{number}{4} (\bibinfo{year}{1979}), \bibinfo{pages}{349}.
\newblock


\bibitem[Jorgensen et~al\mbox{.}(2023)]%
        {jorgensen2023not}
\bibfield{author}{\bibinfo{person}{Mackenzie Jorgensen}, \bibinfo{person}{Hannah Richert}, \bibinfo{person}{Elizabeth Black}, \bibinfo{person}{Natalia Criado}, {and} \bibinfo{person}{Jose Such}.} \bibinfo{year}{2023}\natexlab{}.
\newblock \showarticletitle{Not so fair: The impact of presumably fair machine learning models}. In \bibinfo{booktitle}{\emph{Proceedings of the 2023 AAAI/ACM Conference on AI, Ethics, and Society}}. \bibinfo{pages}{297--311}.
\newblock


\bibitem[Jost and Hunyady(2005)]%
        {jost2005antecedents}
\bibfield{author}{\bibinfo{person}{John~T Jost} {and} \bibinfo{person}{Orsolya Hunyady}.} \bibinfo{year}{2005}\natexlab{}.
\newblock \showarticletitle{Antecedents and consequences of system-justifying ideologies}.
\newblock \bibinfo{journal}{\emph{Current directions in psychological science}} \bibinfo{volume}{14}, \bibinfo{number}{5} (\bibinfo{year}{2005}), \bibinfo{pages}{260--265}.
\newblock


\bibitem[Kasinidou et~al\mbox{.}(2021)]%
        {kasinidou2021agree}
\bibfield{author}{\bibinfo{person}{Maria Kasinidou}, \bibinfo{person}{Styliani Kleanthous}, \bibinfo{person}{P{\i}nar Barlas}, {and} \bibinfo{person}{Jahna Otterbacher}.} \bibinfo{year}{2021}\natexlab{}.
\newblock \showarticletitle{I agree with the decision, but they didn't deserve this: Future developers' perception of fairness in algorithmic decisions}. In \bibinfo{booktitle}{\emph{Proceedings of the 2021 acm conference on fairness, accountability, and transparency}}. \bibinfo{pages}{690--700}.
\newblock


\bibitem[Kasirzadeh(2022)]%
        {kasirzadeh2022algorithmic}
\bibfield{author}{\bibinfo{person}{Atoosa Kasirzadeh}.} \bibinfo{year}{2022}\natexlab{}.
\newblock \showarticletitle{Algorithmic fairness and structural injustice: Insights from feminist political philosophy}. In \bibinfo{booktitle}{\emph{Proceedings of the 2022 AAAI/ACM Conference on AI, Ethics, and Society}}. \bibinfo{pages}{349--356}.
\newblock


\bibitem[Kieslich et~al\mbox{.}(2022)]%
        {kieslich2022artificial}
\bibfield{author}{\bibinfo{person}{Kimon Kieslich}, \bibinfo{person}{Birte Keller}, {and} \bibinfo{person}{Christopher Starke}.} \bibinfo{year}{2022}\natexlab{}.
\newblock \showarticletitle{Artificial intelligence ethics by design. Evaluating public perception on the importance of ethical design principles of artificial intelligence}.
\newblock \bibinfo{journal}{\emph{Big Data \& Society}} \bibinfo{volume}{9}, \bibinfo{number}{1} (\bibinfo{year}{2022}), \bibinfo{pages}{20539517221092956}.
\newblock


\bibitem[Kleinberg et~al\mbox{.}(2016)]%
        {kleinberg2016inherent}
\bibfield{author}{\bibinfo{person}{Jon Kleinberg}, \bibinfo{person}{Sendhil Mullainathan}, {and} \bibinfo{person}{Manish Raghavan}.} \bibinfo{year}{2016}\natexlab{}.
\newblock \showarticletitle{Inherent trade-offs in the fair determination of risk scores}.
\newblock \bibinfo{journal}{\emph{arXiv preprint arXiv:1609.05807}} (\bibinfo{year}{2016}).
\newblock


\bibitem[Klineberg and Kravitz(2003)]%
        {klineberg2003ethnic}
\bibfield{author}{\bibinfo{person}{Stephen~L Klineberg} {and} \bibinfo{person}{David~A Kravitz}.} \bibinfo{year}{2003}\natexlab{}.
\newblock \showarticletitle{Ethnic differences in predictors of support for municipal affirmative action contracting}.
\newblock \bibinfo{journal}{\emph{Social Science Quarterly}} \bibinfo{volume}{84}, \bibinfo{number}{2} (\bibinfo{year}{2003}), \bibinfo{pages}{425--440}.
\newblock


\bibitem[Knowles et~al\mbox{.}(2023)]%
        {knowles2023trustworthy}
\bibfield{author}{\bibinfo{person}{Bran Knowles}, \bibinfo{person}{Jasmine Fledderjohann}, \bibinfo{person}{John~T Richards}, {and} \bibinfo{person}{Kush~R Varshney}.} \bibinfo{year}{2023}\natexlab{}.
\newblock \showarticletitle{Trustworthy AI and the Logics of Intersectional Resistance}. In \bibinfo{booktitle}{\emph{Proceedings of the 2023 ACM Conference on Fairness, Accountability, and Transparency}}. \bibinfo{pages}{172--182}.
\newblock


\bibitem[Knowles et~al\mbox{.}(2014)]%
        {knowles2014deny}
\bibfield{author}{\bibinfo{person}{Eric~D Knowles}, \bibinfo{person}{Brian~S Lowery}, \bibinfo{person}{Rosalind~M Chow}, {and} \bibinfo{person}{Miguel~M Unzueta}.} \bibinfo{year}{2014}\natexlab{}.
\newblock \showarticletitle{Deny, distance, or dismantle? How white Americans manage a privileged identity}.
\newblock \bibinfo{journal}{\emph{Perspectives on Psychological Science}} \bibinfo{volume}{9}, \bibinfo{number}{6} (\bibinfo{year}{2014}), \bibinfo{pages}{594--609}.
\newblock


\bibitem[Kraus et~al\mbox{.}(2022)]%
        {kraus2022framing}
\bibfield{author}{\bibinfo{person}{Michael~W Kraus}, \bibinfo{person}{Sa-kiera~TJ Hudson}, {and} \bibinfo{person}{Jennifer~A Richeson}.} \bibinfo{year}{2022}\natexlab{}.
\newblock \showarticletitle{Framing, context, and the misperception of Black--White wealth inequality}.
\newblock \bibinfo{journal}{\emph{Social Psychological and Personality Science}} \bibinfo{volume}{13}, \bibinfo{number}{1} (\bibinfo{year}{2022}), \bibinfo{pages}{4--13}.
\newblock


\bibitem[Kraus et~al\mbox{.}(2017)]%
        {kraus2017americans}
\bibfield{author}{\bibinfo{person}{Michael~W Kraus}, \bibinfo{person}{Julian~M Rucker}, {and} \bibinfo{person}{Jennifer~A Richeson}.} \bibinfo{year}{2017}\natexlab{}.
\newblock \showarticletitle{Americans misperceive racial economic equality}.
\newblock \bibinfo{journal}{\emph{Proceedings of the National Academy of Sciences}} \bibinfo{volume}{114}, \bibinfo{number}{39} (\bibinfo{year}{2017}), \bibinfo{pages}{10324--10331}.
\newblock


\bibitem[Kravitz and Platania(1993)]%
        {kravitz1993attitudes}
\bibfield{author}{\bibinfo{person}{David~A Kravitz} {and} \bibinfo{person}{Judith Platania}.} \bibinfo{year}{1993}\natexlab{}.
\newblock \showarticletitle{Attitudes and beliefs about affirmative action: Effects of target and of respondent sex and ethnicity.}
\newblock \bibinfo{journal}{\emph{Journal of applied psychology}} \bibinfo{volume}{78}, \bibinfo{number}{6} (\bibinfo{year}{1993}), \bibinfo{pages}{928}.
\newblock


\bibitem[Kubin et~al\mbox{.}(2023)]%
        {kubin2023reducing}
\bibfield{author}{\bibinfo{person}{Emily Kubin}, \bibinfo{person}{Kurt~J Gray}, {and} \bibinfo{person}{Christian von Sikorski}.} \bibinfo{year}{2023}\natexlab{}.
\newblock \showarticletitle{Reducing political dehumanization by pairing facts with personal experiences}.
\newblock \bibinfo{journal}{\emph{Political Psychology}} \bibinfo{volume}{44}, \bibinfo{number}{5} (\bibinfo{year}{2023}), \bibinfo{pages}{1119--1140}.
\newblock


\bibitem[Kubin et~al\mbox{.}(2021)]%
        {kubin2021personal}
\bibfield{author}{\bibinfo{person}{Emily Kubin}, \bibinfo{person}{Curtis Puryear}, \bibinfo{person}{Chelsea Schein}, {and} \bibinfo{person}{Kurt Gray}.} \bibinfo{year}{2021}\natexlab{}.
\newblock \showarticletitle{Personal experiences bridge moral and political divides better than facts}.
\newblock \bibinfo{journal}{\emph{Proceedings of the National Academy of Sciences}} \bibinfo{volume}{118}, \bibinfo{number}{6} (\bibinfo{year}{2021}), \bibinfo{pages}{e2008389118}.
\newblock


\bibitem[Kuo et~al\mbox{.}(2020)]%
        {kuo2020high}
\bibfield{author}{\bibinfo{person}{Entung~Enya Kuo}, \bibinfo{person}{Michael~W Kraus}, {and} \bibinfo{person}{Jennifer~A Richeson}.} \bibinfo{year}{2020}\natexlab{}.
\newblock \showarticletitle{High-status exemplars and the misperception of the Asian-White wealth gap}.
\newblock \bibinfo{journal}{\emph{Social Psychological and Personality Science}} \bibinfo{volume}{11}, \bibinfo{number}{3} (\bibinfo{year}{2020}), \bibinfo{pages}{397--405}.
\newblock


\bibitem[Kusner et~al\mbox{.}(2017)]%
        {kusner2017counterfactual}
\bibfield{author}{\bibinfo{person}{Matt~J Kusner}, \bibinfo{person}{Joshua Loftus}, \bibinfo{person}{Chris Russell}, {and} \bibinfo{person}{Ricardo Silva}.} \bibinfo{year}{2017}\natexlab{}.
\newblock \showarticletitle{Counterfactual fairness}.
\newblock \bibinfo{journal}{\emph{Advances in neural information processing systems}}  \bibinfo{volume}{30} (\bibinfo{year}{2017}).
\newblock


\bibitem[Langer et~al\mbox{.}(2021)]%
        {langer2021look}
\bibfield{author}{\bibinfo{person}{Markus Langer}, \bibinfo{person}{Tim Hunsicker}, \bibinfo{person}{Tina Feldkamp}, \bibinfo{person}{Cornelius~J. König}, {and} \bibinfo{person}{Nina Grgić-Hlača}.} \bibinfo{year}{2021}\natexlab{}.
\newblock \bibinfo{title}{"Look! It's a Computer Program! It's an Algorithm! It's AI!": Does Terminology Affect Human Perceptions and Evaluations of Intelligent Systems?}
\newblock
\newblock
\showeprint[arxiv]{2108.11486}~[cs.HC]


\bibitem[Langer et~al\mbox{.}(2019)]%
        {langer2019highly}
\bibfield{author}{\bibinfo{person}{Markus Langer}, \bibinfo{person}{Cornelius~J K{\"o}nig}, {and} \bibinfo{person}{Maria Papathanasiou}.} \bibinfo{year}{2019}\natexlab{}.
\newblock \showarticletitle{Highly automated job interviews: Acceptance under the influence of stakes}.
\newblock \bibinfo{journal}{\emph{International Journal of Selection and Assessment}} \bibinfo{volume}{27}, \bibinfo{number}{3} (\bibinfo{year}{2019}), \bibinfo{pages}{217--234}.
\newblock


\bibitem[Lee et~al\mbox{.}(2024)]%
        {lee2024algorithms}
\bibfield{author}{\bibinfo{person}{Jinsook Lee}, \bibinfo{person}{Emma Harvey}, \bibinfo{person}{Joyce Zhou}, \bibinfo{person}{Nikhil Garg}, \bibinfo{person}{Thorsten Joachims}, {and} \bibinfo{person}{Rene~F Kizilcec}.} \bibinfo{year}{2024}\natexlab{}.
\newblock \showarticletitle{Algorithms for College Admissions Decision Support: Impacts of Policy Change and Inherent Variability}.
\newblock \bibinfo{journal}{\emph{arXiv preprint arXiv:2407.11199}} (\bibinfo{year}{2024}).
\newblock


\bibitem[Lee(2018)]%
        {lee2018understanding}
\bibfield{author}{\bibinfo{person}{Min~Kyung Lee}.} \bibinfo{year}{2018}\natexlab{}.
\newblock \showarticletitle{Understanding perception of algorithmic decisions: Fairness, trust, and emotion in response to algorithmic management}.
\newblock \bibinfo{journal}{\emph{Big Data \& Society}} \bibinfo{volume}{5}, \bibinfo{number}{1} (\bibinfo{year}{2018}), \bibinfo{pages}{2053951718756684}.
\newblock


\bibitem[Lee et~al\mbox{.}(2019)]%
        {lee2019procedural}
\bibfield{author}{\bibinfo{person}{Min~Kyung Lee}, \bibinfo{person}{Anuraag Jain}, \bibinfo{person}{Hea~Jin Cha}, \bibinfo{person}{Shashank Ojha}, {and} \bibinfo{person}{Daniel Kusbit}.} \bibinfo{year}{2019}\natexlab{}.
\newblock \showarticletitle{Procedural justice in algorithmic fairness: Leveraging transparency and outcome control for fair algorithmic mediation}.
\newblock \bibinfo{journal}{\emph{Proceedings of the ACM on Human-Computer Interaction}} \bibinfo{volume}{3}, \bibinfo{number}{CSCW} (\bibinfo{year}{2019}), \bibinfo{pages}{1--26}.
\newblock


\bibitem[Lepri et~al\mbox{.}(2018)]%
        {lepri2018fair}
\bibfield{author}{\bibinfo{person}{Bruno Lepri}, \bibinfo{person}{Nuria Oliver}, \bibinfo{person}{Emmanuel Letouz{\'e}}, \bibinfo{person}{Alex Pentland}, {and} \bibinfo{person}{Patrick Vinck}.} \bibinfo{year}{2018}\natexlab{}.
\newblock \showarticletitle{Fair, transparent, and accountable algorithmic decision-making processes: The premise, the proposed solutions, and the open challenges}.
\newblock \bibinfo{journal}{\emph{Philosophy \& Technology}} \bibinfo{volume}{31}, \bibinfo{number}{4} (\bibinfo{year}{2018}), \bibinfo{pages}{611--627}.
\newblock


\bibitem[Lin and Chen(2022)]%
        {lin2022artificial}
\bibfield{author}{\bibinfo{person}{Ting-An Lin} {and} \bibinfo{person}{Po-Hsuan~Cameron Chen}.} \bibinfo{year}{2022}\natexlab{}.
\newblock \showarticletitle{Artificial Intelligence in a Structurally Unjust Society}.
\newblock \bibinfo{journal}{\emph{Feminist Philosophy Quarterly}} \bibinfo{volume}{8}, \bibinfo{number}{3/4} (\bibinfo{year}{2022}).
\newblock


\bibitem[London(2003)]%
        {london2003job}
\bibfield{author}{\bibinfo{person}{Manuel London}.} \bibinfo{year}{2003}\natexlab{}.
\newblock \bibinfo{booktitle}{\emph{Job feedback: Giving, seeking, and using feedback for performance improvement}}.
\newblock \bibinfo{publisher}{Psychology Press}.
\newblock


\bibitem[Lord et~al\mbox{.}(1979)]%
        {lord1979biased}
\bibfield{author}{\bibinfo{person}{Charles~G Lord}, \bibinfo{person}{Lee Ross}, {and} \bibinfo{person}{Mark~R Lepper}.} \bibinfo{year}{1979}\natexlab{}.
\newblock \showarticletitle{Biased assimilation and attitude polarization: The effects of prior theories on subsequently considered evidence}.
\newblock \bibinfo{journal}{\emph{Journal of personality and social psychology}} \bibinfo{volume}{37}, \bibinfo{number}{11} (\bibinfo{year}{1979}), \bibinfo{pages}{2098}.
\newblock


\bibitem[Michelle(2010)]%
        {michelle2010new}
\bibfield{author}{\bibinfo{person}{Alexander Michelle}.} \bibinfo{year}{2010}\natexlab{}.
\newblock \bibinfo{title}{The new Jim Crow: Mass incarceration in the age of colorblindness}.
\newblock
\newblock


\bibitem[Minow(2023)]%
        {minow2023equality}
\bibfield{author}{\bibinfo{person}{Martha Minow}.} \bibinfo{year}{2023}\natexlab{}.
\newblock \showarticletitle{Equality, Equity, and Algorithms: Learning from Justice Rosalie Abella}.
\newblock \bibinfo{journal}{\emph{University of Toronto Law Journal}} \bibinfo{volume}{73}, \bibinfo{number}{Supplement 2} (\bibinfo{year}{2023}), \bibinfo{pages}{163--178}.
\newblock


\bibitem[Mishel(2016)]%
        {mishel2016discrimination}
\bibfield{author}{\bibinfo{person}{Emma Mishel}.} \bibinfo{year}{2016}\natexlab{}.
\newblock \showarticletitle{Discrimination against queer women in the US workforce: A r{\'e}sum{\'e} audit study}.
\newblock \bibinfo{journal}{\emph{Socius}}  \bibinfo{volume}{2} (\bibinfo{year}{2016}), \bibinfo{pages}{2378023115621316}.
\newblock


\bibitem[Mitchell et~al\mbox{.}(2021)]%
        {mitchell2021algorithmic}
\bibfield{author}{\bibinfo{person}{Shira Mitchell}, \bibinfo{person}{Eric Potash}, \bibinfo{person}{Solon Barocas}, \bibinfo{person}{Alexander D'Amour}, {and} \bibinfo{person}{Kristian Lum}.} \bibinfo{year}{2021}\natexlab{}.
\newblock \showarticletitle{Algorithmic fairness: Choices, assumptions, and definitions}.
\newblock \bibinfo{journal}{\emph{Annual review of statistics and its application}} \bibinfo{volume}{8}, \bibinfo{number}{1} (\bibinfo{year}{2021}), \bibinfo{pages}{141--163}.
\newblock


\bibitem[Nelson et~al\mbox{.}(2013)]%
        {nelson2013marley}
\bibfield{author}{\bibinfo{person}{Jessica~C Nelson}, \bibinfo{person}{Glenn Adams}, {and} \bibinfo{person}{Phia~S Salter}.} \bibinfo{year}{2013}\natexlab{}.
\newblock \showarticletitle{The Marley hypothesis: Denial of racism reflects ignorance of history}.
\newblock \bibinfo{journal}{\emph{Psychological science}} \bibinfo{volume}{24}, \bibinfo{number}{2} (\bibinfo{year}{2013}), \bibinfo{pages}{213--218}.
\newblock


\bibitem[Obermeyer et~al\mbox{.}(2019)]%
        {obermeyer2019dissecting}
\bibfield{author}{\bibinfo{person}{Ziad Obermeyer}, \bibinfo{person}{Brian Powers}, \bibinfo{person}{Christine Vogeli}, {and} \bibinfo{person}{Sendhil Mullainathan}.} \bibinfo{year}{2019}\natexlab{}.
\newblock \showarticletitle{Dissecting racial bias in an algorithm used to manage the health of populations}.
\newblock \bibinfo{journal}{\emph{Science}} \bibinfo{volume}{366}, \bibinfo{number}{6464} (\bibinfo{year}{2019}), \bibinfo{pages}{447--453}.
\newblock


\bibitem[Onyeador et~al\mbox{.}(2023)]%
        {onyeador2023misperception}
\bibfield{author}{\bibinfo{person}{Ivuoma~Ngozi Onyeador}, \bibinfo{person}{Sa-kiera Tiarra~Jolynn Hudson}, \bibinfo{person}{Julian Rucker}, \bibinfo{person}{Natalie Daumeyer}, \bibinfo{person}{Kate Zendell}, \bibinfo{person}{Eliette Albrecht}, \bibinfo{person}{Michael Kraus}, {and} \bibinfo{person}{Jennifer Richeson}.} \bibinfo{year}{2023}\natexlab{}.
\newblock \showarticletitle{The Misperception of Gender Economic Equality}.
\newblock  (\bibinfo{year}{2023}).
\newblock


\bibitem[Palan and Schitter(2018)]%
        {palan2018prolific}
\bibfield{author}{\bibinfo{person}{Stefan Palan} {and} \bibinfo{person}{Christian Schitter}.} \bibinfo{year}{2018}\natexlab{}.
\newblock \showarticletitle{Prolific. ac—A subject pool for online experiments}.
\newblock \bibinfo{journal}{\emph{Journal of Behavioral and Experimental Finance}}  \bibinfo{volume}{17} (\bibinfo{year}{2018}), \bibinfo{pages}{22--27}.
\newblock


\bibitem[Pessach and Shmueli(2022)]%
        {pessach2022review}
\bibfield{author}{\bibinfo{person}{Dana Pessach} {and} \bibinfo{person}{Erez Shmueli}.} \bibinfo{year}{2022}\natexlab{}.
\newblock \showarticletitle{A review on fairness in machine learning}.
\newblock \bibinfo{journal}{\emph{ACM Computing Surveys (CSUR)}} \bibinfo{volume}{55}, \bibinfo{number}{3} (\bibinfo{year}{2022}), \bibinfo{pages}{1--44}.
\newblock


\bibitem[Pethig and Kroenung(2023)]%
        {pethig2023biased}
\bibfield{author}{\bibinfo{person}{Florian Pethig} {and} \bibinfo{person}{Julia Kroenung}.} \bibinfo{year}{2023}\natexlab{}.
\newblock \showarticletitle{Biased humans, (un)biased algorithms?}
\newblock \bibinfo{journal}{\emph{Journal of Business Ethics}} \bibinfo{volume}{183}, \bibinfo{number}{3} (\bibinfo{year}{2023}), \bibinfo{pages}{637--652}.
\newblock


\bibitem[Pierson(2017)]%
        {pierson2017gender}
\bibfield{author}{\bibinfo{person}{Emma Pierson}.} \bibinfo{year}{2017}\natexlab{}.
\newblock \showarticletitle{Gender differences in beliefs about algorithmic fairness}.
\newblock  (\bibinfo{year}{2017}).
\newblock
\urldef\tempurl%
\url{arXiv:1712.09124}
\showURL{%
\tempurl}


\bibitem[Plane et~al\mbox{.}(2017)]%
        {plane2017exploring}
\bibfield{author}{\bibinfo{person}{Angelisa~C Plane}, \bibinfo{person}{Elissa~M Redmiles}, \bibinfo{person}{Michelle~L Mazurek}, {and} \bibinfo{person}{Michael~Carl Tschantz}.} \bibinfo{year}{2017}\natexlab{}.
\newblock \showarticletitle{Exploring user perceptions of discrimination in online targeted advertising}. In \bibinfo{booktitle}{\emph{proc. of the USENIX Security Symposium}}. \bibinfo{pages}{935--951}.
\newblock


\bibitem[Rahwan(2018)]%
        {rahwan2018society}
\bibfield{author}{\bibinfo{person}{Iyad Rahwan}.} \bibinfo{year}{2018}\natexlab{}.
\newblock \showarticletitle{Society-in-the-loop: programming the algorithmic social contract}.
\newblock \bibinfo{journal}{\emph{Ethics and information technology}} \bibinfo{volume}{20}, \bibinfo{number}{1} (\bibinfo{year}{2018}), \bibinfo{pages}{5--14}.
\newblock


\bibitem[Rios and Stein(2023)]%
        {guardianaffirmatice}
\bibfield{author}{\bibinfo{person}{Edwin Rios} {and} \bibinfo{person}{Chris Stein}.} \bibinfo{year}{2023}\natexlab{}.
\newblock \bibinfo{title}{US supreme court rules against affirmative action in Harvard and UNC cases}.
\newblock
\newblock
\newblock
\shownote{The Guardian. \url{https://tinyurl.com/murubuu2}}.


\bibitem[RJ(1986)]%
        {rj1986interactional}
\bibfield{author}{\bibinfo{person}{BIES RJ}.} \bibinfo{year}{1986}\natexlab{}.
\newblock \showarticletitle{Interactional justice: Communication criteria of fairness}.
\newblock \bibinfo{journal}{\emph{Research on negotiation in organizations}}  \bibinfo{volume}{1} (\bibinfo{year}{1986}), \bibinfo{pages}{43--55}.
\newblock


\bibitem[Rosette et~al\mbox{.}(2018)]%
        {rosette2018intersectionality}
\bibfield{author}{\bibinfo{person}{Ashleigh~Shelby Rosette}, \bibinfo{person}{Rebecca~Ponce de Leon}, \bibinfo{person}{Christy~Zhou Koval}, {and} \bibinfo{person}{David~A Harrison}.} \bibinfo{year}{2018}\natexlab{}.
\newblock \showarticletitle{Intersectionality: Connecting experiences of gender with race at work}.
\newblock \bibinfo{journal}{\emph{Research in Organizational Behavior}}  \bibinfo{volume}{38} (\bibinfo{year}{2018}), \bibinfo{pages}{1--22}.
\newblock


\bibitem[Sankin et~al\mbox{.}(2021)]%
        {markup2021crime}
\bibfield{author}{\bibinfo{person}{Aaron Sankin}, \bibinfo{person}{Dhruv Mehrotra}, \bibinfo{person}{Surya Mattu}, {and} \bibinfo{person}{Annie Gilbertson}.} \bibinfo{year}{2021}\natexlab{}.
\newblock \bibinfo{title}{Crime Prediction Software Promised to Be Free of Biases. New Data Shows It Perpetuates Them}.
\newblock
\newblock
\newblock
\shownote{The Markup. \url{https://tinyurl.com/mujc6hvy}}.


\bibitem[Saxena et~al\mbox{.}(2019)]%
        {saxena2019fairness}
\bibfield{author}{\bibinfo{person}{Nripsuta~Ani Saxena}, \bibinfo{person}{Karen Huang}, \bibinfo{person}{Evan DeFilippis}, \bibinfo{person}{Goran Radanovic}, \bibinfo{person}{David~C Parkes}, {and} \bibinfo{person}{Yang Liu}.} \bibinfo{year}{2019}\natexlab{}.
\newblock \showarticletitle{How do fairness definitions fare? Examining public attitudes towards algorithmic definitions of fairness}. In \bibinfo{booktitle}{\emph{Proceedings of the 2019 AAAI/ACM Conference on AI, Ethics, and Society}}. \bibinfo{pages}{99--106}.
\newblock


\bibitem[Schlicker et~al\mbox{.}(2021)]%
        {schlicker2021expect}
\bibfield{author}{\bibinfo{person}{Nadine Schlicker}, \bibinfo{person}{Markus Langer}, \bibinfo{person}{Sonja~K {\"O}tting}, \bibinfo{person}{Kevin Baum}, \bibinfo{person}{Cornelius~J K{\"o}nig}, {and} \bibinfo{person}{Dieter Wallach}.} \bibinfo{year}{2021}\natexlab{}.
\newblock \showarticletitle{What to expect from opening up ‘black boxes’? Comparing perceptions of justice between human and automated agents}.
\newblock \bibinfo{journal}{\emph{Computers in Human Behavior}}  \bibinfo{volume}{122} (\bibinfo{year}{2021}), \bibinfo{pages}{106837}.
\newblock


\bibitem[Sears et~al\mbox{.}(2024)]%
        {sears2024lgbtq}
\bibfield{author}{\bibinfo{person}{Brad Sears}, \bibinfo{person}{Neko Castleberry}, \bibinfo{person}{Andy Lin}, {and} \bibinfo{person}{Christy Mallory}.} \bibinfo{year}{2024}\natexlab{}.
\newblock \showarticletitle{LGBTQ people’s experiences of workplace discrimination and harassment}.
\newblock  (\bibinfo{year}{2024}).
\newblock


\bibitem[Shin(2020)]%
        {shin2020user}
\bibfield{author}{\bibinfo{person}{Donghee Shin}.} \bibinfo{year}{2020}\natexlab{}.
\newblock \showarticletitle{User perceptions of algorithmic decisions in the personalized AI system: Perceptual evaluation of fairness, accountability, transparency, and explainability}.
\newblock \bibinfo{journal}{\emph{Journal of Broadcasting \& Electronic Media}} \bibinfo{volume}{64}, \bibinfo{number}{4} (\bibinfo{year}{2020}), \bibinfo{pages}{541--565}.
\newblock


\bibitem[Shin and Park(2019)]%
        {shin2019role}
\bibfield{author}{\bibinfo{person}{Donghee Shin} {and} \bibinfo{person}{Yong~Jin Park}.} \bibinfo{year}{2019}\natexlab{}.
\newblock \showarticletitle{Role of fairness, accountability, and transparency in algorithmic affordance}.
\newblock \bibinfo{journal}{\emph{Computers in Human Behavior}}  \bibinfo{volume}{98} (\bibinfo{year}{2019}), \bibinfo{pages}{277--284}.
\newblock


\bibitem[Sidanius et~al\mbox{.}(1996)]%
        {sidanius1996racism}
\bibfield{author}{\bibinfo{person}{Jim Sidanius}, \bibinfo{person}{Felicia Pratto}, {and} \bibinfo{person}{Lawrence Bobo}.} \bibinfo{year}{1996}\natexlab{}.
\newblock \showarticletitle{Racism, conservatism, affirmative action, and intellectual sophistication: A matter of principled conservatism or group dominance?}
\newblock \bibinfo{journal}{\emph{Journal of personality and social psychology}} \bibinfo{volume}{70}, \bibinfo{number}{3} (\bibinfo{year}{1996}), \bibinfo{pages}{476}.
\newblock


\bibitem[Slaughter et~al\mbox{.}(2002)]%
        {slaughter2002black}
\bibfield{author}{\bibinfo{person}{Jerel~E Slaughter}, \bibinfo{person}{Evan~F Sinar}, {and} \bibinfo{person}{Peter~D Bachiochi}.} \bibinfo{year}{2002}\natexlab{}.
\newblock \showarticletitle{Black applicants' reactions to affirmative action plans: Effects of plan content and previous experience with discrimination.}
\newblock \bibinfo{journal}{\emph{Journal of Applied Psychology}} \bibinfo{volume}{87}, \bibinfo{number}{2} (\bibinfo{year}{2002}), \bibinfo{pages}{333}.
\newblock


\bibitem[Smith et~al\mbox{.}(2020)]%
        {smith2020exploring}
\bibfield{author}{\bibinfo{person}{Jessie Smith}, \bibinfo{person}{Nasim Sonboli}, \bibinfo{person}{Casey Fiesler}, {and} \bibinfo{person}{Robin Burke}.} \bibinfo{year}{2020}\natexlab{}.
\newblock \showarticletitle{Exploring user opinions of fairness in recommender systems}.
\newblock \bibinfo{journal}{\emph{arXiv preprint arXiv:2003.06461}} (\bibinfo{year}{2020}).
\newblock


\bibitem[Sniderman and Piazza(1993)]%
        {sniderman1993scar}
\bibfield{author}{\bibinfo{person}{Paul~M Sniderman} {and} \bibinfo{person}{Thomas Piazza}.} \bibinfo{year}{1993}\natexlab{}.
\newblock \bibinfo{booktitle}{\emph{The scar of race}}.
\newblock \bibinfo{publisher}{Harvard University Press}.
\newblock


\bibitem[So and D’Ignazio(2023)]%
        {so2023race}
\bibfield{author}{\bibinfo{person}{Wonyoung So} {and} \bibinfo{person}{Catherine D’Ignazio}.} \bibinfo{year}{2023}\natexlab{}.
\newblock \showarticletitle{Race-neutral vs race-conscious: Using algorithmic methods to evaluate the reparative potential of housing programs}.
\newblock \bibinfo{journal}{\emph{Big Data \& Society}} \bibinfo{volume}{10}, \bibinfo{number}{2} (\bibinfo{year}{2023}), \bibinfo{pages}{20539517231210272}.
\newblock


\bibitem[So et~al\mbox{.}(2022)]%
        {so2022beyond}
\bibfield{author}{\bibinfo{person}{Wonyoung So}, \bibinfo{person}{Pranay Lohia}, \bibinfo{person}{Rakesh Pimplikar}, \bibinfo{person}{AE Hosoi}, {and} \bibinfo{person}{Catherine D'Ignazio}.} \bibinfo{year}{2022}\natexlab{}.
\newblock \showarticletitle{Beyond Fairness: Reparative Algorithms to Address Historical Injustices of Housing Discrimination in the US}. In \bibinfo{booktitle}{\emph{Proceedings of the 2022 ACM Conference on Fairness, Accountability, and Transparency}}. \bibinfo{pages}{988--1004}.
\newblock


\bibitem[Son~Hing et~al\mbox{.}(2002)]%
        {son2002meritocracy}
\bibfield{author}{\bibinfo{person}{Leanne~S Son~Hing}, \bibinfo{person}{D~Ramona Bobocel}, {and} \bibinfo{person}{Mark~P Zanna}.} \bibinfo{year}{2002}\natexlab{}.
\newblock \showarticletitle{Meritocracy and opposition to affirmative action: making concessions in the face of discrimination.}
\newblock \bibinfo{journal}{\emph{Journal of personality and social psychology}} \bibinfo{volume}{83}, \bibinfo{number}{3} (\bibinfo{year}{2002}), \bibinfo{pages}{493}.
\newblock


\bibitem[Srivastava et~al\mbox{.}(2019)]%
        {srivastava2019mathematical}
\bibfield{author}{\bibinfo{person}{Megha Srivastava}, \bibinfo{person}{Hoda Heidari}, {and} \bibinfo{person}{Andreas Krause}.} \bibinfo{year}{2019}\natexlab{}.
\newblock \showarticletitle{Mathematical notions vs. human perception of fairness: A descriptive approach to fairness for machine learning}. In \bibinfo{booktitle}{\emph{Proceedings of the 25th ACM SIGKDD international conference on knowledge discovery \& data mining}}.
\newblock


\bibitem[Starke et~al\mbox{.}(2022)]%
        {starke2022fairness}
\bibfield{author}{\bibinfo{person}{Christopher Starke}, \bibinfo{person}{Janine Baleis}, \bibinfo{person}{Birte Keller}, {and} \bibinfo{person}{Frank Marcinkowski}.} \bibinfo{year}{2022}\natexlab{}.
\newblock \showarticletitle{Fairness perceptions of algorithmic decision-making: A systematic review of the empirical literature}.
\newblock \bibinfo{journal}{\emph{Big Data \& Society}} \bibinfo{volume}{9}, \bibinfo{number}{2} (\bibinfo{year}{2022}), \bibinfo{pages}{20539517221115189}.
\newblock


\bibitem[Theus(2023)]%
        {theus2023striving}
\bibfield{author}{\bibinfo{person}{Anna-Lena Theus}.} \bibinfo{year}{2023}\natexlab{}.
\newblock \showarticletitle{Striving for Affirmative Algorithmic Futures: How the Social Sciences can Promote more Equitable and Just Algorithmic System Design}. In \bibinfo{booktitle}{\emph{Proceedings of the 2023 ACM Conference on Fairness, Accountability, and Transparency}}. \bibinfo{pages}{558--568}.
\newblock


\bibitem[Truxillo and Bauer(2000)]%
        {truxillo2000roles}
\bibfield{author}{\bibinfo{person}{Donald~M Truxillo} {and} \bibinfo{person}{Talya~N Bauer}.} \bibinfo{year}{2000}\natexlab{}.
\newblock \showarticletitle{The Roles of Gender and Affirmative Action Attitude in Reactions to Test Score Use Methods 1}.
\newblock \bibinfo{journal}{\emph{Journal of Applied Social Psychology}} \bibinfo{volume}{30}, \bibinfo{number}{9} (\bibinfo{year}{2000}), \bibinfo{pages}{1812--1828}.
\newblock


\bibitem[Vagias(2006)]%
        {vagias2006likert}
\bibfield{author}{\bibinfo{person}{Wade~M Vagias}.} \bibinfo{year}{2006}\natexlab{}.
\newblock \showarticletitle{Likert-type scale response anchors. clemson international institute for tourism}.
\newblock \bibinfo{journal}{\emph{\& Research Development, Department of Parks, Recreation and Tourism Management, Clemson University}} \bibinfo{volume}{4}, \bibinfo{number}{5} (\bibinfo{year}{2006}).
\newblock


\bibitem[Veselovsky et~al\mbox{.}(2023)]%
        {veselovsky2023prevalence}
\bibfield{author}{\bibinfo{person}{Veniamin Veselovsky}, \bibinfo{person}{Manoel~Horta Ribeiro}, \bibinfo{person}{Philip Cozzolino}, \bibinfo{person}{Andrew Gordon}, \bibinfo{person}{David Rothschild}, {and} \bibinfo{person}{Robert West}.} \bibinfo{year}{2023}\natexlab{}.
\newblock \showarticletitle{Prevalence and prevention of large language model use in crowd work}.
\newblock \bibinfo{journal}{\emph{arXiv preprint arXiv:2310.15683}} (\bibinfo{year}{2023}).
\newblock


\bibitem[Wachter et~al\mbox{.}(2021)]%
        {wachter2021fairness}
\bibfield{author}{\bibinfo{person}{Sandra Wachter}, \bibinfo{person}{Brent Mittelstadt}, {and} \bibinfo{person}{Chris Russell}.} \bibinfo{year}{2021}\natexlab{}.
\newblock \showarticletitle{Why fairness cannot be automated: Bridging the gap between EU non-discrimination law and AI}.
\newblock \bibinfo{journal}{\emph{Computer Law \& Security Review}}  \bibinfo{volume}{41} (\bibinfo{year}{2021}), \bibinfo{pages}{105567}.
\newblock


\bibitem[Wang et~al\mbox{.}(2020)]%
        {wang2020factors}
\bibfield{author}{\bibinfo{person}{Ruotong Wang}, \bibinfo{person}{F~Maxwell Harper}, {and} \bibinfo{person}{Haiyi Zhu}.} \bibinfo{year}{2020}\natexlab{}.
\newblock \showarticletitle{Factors influencing perceived fairness in algorithmic decision-making: Algorithm outcomes, development procedures, and individual differences}. In \bibinfo{booktitle}{\emph{Proceedings of the 2020 CHI conference on human factors in computing systems}}. \bibinfo{pages}{1--14}.
\newblock


\bibitem[Weed(2021)]%
        {nytresume}
\bibfield{author}{\bibinfo{person}{Julie Weed}.} \bibinfo{year}{2021}\natexlab{}.
\newblock \bibinfo{title}{Résumé-Writing Tips to Help You Get Past the A.I. Gatekeepers}.
\newblock
\newblock
\newblock
\shownote{New York Times. \url{https://tinyurl.com/yc2hz9tp}}.


\bibitem[Weerts et~al\mbox{.}(2024)]%
        {weerts2024neutrality}
\bibfield{author}{\bibinfo{person}{Hilde Weerts}, \bibinfo{person}{Rapha{\"e}le Xenidis}, \bibinfo{person}{Fabien Tarissan}, \bibinfo{person}{Henrik~Palmer Olsen}, {and} \bibinfo{person}{Mykola Pechenizkiy}.} \bibinfo{year}{2024}\natexlab{}.
\newblock \showarticletitle{The Neutrality Fallacy: When Algorithmic Fairness Interventions are (Not) Positive Action}. In \bibinfo{booktitle}{\emph{The 2024 ACM Conference on Fairness, Accountability, and Transparency}}. \bibinfo{pages}{2060--2070}.
\newblock


\bibitem[Weinberg(2022)]%
        {weinberg2022rethinking}
\bibfield{author}{\bibinfo{person}{Lindsay Weinberg}.} \bibinfo{year}{2022}\natexlab{}.
\newblock \showarticletitle{Rethinking fairness: an interdisciplinary survey of critiques of hegemonic ML fairness approaches}.
\newblock \bibinfo{journal}{\emph{Journal of Artificial Intelligence Research}}  \bibinfo{volume}{74} (\bibinfo{year}{2022}), \bibinfo{pages}{75--109}.
\newblock


\bibitem[Womick et~al\mbox{.}({[n.\,d.]})]%
        {womick2024moral}
\bibfield{author}{\bibinfo{person}{Jake Womick}, \bibinfo{person}{Daniela Goya-Tocchetto}, \bibinfo{person}{Nicolas~Restrepo Ochoa}, \bibinfo{person}{Carlos Rebollar}, \bibinfo{person}{Kyra Kapsaskis}, \bibinfo{person}{Samuel Pratt}, \bibinfo{person}{B~Keith Payne}, \bibinfo{person}{Stephen Vaisey}, {and} \bibinfo{person}{Kurt Gray}.} \bibinfo{year}{[n.\,d.]}\natexlab{}.
\newblock \showarticletitle{Moral Disagreement across Politics is Explained by Different Assumptions about who is Most Vulnerable to Harm}.
\newblock  (\bibinfo{year}{[n.\,d.]}).
\newblock


\bibitem[Wong(2020)]%
        {wong2020democratizing}
\bibfield{author}{\bibinfo{person}{Pak-Hang Wong}.} \bibinfo{year}{2020}\natexlab{}.
\newblock \showarticletitle{Democratizing algorithmic fairness}.
\newblock \bibinfo{journal}{\emph{Philosophy \& Technology}}  \bibinfo{volume}{33} (\bibinfo{year}{2020}), \bibinfo{pages}{225--244}.
\newblock


\bibitem[Zafar et~al\mbox{.}(2017a)]%
        {zafar2017disp_mistreat}
\bibfield{author}{\bibinfo{person}{Muhammad~Bilal Zafar}, \bibinfo{person}{Isabel Valera}, \bibinfo{person}{Manuel Gomez~Rodriguez}, {and} \bibinfo{person}{Krishna~P Gummadi}.} \bibinfo{year}{2017}\natexlab{a}.
\newblock \showarticletitle{Fairness Beyond Disparate Treatment \& Disparate Impact: Learning Classification without Disparate Mistreatment}. In \bibinfo{booktitle}{\emph{Proceedings of the 26th International World Wide Web Conference}}. \bibinfo{pages}{1171--1180}.
\newblock


\bibitem[Zafar et~al\mbox{.}(2017b)]%
        {zafar2017disp_impact}
\bibfield{author}{\bibinfo{person}{Muhammad~Bilal Zafar}, \bibinfo{person}{Isabel Valera}, \bibinfo{person}{Manuel~Gomez Rogriguez}, {and} \bibinfo{person}{Krishna~P Gummadi}.} \bibinfo{year}{2017}\natexlab{b}.
\newblock \showarticletitle{Fairness Constraints: Mechanisms for Fair Classification}. In \bibinfo{booktitle}{\emph{Proceedings of the 20th International Conference on Artificial Intelligence and Statistics}}. \bibinfo{pages}{962--970}.
\newblock


\bibitem[Zhang and Hosoi(2024)]%
        {zhang2024structural}
\bibfield{author}{\bibinfo{person}{Aurora Zhang} {and} \bibinfo{person}{Anette Hosoi}.} \bibinfo{year}{2024}\natexlab{}.
\newblock \showarticletitle{Structural Interventions and the Dynamics of Inequality}. In \bibinfo{booktitle}{\emph{The 2024 ACM Conference on Fairness, Accountability, and Transparency}}. \bibinfo{pages}{1014--1030}.
\newblock


\bibitem[Zhang(2022)]%
        {zhang2022affirmative}
\bibfield{author}{\bibinfo{person}{Marilyn Zhang}.} \bibinfo{year}{2022}\natexlab{}.
\newblock \showarticletitle{Affirmative Algorithms: Relational Equality as Algorithmic Fairness}. In \bibinfo{booktitle}{\emph{Proceedings of the 2022 ACM Conference on Fairness, Accountability, and Transparency}}. \bibinfo{pages}{495--507}.
\newblock


\bibitem[Zimmermann and Lee-Stronach(2022)]%
        {zimmermann2022proceed}
\bibfield{author}{\bibinfo{person}{Annette Zimmermann} {and} \bibinfo{person}{Chad Lee-Stronach}.} \bibinfo{year}{2022}\natexlab{}.
\newblock \showarticletitle{Proceed with caution}.
\newblock \bibinfo{journal}{\emph{Canadian Journal of Philosophy}} \bibinfo{volume}{52}, \bibinfo{number}{1} (\bibinfo{year}{2022}), \bibinfo{pages}{6--25}.
\newblock


\bibitem[Zink et~al\mbox{.}(2024)]%
        {zink2024race}
\bibfield{author}{\bibinfo{person}{Anna Zink}, \bibinfo{person}{Ziad Obermeyer}, {and} \bibinfo{person}{Emma Pierson}.} \bibinfo{year}{2024}\natexlab{}.
\newblock \showarticletitle{Race adjustments in clinical algorithms can help correct for racial disparities in data quality}.
\newblock \bibinfo{journal}{\emph{Proceedings of the National Academy of Sciences}} \bibinfo{volume}{121}, \bibinfo{number}{34} (\bibinfo{year}{2024}), \bibinfo{pages}{e2402267121}.
\newblock


\end{thebibliography}

\appendix

\newpage

\section{Supplementary Methods}
\label{supp:methods}

Section~\ref{sec:methods} describes our methodology. Below, we present the exact phrasing of our experimental manipulations, vignettes, and measures.

\subsection{Introduction to Bail Decisions}

Participants who took part in the bail decision-making study read the following introduction to bail decisions right after agreeing to the research terms.

\begin{displayquote}
    Bail decisions determine whether a defendant accused of a crime should should remain in custody or be released while awaiting trial or other court proceedings.
    \begin{itemize}
        \item If bail is granted, the defendant can be released from custody by paying a monetary deposit (or providing another form of financial guarantee) and promising to appear in court as required.
        \item If bail is denied, the defendant has to remain in custody until their trial or further court proceedings.
    \end{itemize}
\end{displayquote}

\subsection{Context Manipulations}

Participants assigned to the \emph{contextualized evaluation} condition in the hiring study read the following two paragraphs:

\begin{displayquote}
    Inequality has been a feature of American society since its inception. One setting in which discrimination is still evident to this day involves hiring decisions. Historically, there has been a close connection between how companies make hiring decisions and job candidates’ demographics. \vspace{5pt}
    
    Companies have historically imposed different criteria on job candidates based on their demographics, refusing to hire candidates from historically marginalized groups while prioritizing those from historically privileged groups. These long-standing discriminatory practices have had profound and lasting effects on society. Although some efforts to confront and rectify these disparities exist, some people still face difficulties while others have advantages being hired to this day.
\end{displayquote}

Similarly, those assigned to the same condition but in the bail decision-making study read the following paragraphs:

\begin{displayquote}
    Inequality has been a feature of American society since its inception. One setting in which discrimination is still evident to this day involves bail decisions. Historically, there has been a close connection between how courts make bail decisions and defendants’ demographics. \vspace{5pt}
    
    Courts have historically imposed different criteria on defendants based on their demographics, keeping historically marginalized groups in custody without bail while releasing those from historically privileged groups on bail. These long-standing discriminatory practices have had profound and lasting effects on society. Although some efforts to confront and rectify these disparities exist, some people are still more likely to be kept in custody while others are more likely to be released on bail to this day.
\end{displayquote}

\subsection{Vignettes}

\subsubsection{Hiring Decisions:}

All participants (including those in the control condition) in the hiring decision-making study read the following baseline vignette.

\begin{displayquote}
    An algorithm was developed to assist companies in making hiring decisions. This algorithm was trained on data from several companies' past hiring decisions to learn how to evaluate job candidates. Given a list of job candidates, it determines a score for each candidate based on their resume and cover letter. These scores correspond to an evaluation of the applicant in relation to a job position, with higher scores indicating that the applicant is a better candidate for that particular position. The algorithm then uses these scores to screen job applicants and recommend who should be offered interviews.
\end{displayquote}

Participants randomly assigned to the \emph{affirmative} algorithm type condition additionally read the following paragraph.

\begin{displayquote}
    This algorithm was found to be prioritizing members of demographic groups that have been historically discriminated against in hiring decisions. In other words, given job applicants with similar qualifications, the algorithm was more likely to recommend candidates belonging to historically marginalized groups for interviews.
\end{displayquote}

Those in the \emph{discriminatory} condition read instead the following paragraph:

\begin{displayquote}
    This algorithm was found to be prioritizing members of demographic groups that have been historically favored in hiring decisions. In other words, given job applicants with similar qualifications, the algorithm was more likely to recommend candidates belonging to historically privileged groups for interviews.
\end{displayquote}

Finally, participants assigned to the \emph{fair} algorithm type condition read the following additional information:

\begin{displayquote}
    This algorithm was found to be treating members of all demographic groups equally. In other words, given job applicants with similar qualifications, the algorithm was equally likely to recommend candidates for interviews regardless of their demographic group.
\end{displayquote}

\subsubsection{Bail Decisions:}

All participants who took part in the bail decision-making study read the following baseline vignette:

\begin{displayquote}
   An algorithm was developed to assist courts in making bail decisions. This algorithm was trained on data from several courts' past bail decisions to learn how to evaluate defendants. Given a list of defendants, it determines a score for each defendant based on their current charges and past criminal records. These scores correspond to the risk of the defendant reoffending within 2 years, with higher scores indicating that the applicant is more likely to reoffend within 2 years. The algorithm then uses these scores to screen defendants and recommend who should be released on bail.
\end{displayquote}

Those randomly assigned to the \emph{affirmative} algorithm type condition read the additional paragraph below:

\begin{displayquote}
    This algorithm was found to be prioritizing members of demographic groups that have been historically discriminated against in bail decisions. In other words, given defendants with similar charges and criminal records, the algorithm was more likely to recommend defendants belonging to historically marginalized groups for release on bail.
\end{displayquote}

Similarly, participants in the \emph{discriminatory} condition were shown the following information:

\begin{displayquote}
    This algorithm was found to be prioritizing members of demographic groups that have been historically favored in bail decisions. In other words, given defendants with similar charges and criminal records, the algorithm was more likely to recommend defendants belonging to historically privileged groups for release on bail.
\end{displayquote}

Finally, participants assigned to the \emph{fair} condition read the following paragraph:

\begin{displayquote}
    This algorithm was found to be treating members of all demographic groups equally. In other words, given defendants with similar charges and criminal records, the algorithm was equally likely to recommend defendants for release on bail regardless of their demographic group.
\end{displayquote}

\subsection{Measures}

Participants answered five groups of questions, as described below. Questions groups were presented in random order between participants.

\subsubsection{Perceived Fairness of the Decision-Making Domain:}

Participants assigned to the \emph{contextualized evaluation} and \emph{decontextualized evaluation} context manipulations first evaluated the fairness of the decision-making domain using the following measures:

\begin{itemize}
	\item \textbf{Procedural Fairness: } [Hiring/Bail] decisions in my country 1) are consistent; 2) are free of bias; 3) are based on accurate information; and 4) uphold ethical and moral standards (-3 = Strongly disagree, 3 = Strongly agree; adapted from~\citet{colquitt2001dimensionality}; Cronbach's $\alpha$ = 0.910).
	\item \textbf{Distributive Fairness: } Given similar [qualifications/criminal records], 1) everyone has the same chance to be [hired/granted bail] in my country; 2) people from different demographic groups are equally likely to be [hired/granted bail] in my country ($\alpha$ = 0.910).
	\item \textbf{Interpersonal Fairness: } [Hiring/Bail] decisions in my country 1) treat [job candidates/defendants] with respect; 2) take into consideration the dignity of [job candidates/defendants] ($\alpha$ = 0.890).
	\item \textbf{General Fairness: } [Hiring/Bail] decisions in my country are fair.
\end{itemize}

\subsubsection{Perceptions of the Algorithm:}

Participants evaluated the algorithm depicted in their randomly assigned vignette using the following measures:

\begin{itemize}
	\item \textbf{Fairness: } How fair is it for job candidates that the algorithm makes recommendations? (0 = Not fair at all, 6 = Very fair; adapted from~\citet{lee2018understanding}).
	\item \textbf{Objectivity: } I believe recommendations made by the algorithm 1) are reasonable and logical; 2) objectively consider all the factors; 3) are based on logical analysis; 4) are rational and objective (-3 = Strongly disagree, 3 = Strongly agree; adapted from~\citet{pethig2023biased}; $\alpha$ = 0.950).
	\item \textbf{Trust: } How much do you trust that the algorithm makes good-quality recommendations? (0 = No trust at all, 6 = Extreme trust; adapted from~\citet{lee2018understanding}).
	\item \textbf{Support: } How much do you support or oppose the algorithm making recommendations? (-3 = Strongly oppose, 0 = Neutral, 6 = Strongly support; scale adapted from~\citet{vagias2006likert}).
\end{itemize}

\subsubsection{Perceived Fairness of the Domain After Deployment of the Algorithm:}

All participants then evaluated the fairness of the decision-making context while imagining that the algorithm was deployed in their own country. We employed the same measures as those used to capture people's initial judgments of fairness with minor changes to indicate that this is a hypothetical scenario. For instance, instead of asking whether bail decisions ``were free of bias,'' we asked whether bail decisions ``would be free of bias'' had the algorithm been deployed.

\subsubsection{Open-Ended Questions:}

Participants then answered the following open-ended questions. We report an analysis of the responses to the first question in the main text. We disabled copy-pasting in these questions to mitigate large language model use~\cite{veselovsky2023prevalence}. 

\begin{itemize}
    \item Please, explain which groups come to your mind when you think of historically marginalized groups in your country.
    \item Please, explain which groups come to your mind when you think of historically privileged groups in your country.
    \item Please, explain in your own words what an affirmative algorithm is.
    \item Please, explain in your own words what a fair algorithm is.
    \item Please, explain in your own words what a discriminatory algorithm is.
\end{itemize}

\subsubsection{Personal and Demographic Questions:}

Finally, participants answered several optional questions, ranging from whether they had any experience in professions related to artificial intelligence, law, or human resources to their income level. More relevant to our study, participants indicated their political leaning (``How would you describe your political views?'') using a 5-point scale (Liberal, Somewhat liberal, Moderate, Somewhat conservative, Conservative) alongside the option to not answer the question. Participants also self-reported their racial identity out of the following options: 1) White, 2) Black or African American, 3) Asian, 4) American Indian or Alaska Native, 5) Native Hawaiian or Other Pacific Islander, and 6) Other (with the option to self-describe). Finally, participants also indicated their gender as 1) Female, 2) Male, 3) Non-binary, 4) Transgender, or 5) Other (also with the option to self-describe). Participants were allowed to withhold their racial and gender identity.

\section{ANOVA and Pairwise Comparison Tables}
\label{supp:anova}

Tables~\ref{tab:hiring_DV_experiment}-\ref{tab:hiring_gender_pairwise} present the ANOVA and pairwise comparison tables for our study on hiring decisions. Tables~\ref{tab:bail_DV_experiment}-\ref{tab:bail_gender_pairwise} present the same analysis for the bail decision-making study. These analyses are discussed in detail in Section~\ref{sec:results}.


\begin{table}[ht]
\footnotesize
\centering
\begin{tabular}{lrrrrrr}
  \hline
Parameter & Sum of Squares & df & Mean Squares & $F$ & $p$ & partial $\eta^2$ \\ 
  \hline
  \textbf{Fairness} & & & & & & \\
Context & 2.62 & 2.00 & 1.31 & 0.50 & 0.61 & 0.00 \\ 
  Algorithm Type & 360.09 & 3.00 & 120.03 & 45.63 & 0.00 & 0.19 \\ 
  Context:Algorithm Type & 12.47 & 6.00 & 2.08 & 0.79 & 0.58 & 0.01 \\ 
  Residuals & 1541.39 & 586.00 & 2.63 &  &  &  \\ 
  \\[-1.8ex]
  \textbf{Trust} & & & & & & \\ 
  Context & 10.55 & 2.00 & 5.27 & 2.21 & 0.11 & 0.01 \\ 
  Algorithm Type & 230.41 & 3.00 & 76.80 & 32.23 & 0.00 & 0.14 \\ 
  Context:Algorithm Type & 9.59 & 6.00 & 1.60 & 0.67 & 0.67 & 0.01 \\ 
  Residuals & 1396.49 & 586.00 & 2.38 &  &  &  \\ 
  \\[-1.8ex]
  \textbf{Objectivity} & & & & & & \\ 
  Context & 3.37 & 2.00 & 1.68 & 0.73 & 0.48 & 0.00 \\ 
  Algorithm Type & 318.94 & 3.00 & 106.31 & 45.80 & 0.00 & 0.19 \\ 
  Context:Algorithm Type & 12.87 & 6.00 & 2.14 & 0.92 & 0.48 & 0.01 \\ 
  Residuals & 1360.31 & 586.00 & 2.32 &  &  &  \\ 
  \\[-1.8ex]
  \textbf{Support} & & & & & & \\ 
  Context & 10.72 & 2.00 & 5.36 & 1.93 & 0.15 & 0.01 \\ 
  Algorithm Type & 321.65 & 3.00 & 107.22 & 38.65 & 0.00 & 0.17 \\ 
  Context:Algorithm Type & 11.71 & 6.00 & 1.95 & 0.70 & 0.65 & 0.01 \\ 
  Residuals & 1625.60 & 586.00 & 2.77 &  &  &  \\ 
   \hline
\end{tabular}
    \caption{ANOVA tables for judgments of fairness, trust, objectivity, and support for algorithms deployed in the hiring domain. Independent variables: experimental conditions.}
    \label{tab:hiring_DV_experiment}
\end{table}

\begin{table}[ht]
\footnotesize
\centering
\begin{tabular}{lrrrrl}
  \hline
Contrast & Estimate & $SE$ & df & t-value & $p$ \\ 
  \hline
  \textbf{Fairness} & & & & & \\ 
Control - Affirmative & 0.5451 & 0.1897 & 586 & 2.874 & 0.0252 \\ 
  Control - Discriminatory & 1.4419 & 0.1894 & 586 & 7.614 & $<$.0001 \\ 
  Control - Fair & -0.6707 & 0.1881 & 586 & -3.565 & 0.0024 \\ 
  Affirmative - Discriminatory & 0.8968 & 0.1887 & 586 & 4.753 & $<$.0001 \\ 
  Affirmative - Fair & -1.2158 & 0.1874 & 586 & -6.488 & $<$.0001 \\ 
  Discriminatory - Fair & -2.1126 & 0.1871 & 586 & -11.291 & $<$.0001 \\ 
    \\[-1.8ex]
\textbf{Trust} & & & & & \\ 
  Control - Affirmative & 0.2306 & 0.1805 & 586 & 1.277 & 1.0000 \\
  Control - Discriminatory & 1.2449 & 0.1803 & 586 & 6.906 & $<$.0001 \\ 
  Control - Fair & -0.4397 & 0.1791 & 586 & -2.456 & 0.0861 \\ 
  Affirmative - Discriminatory & 1.0143 & 0.1796 & 586 & 5.648 & $<$.0001 \\ 
  Affirmative - Fair & -0.6703 & 0.1784 & 586 & -3.758 & 0.0011 \\ 
  Discriminatory - Fair & -1.6846 & 0.1781 & 586 & -9.459 & $<$.0001 \\ 
    \\[-1.8ex]
\textbf{Objectivity} & & & & & \\ 
  Control - Affirmative & 0.6582 & 0.1782 & 586 & 3.694 & 0.0014 \\
  Control - Discriminatory & 1.5066 & 0.1779 & 586 & 8.468 & $<$.0001 \\ 
  Control - Fair & -0.4100 & 0.1767 & 586 & -2.320 & 0.1241 \\ 
  Affirmative - Discriminatory & 0.8484 & 0.1772 & 586 & 4.787 & $<$.0001 \\ 
  Affirmative - Fair & -1.0681 & 0.1760 & 586 & -6.068 & $<$.0001 \\ 
  Discriminatory - Fair & -1.9166 & 0.1758 & 586 & -10.903 & $<$.0001 \\
  \\[-1.8ex]
\textbf{Support} & & & & & \\ 
  Control - Affirmative & 0.3425 & 0.1948 & 586 & 1.758 & 0.4753 \\ 
  Control - Discriminatory & 1.4284 & 0.1945 & 586 & 7.344 & $<$.0001 \\ 
  Control - Fair & -0.5949 & 0.1932 & 586 & -3.080 & 0.0130 \\ 
  Affirmative - Discriminatory & 1.0860 & 0.1938 & 586 & 5.605 & $<$.0001 \\ 
  Affirmative - Fair & -0.9374 & 0.1924 & 586 & -4.871 & $<$.0001 \\ 
  Discriminatory - Fair & -2.0234 & 0.1922 & 586 & -10.530 & $<$.0001 \\ 
   \hline
\end{tabular}
    \caption{Pairwise comparisons of judgments of fairness, trust, objectivity, and support between different types of algorithm deployed in the hiring domain.}
    \label{tab:hiring_DV_pairwise}
\end{table}

\begin{table}[ht]
\footnotesize
\centering
\begin{tabular}{lrrrrrr}
  \hline
Parameter & Sum of Squares & df & Mean Squares & $F$ & $p$ & partial $\eta^2$ \\ 
  \hline
\textbf{Fairness} & & & & & & \\ 
Political Leaning & 2.95 & 1.00 & 2.95 & 1.15 & 0.28 & 0.00 \\ 
  Algorithm Type & 361.00 & 3.00 & 120.33 & 46.88 & 0.00 & 0.20 \\ 
  Political Leaning:Algorithm Type & 28.25 & 3.00 & 9.42 & 3.67 & 0.01 & 0.02 \\ 
  Residuals & 1488.80 & 580.00 & 2.57 &  &  &  \\ 
  \\[-1.8ex]
\textbf{Trust} & & & & & & \\
  Political Leaning & 1.31 & 1.00 & 1.31 & 0.56 & 0.45 & 0.00 \\ 
  Algorithm Type & 237.13 & 3.00 & 79.04 & 33.87 & 0.00 & 0.15 \\ 
  Political Leaning:Algorithm Type & 30.92 & 3.00 & 10.31 & 4.42 & 0.00 & 0.02 \\ 
  Residuals & 1353.73 & 580.00 & 2.33 &  &  &  \\
  \\[-1.8ex]
\textbf{Objectivity} & & & & & & \\
  Political Leaning & 1.93 & 1.00 & 1.93 & 0.85 & 0.36 & 0.00 \\ 
  Algorithm Type & 319.70 & 3.00 & 106.57 & 47.19 & 0.00 & 0.20 \\ 
  Political Leaning:Algorithm Type & 26.15 & 3.00 & 8.72 & 3.86 & 0.01 & 0.02 \\ 
  Residuals & 1309.87 & 580.00 & 2.26 &  &  &  \\
  \\[-1.8ex]
\textbf{Support} & & & & & & \\
  Political Leaning & 0.00 & 1.00 & 0.00 & 0.00 & 0.97 & 0.00 \\ 
  Algorithm Type & 331.64 & 3.00 & 110.55 & 41.37 & 0.00 & 0.18 \\ 
  Political Leaning:Algorithm Type & 49.02 & 3.00 & 16.34 & 6.11 & 0.00 & 0.03 \\ 
  Residuals & 1549.98 & 580.00 & 2.67 &  &  &  \\ 
   \hline
\end{tabular}
    \caption{ANCOVA tables for judgments of fairness, trust, objectivity, and support for algorithms deployed in the hiring domain. Independent variables: algorithm type and political orientation.}
    \label{tab:hiring_DV_Political Leaning}
\end{table}

\begin{table}[ht]
\footnotesize
\centering
\begin{tabular}{lrrrrl}
  \hline
Contrast & Estimate & $SE$ & df & t-value & $p$ \\ 
  \hline
\textbf{Fairness} & & & & & \\
Control - Affirmative & -0.1893 & 0.1370 & 580 & -1.382 & 1.0000 \\ 
  Control - Discriminatory & 0.2272 & 0.1396 & 580 & 1.627 & 0.6254 \\ 
  Control - Fair & -0.1522 & 0.1394 & 580 & -1.092 & 1.0000 \\ 
  Affirmative - Discriminatory & 0.4165 & 0.1385 & 580 & 3.008 & 0.0165 \\ 
  Affirmative - Fair & 0.0371 & 0.1382 & 580 & 0.268 & 1.0000 \\ 
  Discriminatory - Fair & -0.3794 & 0.1408 & 580 & -2.694 & 0.0436 \\
  \\[-1.8ex]
\textbf{Trust} & & & & & \\
  Control - Affirmative & -0.3517 & 0.1306 & 580 & -2.693 & 0.0437 \\ 
  Control - Discriminatory & 0.0721 & 0.1331 & 580 & 0.542 & 1.0000 \\ 
  Control - Fair & -0.2211 & 0.1329 & 580 & -1.664 & 0.5802 \\ 
  Affirmative - Discriminatory & 0.4238 & 0.1320 & 580 & 3.210 & 0.0084 \\ 
  Affirmative - Fair & 0.1307 & 0.1318 & 580 & 0.991 & 1.0000 \\ 
  Discriminatory - Fair & -0.2932 & 0.1343 & 580 & -2.183 & 0.1764 \\
  \\[-1.8ex]
\textbf{Objectivity} & & & & & \\
  Control - Affirmative & -0.3050 & 0.1285 & 580 & -2.374 & 0.1076 \\ 
  Control - Discriminatory & 0.0881 & 0.1309 & 580 & 0.673 & 1.0000 \\ 
  Control - Fair & -0.2020 & 0.1307 & 580 & -1.545 & 0.7373 \\ 
  Affirmative - Discriminatory & 0.3931 & 0.1299 & 580 & 3.027 & 0.0155 \\ 
  Affirmative - Fair & 0.1030 & 0.1297 & 580 & 0.795 & 1.0000 \\ 
  Discriminatory - Fair & -0.2901 & 0.1321 & 580 & -2.196 & 0.1709 \\
  \\[-1.8ex]
\textbf{Support} & & & & & \\
  Control - Affirmative & -0.4062 & 0.1398 & 580 & -2.906 & 0.0228 \\ 
  Control - Discriminatory & 0.1400 & 0.1424 & 580 & 0.983 & 1.0000 \\ 
  Control - Fair & -0.2613 & 0.1422 & 580 & -1.838 & 0.3995 \\ 
  Affirmative - Discriminatory & 0.5462 & 0.1413 & 580 & 3.866 & 0.0007 \\ 
  Affirmative - Fair & 0.1448 & 0.1410 & 580 & 1.027 & 1.0000 \\ 
  Discriminatory - Fair & -0.4013 & 0.1437 & 580 & -2.793 & 0.0324 \\ 
   \hline
\end{tabular}
    \caption{Pairwise comparisons of the slopes of judgments of fairness, trust, objectivity, and support in relation to political orientation between different types of algorithm deployed in the hiring domain.}
    \label{tab:hiring_Political Leaning_pairwise}
\end{table}

\begin{table}[ht]
\footnotesize
\centering
\begin{tabular}{lrrrrrr}
  \hline
Parameter & Sum of Squares & df & Mean Squares & $F$ & $p$ & partial $\eta^2$ \\ 
  \hline
\textbf{Fairness} & & & & & & \\ 
Racial Group & 0.35 & 1.00 & 0.35 & 0.13 & 0.72 & 0.00 \\ 
  Algorithm Type & 359.19 & 3.00 & 119.73 & 45.51 & 0.00 & 0.19 \\ 
  Racial Group:Algorithm Type & 4.02 & 3.00 & 1.34 & 0.51 & 0.68 & 0.00 \\ 
  Residuals & 1552.11 & 590.00 & 2.63 &  &  &  \\ 
  \\[-1.8ex]
\textbf{Trust} & & & & & & \\ 
  Racial Group & 1.76 & 1.00 & 1.76 & 0.73 & 0.39 & 0.00 \\ 
  Algorithm Type & 230.99 & 3.00 & 77.00 & 32.13 & 0.00 & 0.14 \\ 
  Racial Group:Algorithm Type & 1.12 & 3.00 & 0.37 & 0.16 & 0.93 & 0.00 \\ 
  Residuals & 1413.76 & 590.00 & 2.40 &  &  &  \\ 
  \\[-1.8ex]
\textbf{Objectivity} & & & & & & \\ 
  Racial Group & 0.57 & 1.00 & 0.57 & 0.24 & 0.62 & 0.00 \\ 
  Algorithm Type & 320.48 & 3.00 & 106.83 & 46.03 & 0.00 & 0.19 \\ 
  Racial Group:Algorithm Type & 6.77 & 3.00 & 2.26 & 0.97 & 0.41 & 0.00 \\ 
  Residuals & 1369.21 & 590.00 & 2.32 &  &  &  \\
  \\[-1.8ex]
\textbf{Support} & & & & & & \\ 
  Racial Group & 4.52 & 1.00 & 4.52 & 1.63 & 0.20 & 0.00 \\ 
  Algorithm Type & 320.44 & 3.00 & 106.81 & 38.49 & 0.00 & 0.16 \\ 
  Racial Group:Algorithm Type & 6.27 & 3.00 & 2.09 & 0.75 & 0.52 & 0.00 \\ 
  Residuals & 1637.24 & 590.00 & 2.77 &  &  &  \\ 
   \hline
\end{tabular}
    \caption{ANOVA tables for judgments of fairness, trust, objectivity, and support for algorithms deployed in the hiring domain. Independent variables: algorithm type and participants' racial group.}
    \label{tab:hiring_DV_race}
\end{table}

\begin{table}[ht]
\footnotesize
\centering
\begin{tabular}{llrrrrl}
  \hline
Algorithm Type & Contrast & Estimate & $SE$ & df & t-value & $p$ \\ 
  \hline
\textbf{Fairness} & & & & & & \\ 
Control & (Non-White) - White & 0.1648 & 0.2873 & 590 & 0.574 & 0.5664 \\ 
  Affirmative & (Non-White) - White & 0.2583 & 0.2833 & 590 & 0.912 & 0.3623 \\ 
  Discriminatory & (Non-White) - White & -0.2014 & 0.2855 & 590 & -0.705 & 0.4808 \\ 
  Fair & (Non-White) - White & -0.0118 & 0.2707 & 590 & -0.043 & 0.9654 \\
  \\[-1.8ex]
\textbf{Trust} & & & & & & \\ 
  Control & (Non-White) - White & 0.0711 & 0.2742 & 590 & 0.259 & 0.7954 \\ 
  Affirmative & (Non-White) - White & -0.0150 & 0.2704 & 590 & -0.056 & 0.9556 \\ 
  Discriminatory & (Non-White) - White & 0.2208 & 0.2725 & 590 & 0.810 & 0.4180 \\ 
  Fair & (Non-White) - White & 0.1776 & 0.2583 & 590 & 0.687 & 0.4920 \\
  \\[-1.8ex]
\textbf{Objectivity} & & & & & & \\ 
  Control & (Non-White) - White & 0.2677 & 0.2698 & 590 & 0.992 & 0.3216 \\ 
  Affirmative & (Non-White) - White & 0.0193 & 0.2661 & 590 & 0.073 & 0.9421 \\ 
  Discriminatory & (Non-White) - White & -0.1842 & 0.2682 & 590 & -0.687 & 0.4925 \\ 
  Fair & (Non-White) - White & -0.3314 & 0.2542 & 590 & -1.304 & 0.1929 \\
    \\[-1.8ex]
\textbf{Support} & & & & & & \\ 
  Control & (Non-White) - White & 0.2962 & 0.2951 & 590 & 1.004 & 0.3160 \\ 
  Affirmative & (Non-White) - White & 0.4933 & 0.2910 & 590 & 1.695 & 0.0905 \\ 
  Discriminatory & (Non-White) - White & -0.0132 & 0.2932 & 590 & -0.045 & 0.9641 \\ 
  Fair & (Non-White) - White & -0.0192 & 0.2780 & 590 & -0.069 & 0.9450 \\ 
   \hline
\end{tabular}
    \caption{Pairwise comparisons of judgments of fairness, trust, objectivity, and support between different racial groups for each type of algorithm deployed in the hiring domain.}
    \label{tab:hiring_race_pairwise}
\end{table}

\begin{table}[ht]
\footnotesize
\centering
\begin{tabular}{lrrrrrr}
  \hline
Parameter & Sum of Squares & df & Mean Squares & $F$ & $p$ & partial $\eta^2$ \\ 
  \hline
\textbf{Fairness} & & & & & & \\ 
Gender Group & 4.29 & 1.00 & 4.29 & 1.64 & 0.20 & 0.00 \\ 
  Algorithm Type & 357.56 & 3.00 & 119.19 & 45.69 & 0.00 & 0.19 \\ 
  Gender Group:Algorithm Type & 13.09 & 3.00 & 4.36 & 1.67 & 0.17 & 0.01 \\ 
  Residuals & 1539.10 & 590.00 & 2.61 &  &  &  \\
  \\[-1.8ex]
\textbf{Trust} & & & & & & \\ 

  Gender Group & 15.10 & 1.00 & 15.10 & 6.41 & 0.01 & 0.01 \\ 
  Algorithm Type & 230.54 & 3.00 & 76.85 & 32.65 & 0.00 & 0.14 \\ 
  Gender Group:Algorithm Type & 12.71 & 3.00 & 4.24 & 1.80 & 0.15 & 0.01 \\ 
  Residuals & 1388.82 & 590.00 & 2.35 &  &  &  \\
  \\[-1.8ex]
\textbf{Objectivity} & & & & & & \\ 

  Gender Group & 5.33 & 1.00 & 5.33 & 2.30 & 0.13 & 0.00 \\ 
  Algorithm Type & 317.05 & 3.00 & 105.68 & 45.73 & 0.00 & 0.19 \\ 
  Gender Group:Algorithm Type & 7.78 & 3.00 & 2.59 & 1.12 & 0.34 & 0.01 \\ 
  Residuals & 1363.43 & 590.00 & 2.31 &  &  &  \\
  \\[-1.8ex]
\textbf{Support} & & & & & & \\ 

  Gender Group & 6.62 & 1.00 & 6.62 & 2.40 & 0.12 & 0.00 \\ 
  Algorithm Type & 321.35 & 3.00 & 107.12 & 38.85 & 0.00 & 0.16 \\ 
  Gender Group:Algorithm Type & 14.75 & 3.00 & 4.92 & 1.78 & 0.15 & 0.01 \\ 
  Residuals & 1626.66 & 590.00 & 2.76 &  &  &  \\ 
   \hline
\end{tabular}
    \caption{ANOVA tables for judgments of fairness, trust, objectivity, and support for algorithms deployed in the hiring domain. Independent variables: algorithm type and participants' gender group.}
    \label{tab:hiring_DV_gender}
\end{table}

\begin{table}[ht]
\footnotesize
\centering
\begin{tabular}{llrrrrl}
  \hline
Algorithm Type & Contrast & Estimate & $SE$ & df & t-value & $p$ \\ 
  \hline
\textbf{Fairness} & & & & & & \\ 

Control & Men - Non-Men & 0.0372 & 0.2676 & 590 & 0.139 & 0.8894 \\ 
  Affirmative & Men - Non-Men & -0.1309 & 0.2691 & 590 & -0.486 & 0.6268 \\ 
  Discriminatory & Men - Non-Men & 0.0934 & 0.2638 & 590 & 0.354 & 0.7235 \\ 
  Fair & Men - Non-Men & 0.6533 & 0.2607 & 590 & 2.506 & 0.0125 \\
  \\[-1.8ex]
\textbf{Trust} & & & & & & \\ 

  Control & Men - Non-Men & 0.0214 & 0.2542 & 590 & 0.084 & 0.9328 \\ 
  Affirmative & Men - Non-Men & 0.1594 & 0.2556 & 590 & 0.624 & 0.5331 \\ 
  Discriminatory & Men - Non-Men & 0.2795 & 0.2506 & 590 & 1.115 & 0.2652 \\ 
  Fair & Men - Non-Men & 0.7898 & 0.2476 & 590 & 3.190 & 0.0015 \\
  \\[-1.8ex]
\textbf{Objectivity} & & & & & & \\ 

  Control & Men - Non-Men & 0.0262 & 0.2518 & 590 & 0.104 & 0.9171 \\ 
  Affirmative & Men - Non-Men & 0.1299 & 0.2533 & 590 & 0.513 & 0.6083 \\ 
  Discriminatory & Men - Non-Men & 0.0158 & 0.2483 & 590 & 0.064 & 0.9492 \\ 
  Fair & Men - Non-Men & 0.5698 & 0.2453 & 590 & 2.323 & 0.0205 \\
  \\[-1.8ex]
\textbf{Support} & & & & & & \\ 

  Control & Men - Non-Men & 0.0162 & 0.2751 & 590 & 0.059 & 0.9532 \\ 
  Affirmative & Men - Non-Men & -0.0394 & 0.2766 & 590 & -0.142 & 0.8868 \\ 
  Discriminatory & Men - Non-Men & 0.1021 & 0.2712 & 590 & 0.376 & 0.7067 \\ 
  Fair & Men - Non-Men & 0.7380 & 0.2680 & 590 & 2.754 & 0.0061 \\ 
   \hline
\end{tabular}
    \caption{Pairwise comparisons of judgments of fairness, trust, objectivity, and support between different gender groups for each type of algorithm deployed in the hiring domain.}
    \label{tab:hiring_gender_pairwise}
\end{table}



\begin{table}[ht]
\footnotesize
\centering
\begin{tabular}{lrrrrrr}
  \hline
Parameter & Sum of Squares & df & Mean Squares & $F$ & $p$ & partial $\eta^2$ \\ 
  \hline
  \textbf{Fairness} & & & & & & \\
  Context & 2.18 & 2.00 & 1.09 & 0.44 & 0.65 & 0.00 \\ 
  Algorithm Type & 460.56 & 3.00 & 153.52 & 61.41 & 0.00 & 0.24 \\ 
  Context:Algorithm Type & 17.93 & 6.00 & 2.99 & 1.20 & 0.31 & 0.01 \\ 
  Residuals & 1457.50 & 583.00 & 2.50 &  &  &  \\
  \\[-1.8ex]
  \textbf{Trust} & & & & & & \\ 
  Context & 3.42 & 2.00 & 1.71 & 0.71 & 0.49 & 0.00 \\ 
  Algorithm Type & 271.19 & 3.00 & 90.40 & 37.53 & 0.00 & 0.16 \\ 
  Context:Algorithm Type & 6.27 & 6.00 & 1.04 & 0.43 & 0.86 & 0.00 \\ 
  Residuals & 1404.06 & 583.00 & 2.41 &  &  &  \\
  \\[-1.8ex]
  \textbf{Objectivity} & & & & & & \\ 
  Context & 2.03 & 2.00 & 1.01 & 0.44 & 0.65 & 0.00 \\ 
  Algorithm Type & 297.95 & 3.00 & 99.32 & 42.87 & 0.00 & 0.18 \\ 
  Context:Algorithm Type & 11.76 & 6.00 & 1.96 & 0.85 & 0.53 & 0.01 \\ 
  Residuals & 1350.77 & 583.00 & 2.32 &  &  &  \\
  \\[-1.8ex]
  \textbf{Support} & & & & & & \\ 
  Context & 7.80 & 2.00 & 3.90 & 1.47 & 0.23 & 0.01 \\ 
  Algorithm Type & 383.58 & 3.00 & 127.86 & 48.35 & 0.00 & 0.20 \\ 
  Context:Algorithm Type & 18.77 & 6.00 & 3.13 & 1.18 & 0.31 & 0.01 \\ 
  Residuals & 1541.58 & 583.00 & 2.64 &  &  &  \\ 
   \hline
\end{tabular}
    \caption{ANOVA tables for judgments of fairness, trust, objectivity, and support for algorithms deployed in the bail domain. Independent variables: experimental conditions.}
    \label{tab:bail_DV_experiment}
\end{table}

\begin{table}[ht]
\footnotesize
\centering
\begin{tabular}{lrrrrl}
  \hline
Contrast & Estimate & $SE$ & df & t-value & $p$ \\ 
  \hline
  \textbf{Fairness} & & & & & \\
Control - Affirmative & 0.8453 & 0.1908 & 583 & 4.429 & 0.0001 \\ 
  Control - Discriminatory & 1.0918 & 0.1894 & 583 & 5.764 & $<$.0001 \\ 
  Control - Fair & -1.1448 & 0.1882 & 583 & -6.084 & $<$.0001 \\ 
  Affirmative - Discriminatory & 0.2465 & 0.1868 & 583 & 1.320 & 1.0000 \\ 
  Affirmative - Fair & -1.9901 & 0.1855 & 583 & -10.730 & $<$.0001 \\ 
  Discriminatory - Fair & -2.2366 & 0.1840 & 583 & -12.154 & $<$.0001 \\
  \\[-1.8ex]
  \textbf{Trust} & & & & & \\
  Control - Affirmative & 0.5888 & 0.1873 & 583 & 3.143 & 0.0105 \\ 
  Control - Discriminatory & 0.9158 & 0.1859 & 583 & 4.926 & $<$.0001 \\ 
  Control - Fair & -0.8812 & 0.1847 & 583 & -4.772 & $<$.0001 \\ 
  Affirmative - Discriminatory & 0.3270 & 0.1833 & 583 & 1.784 & 0.4499 \\ 
  Affirmative - Fair & -1.4700 & 0.1820 & 583 & -8.075 & $<$.0001 \\ 
  Discriminatory - Fair & -1.7970 & 0.1806 & 583 & -9.950 & $<$.0001 \\
  \\[-1.8ex]
  \textbf{Objectivity} & & & & & \\
  Control - Affirmative & 0.9839 & 0.1837 & 583 & 5.355 & $<$.0001 \\ 
  Control - Discriminatory & 1.1754 & 0.1824 & 583 & 6.446 & $<$.0001 \\ 
  Control - Fair & -0.5719 & 0.1811 & 583 & -3.157 & 0.0101 \\ 
  Affirmative - Discriminatory & 0.1916 & 0.1798 & 583 & 1.066 & 1.0000 \\ 
  Affirmative - Fair & -1.5557 & 0.1786 & 583 & -8.713 & $<$.0001 \\ 
  Discriminatory - Fair & -1.7473 & 0.1771 & 583 & -9.864 & $<$.0001 \\
  \\[-1.8ex]
  \textbf{Support} & & & & & \\
  Control - Affirmative & 0.9009 & 0.1963 & 583 & 4.590 & $<$.0001 \\ 
  Control - Discriminatory & 1.2689 & 0.1948 & 583 & 6.514 & $<$.0001 \\ 
  Control - Fair & -0.8131 & 0.1935 & 583 & -4.202 & 0.0002 \\ 
  Affirmative - Discriminatory & 0.3680 & 0.1921 & 583 & 1.916 & 0.3351 \\ 
  Affirmative - Fair & -1.7140 & 0.1907 & 583 & -8.986 & $<$.0001 \\ 
  Discriminatory - Fair & -2.0820 & 0.1892 & 583 & -11.002 & $<$.0001 \\ 
   \hline
\end{tabular}
    \caption{Pairwise comparisons of judgments of fairness, trust, objectivity, and support between different types of algorithm deployed in the bail domain.}
    \label{tab:bail_DV_pairwise}
\end{table}

\begin{table}[ht]
\footnotesize
\centering
\begin{tabular}{lrrrrrr}
  \hline
Parameter & Sum of Squares & df & Mean Squares & $F$ & $p$ & partial $\eta^2$ \\ 
  \hline
\textbf{Fairness} & & & & & & \\ 
Political Leaning & 17.95 & 1.00 & 17.95 & 7.22 & 0.01 & 0.01 \\ 
  Algorithm Type & 451.53 & 3.00 & 150.51 & 60.55 & 0.00 & 0.24 \\ 
  Political Leaning:Algorithm Type & 13.92 & 3.00 & 4.64 & 1.87 & 0.13 & 0.01 \\ 
  Residuals & 1426.77 & 574.00 & 2.49 &  &  &  \\
  \\[-1.8ex]
\textbf{Trust} & & & & & & \\ 
  Political Leaning & 24.30 & 1.00 & 24.30 & 10.46 & 0.00 & 0.02 \\ 
  Algorithm Type & 265.99 & 3.00 & 88.66 & 38.15 & 0.00 & 0.17 \\ 
  Political Leaning:Algorithm Type & 17.08 & 3.00 & 5.69 & 2.45 & 0.06 & 0.01 \\ 
  Residuals & 1334.00 & 574.00 & 2.32 &  &  &  \\ 
  \\[-1.8ex]
\textbf{Objectivity} & & & & & & \\ 
  Political Leaning & 13.65 & 1.00 & 13.65 & 6.09 & 0.01 & 0.01 \\ 
  Algorithm Type & 286.66 & 3.00 & 95.55 & 42.60 & 0.00 & 0.18 \\ 
  Political Leaning:Algorithm Type & 15.98 & 3.00 & 5.33 & 2.37 & 0.07 & 0.01 \\ 
  Residuals & 1287.45 & 574.00 & 2.24 &  &  &  \\ 
  \\[-1.8ex]
\textbf{Support} & & & & & & \\ 
  Political Leaning & 9.59 & 1.00 & 9.59 & 3.64 & 0.06 & 0.01 \\ 
  Algorithm Type & 377.14 & 3.00 & 125.71 & 47.76 & 0.00 & 0.20 \\ 
  Political Leaning:Algorithm Type & 9.13 & 3.00 & 3.04 & 1.16 & 0.33 & 0.01 \\ 
  Residuals & 1510.83 & 574.00 & 2.63 &  &  &  \\ 
   \hline
\end{tabular}
    \caption{ANCOVA tables for judgments of fairness, trust, objectivity, and support for algorithms deployed in the bail domain. Independent variables: algorithm type and political orientation.}
    \label{tab:bail_DV_Political Leaning}
\end{table}

\begin{table}[ht]
\footnotesize
\centering
\begin{tabular}{lrrrrl}
  \hline
Contrast & Estimate & $SE$ & df & t-value & $p$ \\ 
  \hline
\textbf{Fairness} & & & & & \\
Control - Affirmative & -0.1863 & 0.1313 & 574 & -1.419 & 0.9384 \\ 
  Control - Discriminatory & 0.0929 & 0.1254 & 574 & 0.741 & 1.0000 \\ 
  Control - Fair & -0.1123 & 0.1279 & 574 & -0.878 & 1.0000 \\ 
  Affirmative - Discriminatory & 0.2792 & 0.1279 & 574 & 2.183 & 0.1769 \\ 
  Affirmative - Fair & 0.0741 & 0.1304 & 574 & 0.568 & 1.0000 \\ 
  Discriminatory - Fair & -0.2052 & 0.1245 & 574 & -1.648 & 0.5990 \\ 
  \\[-1.8ex]
\textbf{Trust} & & & & & \\
  Control - Affirmative & -0.1326 & 0.1270 & 574 & -1.044 & 1.0000 \\ 
  Control - Discriminatory & 0.1760 & 0.1212 & 574 & 1.451 & 0.8833 \\ 
  Control - Fair & -0.0771 & 0.1237 & 574 & -0.624 & 1.0000 \\ 
  Affirmative - Discriminatory & 0.3085 & 0.1237 & 574 & 2.494 & 0.0775 \\ 
  Affirmative - Fair & 0.0554 & 0.1261 & 574 & 0.439 & 1.0000 \\ 
  Discriminatory - Fair & -0.2531 & 0.1204 & 574 & -2.103 & 0.2155 \\ 
  \\[-1.8ex]
\textbf{Objectivity} & & & & & \\
  Control - Affirmative & -0.1715 & 0.1247 & 574 & -1.375 & 1.0000 \\ 
  Control - Discriminatory & 0.1259 & 0.1191 & 574 & 1.057 & 1.0000 \\ 
  Control - Fair & -0.1108 & 0.1215 & 574 & -0.912 & 1.0000 \\ 
  Affirmative - Discriminatory & 0.2975 & 0.1215 & 574 & 2.448 & 0.0881 \\ 
  Affirmative - Fair & 0.0607 & 0.1239 & 574 & 0.490 & 1.0000 \\ 
  Discriminatory - Fair & -0.2367 & 0.1182 & 574 & -2.002 & 0.2744 \\
  \\[-1.8ex]
\textbf{Support} & & & & & \\
  Control - Affirmative & -0.1667 & 0.1351 & 574 & -1.234 & 1.0000 \\ 
  Control - Discriminatory & 0.0654 & 0.1290 & 574 & 0.507 & 1.0000 \\ 
  Control - Fair & -0.0791 & 0.1316 & 574 & -0.601 & 1.0000 \\ 
  Affirmative - Discriminatory & 0.2321 & 0.1316 & 574 & 1.763 & 0.4706 \\ 
  Affirmative - Fair & 0.0876 & 0.1342 & 574 & 0.653 & 1.0000 \\ 
  Discriminatory - Fair & -0.1445 & 0.1281 & 574 & -1.128 & 1.0000 \\ 
   \hline
\end{tabular}
    \caption{Pairwise comparisons of the slopes of judgments of fairness, trust, objectivity, and support in relation to political orientation between different types of algorithm deployed in the bail domain.}
    \label{tab:bail_Political Leaning_pairwise}
\end{table}

\begin{table}[ht]
\footnotesize
\centering
\begin{tabular}{lrrrrrr}
  \hline
Parameter & Sum of Squares & df & Mean Squares & $F$ & $p$ & partial $\eta^2$ \\ 
  \hline
\textbf{Fairness} & & & & & & \\
Racial Group & 1.25 & 1.00 & 1.25 & 0.50 & 0.48 & 0.00 \\ 
  Algorithm Type & 463.52 & 3.00 & 154.51 & 62.53 & 0.00 & 0.24 \\ 
  Racial Group:Algorithm Type & 26.02 & 3.00 & 8.67 & 3.51 & 0.02 & 0.02 \\ 
  Residuals & 1450.35 & 587.00 & 2.47 &  &  &  \\ 
  \\[-1.8ex]
\textbf{Trust} & & & & & & \\
  Racial Group & 1.66 & 1.00 & 1.66 & 0.71 & 0.40 & 0.00 \\ 
  Algorithm Type & 274.30 & 3.00 & 91.43 & 38.93 & 0.00 & 0.17 \\ 
  Racial Group:Algorithm Type & 33.56 & 3.00 & 11.19 & 4.76 & 0.00 & 0.02 \\ 
  Residuals & 1378.52 & 587.00 & 2.35 &  &  &  \\
  \\[-1.8ex]
\textbf{Objectivity} & & & & & & \\
  Racial Group & 1.52 & 1.00 & 1.52 & 0.67 & 0.41 & 0.00 \\ 
  Algorithm Type & 299.55 & 3.00 & 99.85 & 44.17 & 0.00 & 0.18 \\ 
  Racial Group:Algorithm Type & 36.08 & 3.00 & 12.03 & 5.32 & 0.00 & 0.03 \\ 
  Residuals & 1326.96 & 587.00 & 2.26 &  &  &  \\ 
  \\[-1.8ex]
\textbf{Support} & & & & & & \\
  Racial Group & 1.18 & 1.00 & 1.18 & 0.45 & 0.50 & 0.00 \\ 
  Algorithm Type & 387.28 & 3.00 & 129.09 & 49.29 & 0.00 & 0.20 \\ 
  Racial Group:Algorithm Type & 29.46 & 3.00 & 9.82 & 3.75 & 0.01 & 0.02 \\ 
  Residuals & 1537.51 & 587.00 & 2.62 &  &  &  \\ 
   \hline
\end{tabular}
    \caption{ANOVA tables for judgments of fairness, trust, objectivity, and support for algorithms deployed in the bail domain. Independent variables: algorithm type and participants' racial group.}
    \label{tab:bail_DV_race}
\end{table}

\begin{table}[ht]
\footnotesize
\centering
\begin{tabular}{llrrrrl}
  \hline
Algorithm Type & Contrast & Estimate & $SE$ & df & t-value & $p$ \\ 
  \hline
\textbf{Fairness} & & & & & & \\
Control & (Non-White) - White & 0.2176 & 0.2755 & 587 & 0.790 & 0.4298 \\ 
  Affirmative & (Non-White) - White & 0.7900 & 0.2846 & 587 & 2.776 & 0.0057 \\ 
  Discriminatory & (Non-White) - White & -0.4254 & 0.2675 & 587 & -1.590 & 0.1124 \\ 
  Fair & (Non-White) - White & -0.1151 & 0.2751 & 587 & -0.419 & 0.6757 \\
  \\[-1.8ex]
\textbf{Trust} & & & & & & \\
  Control & (Non-White) - White & 0.1111 & 0.2686 & 587 & 0.414 & 0.6793 \\ 
  Affirmative & (Non-White) - White & 0.9411 & 0.2775 & 587 & 3.392 & 0.0007 \\ 
  Discriminatory & (Non-White) - White & -0.4738 & 0.2608 & 587 & -1.817 & 0.0698 \\ 
  Fair & (Non-White) - White & -0.0392 & 0.2682 & 587 & -0.146 & 0.8839 \\
  \\[-1.8ex]
\textbf{Objectivity} & & & & & & \\
  Control & (Non-White) - White & -0.1956 & 0.2635 & 587 & -0.742 & 0.4583 \\ 
  Affirmative & (Non-White) - White & 1.0595 & 0.2722 & 587 & 3.892 & 0.0001 \\ 
  Discriminatory & (Non-White) - White & -0.1722 & 0.2559 & 587 & -0.673 & 0.5012 \\ 
  Fair & (Non-White) - White & -0.1823 & 0.2631 & 587 & -0.693 & 0.4886 \\ 
  Control & (Non-White) - White & -0.0595 & 0.2837 & 587 & -0.210 & 0.8340 \\
  \\[-1.8ex]
\textbf{Support} & & & & & & \\
  Affirmative & (Non-White) - White & 0.9400 & 0.2930 & 587 & 3.208 & 0.0014 \\ 
  Discriminatory & (Non-White) - White & -0.2973 & 0.2755 & 587 & -1.079 & 0.2809 \\ 
  Fair & (Non-White) - White & -0.1255 & 0.2832 & 587 & -0.443 & 0.6579 \\ 
   \hline
\end{tabular}
    \caption{Pairwise comparisons of judgments of fairness, trust, objectivity, and support between different racial groups for each type of algorithm deployed in the bail domain.}
    \label{tab:bail_race_pairwise}
\end{table}

\begin{table}[ht]
\footnotesize
\centering
\begin{tabular}{lrrrrrr}
  \hline
Parameter & Sum of Squares & df & Mean Squares & $F$ & $p$ & partial $\eta^2$ \\ 
  \hline
\textbf{Fairness} & & & & & & \\
Gender Group & 6.60 & 1.00 & 6.60 & 2.64 & 0.10 & 0.00 \\ 
  Algorithm Type & 458.17 & 3.00 & 152.72 & 61.04 & 0.00 & 0.24 \\ 
  Gender Group:Algorithm Type & 2.31 & 3.00 & 0.77 & 0.31 & 0.82 & 0.00 \\ 
  Residuals & 1468.70 & 587.00 & 2.50 &  &  &  \\
  \\[-1.8ex]
\textbf{Trust} & & & & & & \\
  Gender Group & 9.52 & 1.00 & 9.52 & 3.99 & 0.05 & 0.01 \\ 
  Algorithm Type & 269.33 & 3.00 & 89.78 & 37.64 & 0.00 & 0.16 \\ 
  Gender Group:Algorithm Type & 4.18 & 3.00 & 1.39 & 0.58 & 0.63 & 0.00 \\ 
  Residuals & 1400.04 & 587.00 & 2.39 &  &  &  \\
  \\[-1.8ex]
\textbf{Objectivity} & & & & & & \\
  Gender Group & 10.78 & 1.00 & 10.78 & 4.70 & 0.03 & 0.01 \\ 
  Algorithm Type & 294.08 & 3.00 & 98.03 & 42.69 & 0.00 & 0.18 \\ 
  Gender Group:Algorithm Type & 5.86 & 3.00 & 1.95 & 0.85 & 0.47 & 0.00 \\ 
  Residuals & 1347.92 & 587.00 & 2.30 &  &  &  \\
  \\[-1.8ex]
\textbf{Support} & & & & & & \\
  Gender Group & 5.35 & 1.00 & 5.35 & 2.02 & 0.16 & 0.00 \\ 
  Algorithm Type & 382.35 & 3.00 & 127.45 & 48.01 & 0.00 & 0.20 \\ 
  Gender Group:Algorithm Type & 4.53 & 3.00 & 1.51 & 0.57 & 0.64 & 0.00 \\ 
  Residuals & 1558.27 & 587.00 & 2.65 &  &  &  \\ 
   \hline
\end{tabular}
    \caption{ANOVA tables for judgments of fairness, trust, objectivity, and support for algorithms deployed in the bail domain. Independent variables: algorithm type and participants' gender group.}
    \label{tab:bail_DV_gender}
\end{table}

\begin{table}[ht]
\footnotesize
\centering
\begin{tabular}{llrrrrl}
  \hline
Algorithm Type & Contrast & Estimate & $SE$ & df & t-value & $p$ \\ 
  \hline
\textbf{Fairness} & & & & & & \\
Control & Men - Non-Men & 0.2240 & 0.2666 & 587 & 0.840 & 0.4011 \\ 
  Affirmative & Men - Non-Men & 0.3919 & 0.2600 & 587 & 1.507 & 0.1323 \\ 
  Discriminatory & Men - Non-Men & 0.1924 & 0.2584 & 587 & 0.745 & 0.4567 \\ 
  Fair & Men - Non-Men & 0.0435 & 0.2549 & 587 & 0.171 & 0.8645 \\ 
  \\[-1.8ex]
\textbf{Trust} & & & & & & \\
  Control & Men - Non-Men & 0.0429 & 0.2603 & 587 & 0.165 & 0.8691 \\ 
  Affirmative & Men - Non-Men & 0.4459 & 0.2539 & 587 & 1.756 & 0.0795 \\ 
  Discriminatory & Men - Non-Men & 0.3874 & 0.2523 & 587 & 1.535 & 0.1253 \\ 
  Fair & Men - Non-Men & 0.1306 & 0.2489 & 587 & 0.525 & 0.6001 \\
  \\[-1.8ex]
\textbf{Objectivity} & & & & & & \\
  Control & Men - Non-Men & 0.0896 & 0.2554 & 587 & 0.351 & 0.7257 \\ 
  Affirmative & Men - Non-Men & 0.5541 & 0.2491 & 587 & 2.224 & 0.0265 \\ 
  Discriminatory & Men - Non-Men & 0.3571 & 0.2476 & 587 & 1.443 & 0.1497 \\ 
  Fair & Men - Non-Men & 0.0762 & 0.2442 & 587 & 0.312 & 0.7553 \\
  \\[-1.8ex]
\textbf{Support} & & & & & & \\
  Control & Men - Non-Men & 0.1775 & 0.2746 & 587 & 0.646 & 0.5183 \\ 
  Affirmative & Men - Non-Men & 0.4054 & 0.2679 & 587 & 1.514 & 0.1307 \\ 
  Discriminatory & Men - Non-Men & -0.0759 & 0.2662 & 587 & -0.285 & 0.7758 \\ 
  Fair & Men - Non-Men & 0.2534 & 0.2626 & 587 & 0.965 & 0.3350 \\ 
   \hline
\end{tabular}
    \caption{Pairwise comparisons of judgments of fairness, trust, objectivity, and support between different gender groups for each type of algorithm deployed in the hiring domain.}
    \label{tab:bail_gender_pairwise}
\end{table}

\section{Supplementary Analysis}
\label{supp:analysis}

We present several supplementary analyses. First, we discuss the perceived fairness of the decision-making domains and explore whether it was impacted by our context manipulations. Second, we replicate our analysis on racial differences by using more fine-grained racial categories. Finally, we present a complete analysis of participants' open-ended responses to who they deem marginalized.

\subsection{Perceived Fairness of the Decision-Making Domain}

Analyzing the perceived fairness of the decision-making domains, we found that participants on average considered both domains unfair. Participants at least somewhat agreed that hiring decisions were unfair with respect to its procedural (\msd{-0.54}{1.34}), distributional (\msd{-0.73}{1.61}), and interpersonal (\msd{-0.10}{1.51}) dimensions, as well as in a more general sense (\msd{-0.53}{1.52}). In comparison, bail decisions were considered even more procedurally (\msd{-0.72}{1.48}), distributionally (\msd{-0.80}{1.68}), interpersonally (\msd{-0.37}{1.49}), and generally (\msd{-0.69}{1.63}) unfair. 

Explicitly telling participants that hiring decisions are unjust against marginalized groups (i.e., the \emph{contextualized evaluation} context manipulation) only decreased perceived procedural fairness (\ftest{1}{401}{4.75}, \pvalue{.05}, \etasq{0.01}), having no impact on other judgments of fairness (see Tables~\ref{tab:hiring_pre} and~\ref{tab:hiring_pre_pairwise}). On the other hand, Tables~\ref{tab:bail_pre} and \ref{tab:bail_pre_pairwise} shows that being explicit about injustice in bail decision-making made participants consider algorithm more procedurally (\ftest{1}{400}{11.12}, \pvalue{.001}, \etasq{0.03}), distributionally (\ftest{1}{400}{9.41}, \pvalue{.01}, \etasq{0.02}), and interpersonally (\ftest{1}{400}{14.81}, \pvalue{.001}, \etasq{0.04}) unfair. We also found consistent results with respect to general fairness (\ftest{1}{400}{12.39}, \pvalue{.001}, \etasq{0.03}).

\begin{table}[ht]
\footnotesize
\centering
\begin{tabular}{lrrrrrr}
  \hline
Parameter & Sum of Squares & df & Mean Squares & $F$ & $p$ & $\eta^2$ \\ 
  \hline
  \textbf{Procedural Fairness} & & & & & & \\ 
Context & 8.43 & 1.00 & 8.43 & 4.75 & 0.03 & 0.01 \\ 
  Residuals & 710.98 & 401.00 & 1.77 &  &  &  \\ 
  \\[-1.8ex]
  \textbf{Distributional Fairness} & & & & & & \\ 
  Context & 0.03 & 1.00 & 0.03 & 0.01 & 0.91 & 0.00 \\ 
  Residuals & 1039.42 & 401.00 & 2.59 &  &  &  \\
  \\[-1.8ex]
  \textbf{Interpersonal Fairness} & & & & & & \\ 
  Context & 2.41 & 1.00 & 2.41 & 1.06 & 0.30 & 0.00 \\ 
  Residuals & 913.66 & 401.00 & 2.28 &  &  &  \\ 
  \\[-1.8ex]
  \textbf{General Fairness} & & & & & & \\ 
  Context & 4.74 & 1.00 & 4.74 & 2.06 & 0.15 & 0.01 \\ 
  Residuals & 923.74 & 401.00 & 2.30 &  &  &  \\ 
   \hline
\end{tabular}
    \caption{ANOVA tables for judgments of fairness of hiring decisions. Independent variable: context manipulation.}
    \label{tab:hiring_pre}
\end{table}

\begin{table}[ht]
\footnotesize
\centering
\begin{tabular}{lrrrrl}
  \hline
Contrast & Estimate & $SE$ & df & t-value & $p$ \\ 
  \hline
  \textbf{Procedural Fairness} & & & & & \\ 
Decontextualized Evaluation - Contextualized Evaluation & 0.2893 & 0.1327 & 401 & 2.180 & 0.0298 \\
  \\[-1.8ex]
  \textbf{Distributional Fairness} & & & & & \\
  Decontextualized Evaluation - Contextualized Evaluation & 0.0178 & 0.1604 & 401 & 0.111 & 0.9118 \\ 
\\[-1.8ex]
  \textbf{Interpersonal Fairness} & & & & &  \\
  Decontextualized Evaluation - Contextualized Evaluation & 0.1547 & 0.1504 & 401 & 1.029 & 0.3042 \\
    \\[-1.8ex]
  \textbf{General Fairness} & & & & & \\
  Decontextualized Evaluation - Contextualized Evaluation & 0.2169 & 0.1512 & 401 & 1.434 & 0.1524 \\ 
   \hline
\end{tabular}
    \caption{Pairwise comparisons of judgments of fairness of hiring decisions between different context manipulations.}
    \label{tab:hiring_pre_pairwise}
\end{table}

\begin{table}[ht]
\footnotesize
\centering
\begin{tabular}{lrrrrrr}
  \hline
Parameter & Sum of Squares & df & Mean Squares & $F$ & $p$ & $\eta^2$ \\ 
  \hline
  \textbf{Procedural Fairness} & & & & & \\ 
Context & 23.62 & 1.00 & 23.62 & 11.12 & 0.00 & 0.03 \\ 
  Residuals & 850.10 & 400.00 & 2.13 &  &  &  \\ 
  \\[-1.8ex]
  \textbf{Distributional Fairness} & & & & & \\
  Context & 26.19 & 1.00 & 26.19 & 9.41 & 0.00 & 0.02 \\ 
  Residuals & 1112.99 & 400.00 & 2.78 &  &  &  \\
    \\[-1.8ex]
  \textbf{Interpersonal Fairness} & & & & & \\
  Context & 31.90 & 1.00 & 31.90 & 14.81 & 0.00 & 0.04 \\ 
  Residuals & 861.23 & 400.00 & 2.15 &  &  &  \\
    \\[-1.8ex]
  \textbf{General Fairness} & & & & & \\
  Context & 31.97 & 1.00 & 31.97 & 12.39 & 0.00 & 0.03 \\ 
  Residuals & 1032.53 & 400.00 & 2.58 &  &  &  \\ 
   \hline
\end{tabular}
   \caption{ANOVA tables for judgments of fairness of bail decisions. Independent variable: context manipulation.}
    \label{tab:bail_pre}
\end{table}

\begin{table}[ht]
\footnotesize
\centering
\begin{tabular}{lrrrrl}
  \hline
Contrast & Estimate & $SE$ & df & t-value & $p$ \\ 
  \hline
  \textbf{Procedural Fairness} & & & & \\
Decontextualized Evaluation - Contextualized Evaluation & 0.4848 & 0.1454 & 400 & 3.334 & 0.0009 \\
  \\[-1.8ex]
  \textbf{Distributional Fairness} &  & & & \\
  Decontextualized Evaluation - Contextualized Evaluation & 0.5104 & 0.1664 & 400 & 3.068 & 0.0023 \\
    \\[-1.8ex]
  \textbf{Interpersonal Fairness} & & & & \\
  Decontextualized Evaluation - Contextualized Evaluation & 0.5634 & 0.1464 & 400 & 3.849 & 0.0001 \\
    \\[-1.8ex]
  \textbf{General Fairness} & & & & \\
  Decontextualized Evaluation - Contextualized Evaluation & 0.5641 & 0.1603 & 400 & 3.519 & 0.0005 \\ 
   \hline\
\end{tabular}
    \caption{Pairwise comparisons of judgments of fairness of bail decisions between different context manipulations.}
    \label{tab:bail_pre_pairwise}
\end{table}

\subsection{Fine-Grained Racial Groups}

\begin{figure}[t!]
    \centering
    \includegraphics[width=\linewidth]{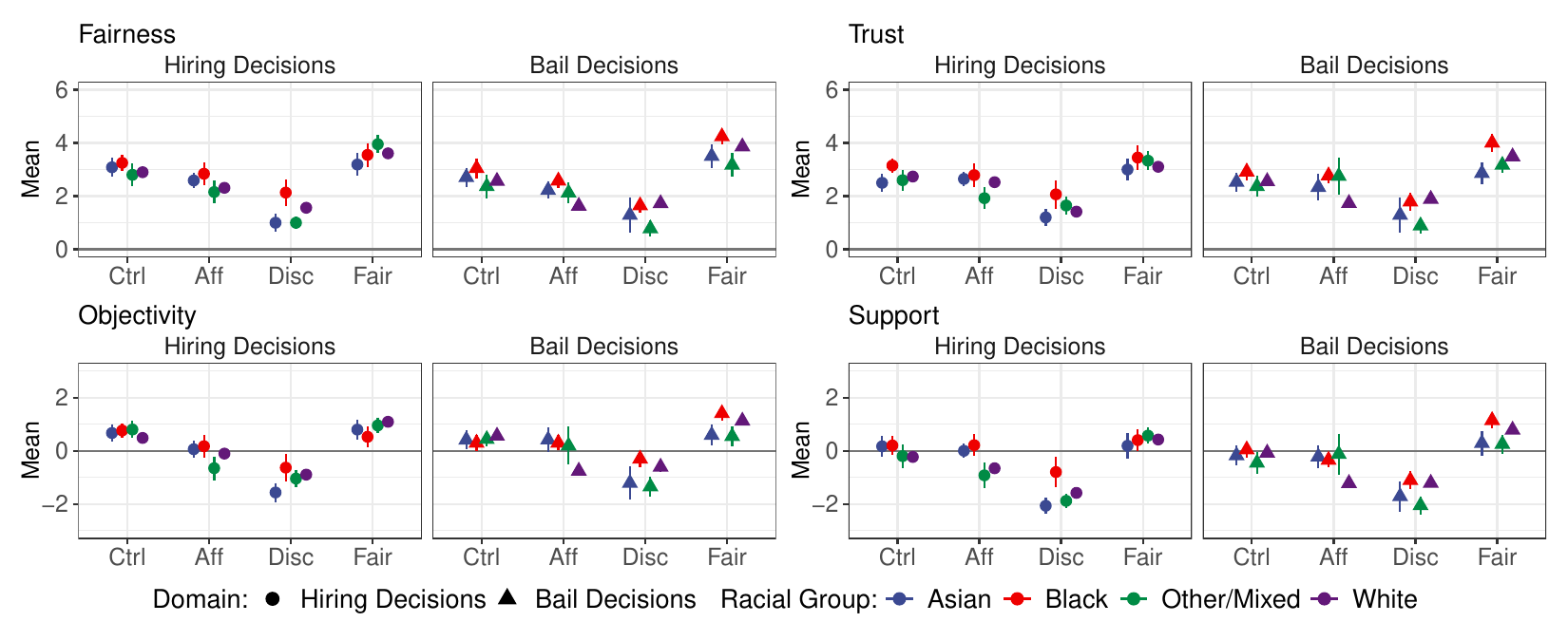}
    \caption{Participants' mean judgments of fairness, trust, objectivity, support concerning different types of algorithms depending on participant's self-reported race. The x-axis refers to the \emph{algorithm type} experimental manipulations: control (Ctrl), affirmative (Aff), discriminatory (Disc), and fair (Fair).}
    \label{fig:algorithm_multiple_race}
    \Description{
    This figure presents perceived fairness, trust, objectivity, and support across two domains: Hiring Decisions and Bail Decisions. The x-axis depicts four algorithm conditions: Control, Affirmative, Discriminatory, and Fair, while the y-axis indicates mean values for each metric. The data is segmented by racial groups, including Asian, Black, Other/Mixed, and White, emphasizing differences in perceptions based on racial identity across algorithmic conditions.}
\end{figure}

Our analysis grouped participants who did not self-identify as White for better statistical power given our small number of participants from racial minorities. Although our approach is in line with findings from prior work on racial injustice and affirmative action~\cite{clawson2003support,klineberg2003ethnic,golden2001reactions,allen2003examining, kraus2017americans,kraus2022framing,kuo2020high,goldsmith2004schools}, we acknowledge that not all groups under the ``Non-White'' group have the same experiences with and opinions about the criminal justice domain. The same applies for our category of Non-Men, which also includes non-binary and trans individuals.

Figure~\ref{fig:algorithm_multiple_race} shows judgments of the algorithms after grouping participants as White (66.97\%), Asian (8.97\%), Black (14.42\%) and Other/Mixed (including all others respondents). We do not present ANOVA tables because most interactions between racial group and algorithm type are either non-significant or only borderline significant due to our imbalanced sample. Nonetheless, we see that that Asian participants' views of the affirmative algorithm are closer to opinions of Black individuals than White participants. We also observe that Black participants are relatively more critical of discriminatory hiring algorithm.

Given the racial imbalance of our samples, we call for future work to recruit participants strategically to study how opinions about affirmative algorithms vary across racial groups. Similarly, we also propose that future work could explore the opinions of trans and non-binary individuals concerning affirmative algorithm, which is currently not possible given our low number of participants ($n$ = 48, 4.02\%).

\subsection{Analysis of Participants' Beliefs of Who is Marginalized}

The first author identified groups of people that participants considered marginalized as they emerged during the coding process. The groups were first identified in the hiring domain and then applied to responses in the bail study. While coding the bail study, the coder also observed that ten responses could have been categorized using other groups (e.g., ``democrats and republicans'' as political groups), but chose to not create new categories for conciseness. The following groups emerged from participants' open-ended responses:

\begin{itemize}
    \item \textbf{Racial Groups (81.6\%): } Mentions to racial groups, such as African Americans, Asian Americans, Native Americans, as well as ethnic minorities, like Jewish individuals.
    \item \textbf{Cisgender Groups (23.1\%): } References to men and women, i.e., cisgender binary individuals.
    \item \textbf{Low-Income Groups (15.6\%): } Mentions to ``poor people,'' ``the poor,'' ''lower class,'' and other similar groups.
    \item \textbf{LGTBQ+ (12.8\%): } References to non-binary and trans individuals and sexual orientation minorities.
    \item \textbf{Immigrants (8.9\%): } Mentions to immigrants, such as Mexicans and "immigrants" in a more general sense.
    \item \textbf{Minorities (7.6\%): } Broader references to minority groups, without explicitly mentioning specific identity axes or groups.
    \item \textbf{Disability Groups (5.4\%): } Mentions to individuals with mental and physical disabilities.
    \item \textbf{Religious Groups (2.3\%): } References to individuals who follows specific religions, e.g., ``Muslims'' and ``Christians.''
    \item \textbf{Age Groups (1.6\%): } Mentions to both older adults and children.
    \item \textbf{Education Attainment (0.9\%): } References to people with low educational level.
    \item \textbf{Pregnant (0.3\%): } Mentions to pregnant people.
    \item \textbf{Criminal History (0.3\%): } References to individuals with prior criminal history.
    \item \textbf{Refugees (0.3\%): } Mentions to refugees.    
\end{itemize}

Table~\ref{tab:marginalized} presents the distributions of each groups between the studies, as well as chi-square tests exploring differences between the two experiments. The significant results are discussed in the main text. Finally, Tables~\ref{tab:hiring_marginalized} and~\ref{tab:bail_marginalized} present logistic regressions modeling the likelihood that specific groups were mentioned depending on participants' political leaning and identity---while controlling for experimental assignment---as discussed in the main text.

\begin{table}
\footnotesize
\centering
\begin{tabular}{l|lrr|l}
                                      &               & Hiring Decision-Making Study & Bail Decision-Making Study & $\chi^2$-test                             \\
                                      \hline
\multirow{2}{*}{Racial Groups}        & Mentioned     & 473                          & 501                        & \multirow{2}{*}{$x^2$(1) = 4.851, $p$ =  0.028}  \\
                                      & Not Mentioned & 125                          & 94                         &                                             \\
\\[-1.8ex]\multirow{2}{*}{Cisgender Groups}        & Mentioned     & 183                          & 92                         & \multirow{2}{*}{$x^2$(1) = 37.692, $p $ \textless .001}   \\
                                      & Not Mentioned & 415                          & 503                        &                                             \\
\\[-1.8ex]\multirow{2}{*}{Low-Income Groups}    & Mentioned     & 62                           & 124                        & \multirow{2}{*}{$x^2$(1) = 24.066, $p $ \textless .001}   \\
                                      & Not Mentioned & 536                          & 471                        &                                             \\
\\[-1.8ex]\multirow{2}{*}{LGBTQ+}               & Mentioned     & 92                           & 61                         & \multirow{2}{*}{$x^2$(1) = 6.576, $p$ =  0.010}  \\
                                      & Not Mentioned & 506                          & 534                        &                                             \\
\\[-1.8ex]\multirow{2}{*}{Immigrants}           & Mentioned     & 55                           & 51                         & \multirow{2}{*}{$x^2$(1) = 0.077, $p$ =  0.781}  \\
                                      & Not Mentioned & 543                          & 544                        &                                             \\
\\[-1.8ex]\multirow{2}{*}{Minorities}           & Mentioned     & 54                           & 37                         & \multirow{2}{*}{$x^2$(1) = 2.959, $p$ =  0.085}  \\
                                      & Not Mentioned & 544                          & 558                        &                                             \\
\\[-1.8ex]\multirow{2}{*}{Disability Groups}    & Mentioned     & 36                           & 29                         & \multirow{2}{*}{$x^2$(1) = 0.554, $p$ =  0.457}  \\
                                      & Not Mentioned & 562                          & 566                        &                                             \\
\\[-1.8ex]\multirow{2}{*}{Religious Groups}     & Mentioned     & 11                           & 17                         & \multirow{2}{*}{$x^2$(1) = 0.554, $p$ =  0.457}  \\
                                      & Not Mentioned & 587                          & 578                        &                                             \\
\\[-1.8ex]\multirow{2}{*}{Age Groups}           & Mentioned     & 11                           & 8                          & \multirow{2}{*}{$x^2$(1) = 0.940, $p$ =  0.332}  \\
                                      & Not Mentioned & 587                          & 587                        &                                             \\
\\[-1.8ex]\multirow{2}{*}{Education Attainment} & Mentioned     & 8                            & 1                          & \multirow{2}{*}{$x^2$(1) = 0.204, $p$ =  0.652}  \\
                                      & Not Mentioned & 590                          & 594                        &                                             \\
\\[-1.8ex]\multirow{2}{*}{Pregnancy}            & Mentioned     & 3                            & 1                          & \multirow{2}{*}{$x^2$(1) = 0.246, $p$ =  0.620}  \\
                                      & Not Mentioned & 595                          & 594                        &                                             \\
\\[-1.8ex]\multirow{2}{*}{Criminal History}     & Mentioned     & 2                            & 2                          & \multirow{2}{*}{$x^2$(1) = 0.000, $p$ =  1.000}  \\
                                      & Not Mentioned & 597                          & 593                        &                                             \\
\\[-1.8ex]\multirow{2}{*}{Refugees}             & Mentioned     & 1                            & 2                          & \multirow{2}{*}{$x^2$(1) = 0.000, $p$ =  1.000}  \\
                                      & Not Mentioned & 597                          & 593                        & \\ 
                                      \hline
\end{tabular}
\caption{Groups mentioned as marginalized in each experiment, alongside $\chi^2$ tests comparing the distributions between the two studies.}
\label{tab:marginalized}
\end{table}

\begin{table}[!htbp] \centering 
\footnotesize
\begin{tabular}{@{\extracolsep{5pt}}lcccc} 
\\[-1.8ex]\hline 
\hline \\[-1.8ex] 
 & \multicolumn{4}{c}{\textit{Logit(Group Mentioned)}} \\ 
\cline{2-5} 
\\[-1.8ex] & Racial Groups & Cisgender Groups & Low-Income Groups & LGBTQ+ \\ 
\\[-1.8ex] & (1) & (2) & (3) & (4)\\ 
\hline \\[-1.8ex] 
 Political Leaning & 0.461$^{***}$ & 0.415$^{***}$ & $-$0.058 & 0.745$^{***}$ \\ 
  & (0.081) & (0.078) & (0.101) & (0.122) \\ 

 Racial Group = White & 0.138 & 0.364 & 0.339 & 0.331 \\ 
  & (0.233) & (0.213) & (0.314) & (0.270) \\ 

 Gender Group = Non-Men & $-$0.103 & 1.129$^{***}$ & 0.015 & 1.088$^{***}$ \\ 
  & (0.217) & (0.202) & (0.276) & (0.270) \\ 

 Algorithm Type = Affirmative & 0.657 & $-$0.338 & $-$2.080 & $-$0.186 \\ 
  & (0.492) & (0.467) & (1.090) & (0.516) \\ 

 Algorithm Type = Discriminatory & 0.088 & $-$0.671 & $-$0.723 & $-$0.725 \\ 
  & (0.483) & (0.526) & (0.727) & (0.599) \\ 

 Algorithm Type = Fair & 0.548 & 0.233 & $-$0.188 & $-$0.651 \\ 
  & (0.485) & (0.475) & (0.628) & (0.625) \\ 

 Context = Decontextualized Evaluation & 0.603 & 0.812 & $-$0.168 & 0.029 \\ 
  & (0.500) & (0.466) & (0.628) & (0.554) \\ 

 Context = Contextualized Evaluation & 0.641 & $-$0.384 & 0.360 & $-$1.155 \\ 
  & (0.495) & (0.482) & (0.552) & (0.623) \\ 

 Algorithm Type = Affirmative:Context = Decontextualized Evaluation & 0.720 & $-$0.756 & 1.452 & 0.018 \\ 
  & (0.871) & (0.669) & (1.332) & (0.762) \\ 

 Algorithm Type = Discriminatory:Context = Decontextualized Evaluation & $-$0.637 & 0.473 & 0.987 & $-$0.148 \\ 
  & (0.691) & (0.688) & (0.951) & (0.845) \\ 

 Algorithm Type = Fair:Context = Decontextualized Evaluation & $-$0.963 & $-$0.841 & $-$0.134 & $-$0.055 \\ 
  & (0.697) & (0.653) & (0.919) & (0.852) \\ 

 Algorithm Type = Affirmative:Context = Contextualized Evaluation & 0.161 & 0.952 & 1.245 & $-$0.013 \\ 
  & (0.822) & (0.675) & (1.265) & (0.898) \\ 

 Algorithm Type = Discriminatory:Context = Contextualized Evaluation & 0.144 & 1.121 & 0.179 & 1.413 \\ 
  & (0.732) & (0.709) & (0.925) & (0.880) \\ 

 Algorithm Type = Fair:Context = Contextualized Evaluation & $-$0.746 & $-$0.205 & $-$0.539 & 1.085 \\ 
  & (0.711) & (0.691) & (0.869) & (0.922) \\ 

 Constant & 0.648 & $-$2.024$^{***}$ & $-$2.073$^{***}$ & $-$2.753$^{***}$ \\ 
  & (0.370) & (0.390) & (0.491) & (0.479) \\ 

\hline \\[-1.8ex] 
Observations & 588 & 588 & 588 & 588 \\ 
\hline 
\hline \\[-1.8ex]  
\end{tabular} 
  \caption{Logistic regressions of the likelihood that groups were mentioned as marginalized in the hiring decision-making study depending on participants' political leaning and racial/gender groups. We also control for potential effects of experimental conditions. $^{*}$p$<$0.05; $^{**}$p$<$0.01; $^{***}$p$<$0.001} 
  \label{tab:hiring_marginalized} 
\end{table}

\begin{table}[!htbp] \centering 
\footnotesize
\begin{tabular}{@{\extracolsep{5pt}}lcccc} 
\\[-1.8ex]\hline 
\hline \\[-1.8ex] 
 & \multicolumn{4}{c}{\textit{Logit(Group Mentioned)}} \\ 
\cline{2-5} 
\\[-1.8ex] & Racial Groups & Cisgender Groups & Low-Income Groups & LGBTQ+ \\ 
\\[-1.8ex] & (1) & (2) & (3) & (4)\\ 
\hline \\[-1.8ex] 
 Political Leaning & 0.423$^{***}$ & 0.445$^{***}$ & $-$0.020 & 0.433$^{***}$ \\ 
  & (0.084) & (0.097) & (0.072) & (0.113) \\ 

 Racial Group = White & $-$0.604$^{*}$ & 0.178 & 0.323 & $-$0.012 \\ 
  & (0.283) & (0.269) & (0.233) & (0.303) \\ 

 Gender Group = Non-Men & 0.041 & 0.633$^{*}$ & 0.078 & 0.255 \\ 
  & (0.241) & (0.250) & (0.211) & (0.284) \\ 

 Algorithm Type = Affirmative & 0.845 & 0.021 & 0.584 & $-$0.130 \\ 
  & (0.632) & (0.617) & (0.606) & (0.817) \\ 

 Algorithm Type = Discriminatory & 0.149 & $-$0.388 & 0.425 & $-$0.471 \\ 
  & (0.528) & (0.606) & (0.585) & (0.812) \\ 

 Algorithm Type = Fair & 0.490 & $-$1.131 & 0.216 & $-$0.010 \\ 
  & (0.553) & (0.676) & (0.599) & (0.759) \\ 

 Context = Decontextualized Evaluation & 0.100 & $-$0.263 & 0.363 & $-$0.274 \\ 
  & (0.537) & (0.597) & (0.593) & (0.781) \\ 

 Context = Contextualized Evaluation & 0.722 & $-$0.645 & 1.213$^{*}$ & 0.395 \\ 
  & (0.605) & (0.653) & (0.575) & (0.747) \\ 

 Algorithm Type = Affirmative:Context = Decontextualized Evaluation & $-$0.475 & $-$0.394 & $-$0.548 & 0.285 \\ 
  & (0.806) & (0.814) & (0.762) & (1.042) \\ 

 Algorithm Type = Discriminatory:Context = Decontextualized Evaluation & 0.791 & $-$0.612 & $-$0.605 & 0.646 \\ 
  & (0.887) & (0.933) & (0.810) & (1.082) \\ 

 Algorithm Type = Fair:Context = Decontextualized Evaluation & $-$0.351 & 1.473 & 0.103 & $-$0.440 \\ 
  & (0.774) & (0.872) & (0.781) & (1.157) \\ 

 Algorithm Type = Affirmative:Context = Contextualized Evaluation & $-$0.438 & 0.571 & $-$2.526$^{**}$ & 0.538 \\ 
  & (0.946) & (0.879) & (0.909) & (1.017) \\ 

 Algorithm Type = Discriminatory:Context = Contextualized Evaluation & $-$0.394 & 0.376 & $-$0.773 & $-$0.015 \\ 
  & (0.796) & (0.848) & (0.723) & (1.013) \\ 

 Algorithm Type = Fair:Context = Contextualized Evaluation & $-$0.588 & 1.201 & $-$1.443 & $-$0.448 \\ 
  & (0.827) & (0.902) & (0.775) & (0.973) \\ 

 Constant & 1.614$^{***}$ & $-$2.049$^{***}$ & $-$2.025$^{***}$ & $-$2.536$^{***}$ \\ 
  & (0.461) & (0.534) & (0.521) & (0.673) \\ 

\hline \\[-1.8ex] 
Observations & 582 & 582 & 582 & 582 \\ 

\hline \\[-1.8ex] 
Observations & 588 & 588 & 588 & 588 \\ 
\hline 
\hline \\[-1.8ex] 
\end{tabular} 
  \caption{Logistic regressions of the likelihood that groups were mentioned as marginalized in the bail decision-making study depending on participants' political leaning and racial/gender groups. We also control for potential effects of experimental conditions. $^{*}$p$<$0.05; $^{**}$p$<$0.01; $^{***}$p$<$0.001} 
  \label{tab:bail_marginalized} 
\end{table}

\end{document}